# GWTC-4.0: Methods for Identifying and Characterizing Gravitational-wave Transients

The LIGO Scientific Collaboration, the Virgo Collaboration, and the KAGRA Collaboration
(See the end matter for the full list of authors)

## ABSTRACT

The Gravitational-Wave Transient Catalog (GWTC) is a collection of candidate gravitational-wave transient signals identified and characterized by the LIGO–Virgo–KAGRA Collaboration. Producing the contents of the GWTC from detector data requires complex analysis methods. These comprise techniques to model the signal; identify the transients in the data; evaluate the quality of the data and mitigate possible instrumental issues; infer the parameters of each transient; compare the data with the waveform models for compact binary coalescences; and handle the large amount of results associated with all these different analyses. In this paper, we describe the methods employed to produce the catalog's fourth release, GWTC-4.0, focusing on the analysis of the first part of the fourth observing run of Advanced LIGO, Advanced Virgo and KAGRA.

*Keywords:* Gravitational wave astronomy (675); Gravitational wave detectors (676); Gravitational wave sources (677); Stellar mass black holes (1611); Neutron stars (1108)

## 1. INTRODUCTION

Interferometric gravitational-wave (GW) detectors produce a calibrated discrete digital time series $h(t)$ known as the *strain* (a dimensionless measure of the relative difference in arm length of the interferometers). The Advanced Laser Interferometer Gravitational-Wave Observatory (LIGO; Aasi et al. 2015) and Advanced Virgo (Acernese et al. 2015) detectors are the most sensitive to date. Alongside the developing KAGRA detector (Akutsu et al. 2019) the LIGO–Virgo–KAGRA Collaboration (LVK) has recently undertaken the fourth observing run (O4) observing run. However, the data produced are dominated by detector noise, with only occasional occurrences of detectable transient GW signals (Abbott et al. 2020a,b); to date, all such observed signals likely arise from compact binary coalescences (CBCs) involving black holes (BHs) and neutron stars (NSs). This paper describes the methodology used to analyze the calibrated strain data up to the first part of the fourth observing run (O4a) and produce version 4.0 of the Gravitational Wave Transient Catalog (GWTC), hereafter referred to as GWTC-4.0. For a general introduction to GWTC-4.0, see Abac et al. (2025a) which also contains a description of the observed source classes and data analysis nomenclature which should be read as a background to this methodology paper. The scientific results of GWTC-4.0 are presented in Abac et al. (2025b). Data analysis methods not directly related to producing catalog results, such as searches for continuous GWs or subsolar-mass CBCs, will be described elsewhere.

There are many interconnected elements to the data processing methodology described in this work. To provide a visual guide and summary, Figure 1 shows a diagram of the data-processing workflow. In Section 2, we introduce the fundamental concepts behind modeling GW waveforms from CBC sources and describe the waveform approximants used in later analyses. The data analysis process then starts with the calibrated strain data $h(t)$ and associated auxiliary data, which is produced by the LVK detectors and is the input to the analysis. The strain data and auxiliary data are inputs to the process of searching for signals and compiling a list of candidates from the strain data which we describe in Section 3. Section 4 discusses how we assess data quality around candidates and mitigate the impact of potential instrumental issues. The strain data, candidate lists, and data-quality information are then inputs into the methods used to infer the properties of the signals and their sources which we describe in Section 5. In Section 6 we detail consistency tests performed on selected candidates to evaluate how well CBC waveform models match the data. Section 7 describes the technologies used to manage the flow of information throughout the analysis process as shown in Figure 1. Finally, we conclude in Section 8.

## 2. MODELING COMPACT BINARY COALESCING SIGNALS

In this section, we will describe the modeling of CBC waveform signals, including binary black hole (BBH), binary

Corresponding author: LSC P&P Committee, via LVK Publications as proxy
lvc.publications@ligo.org

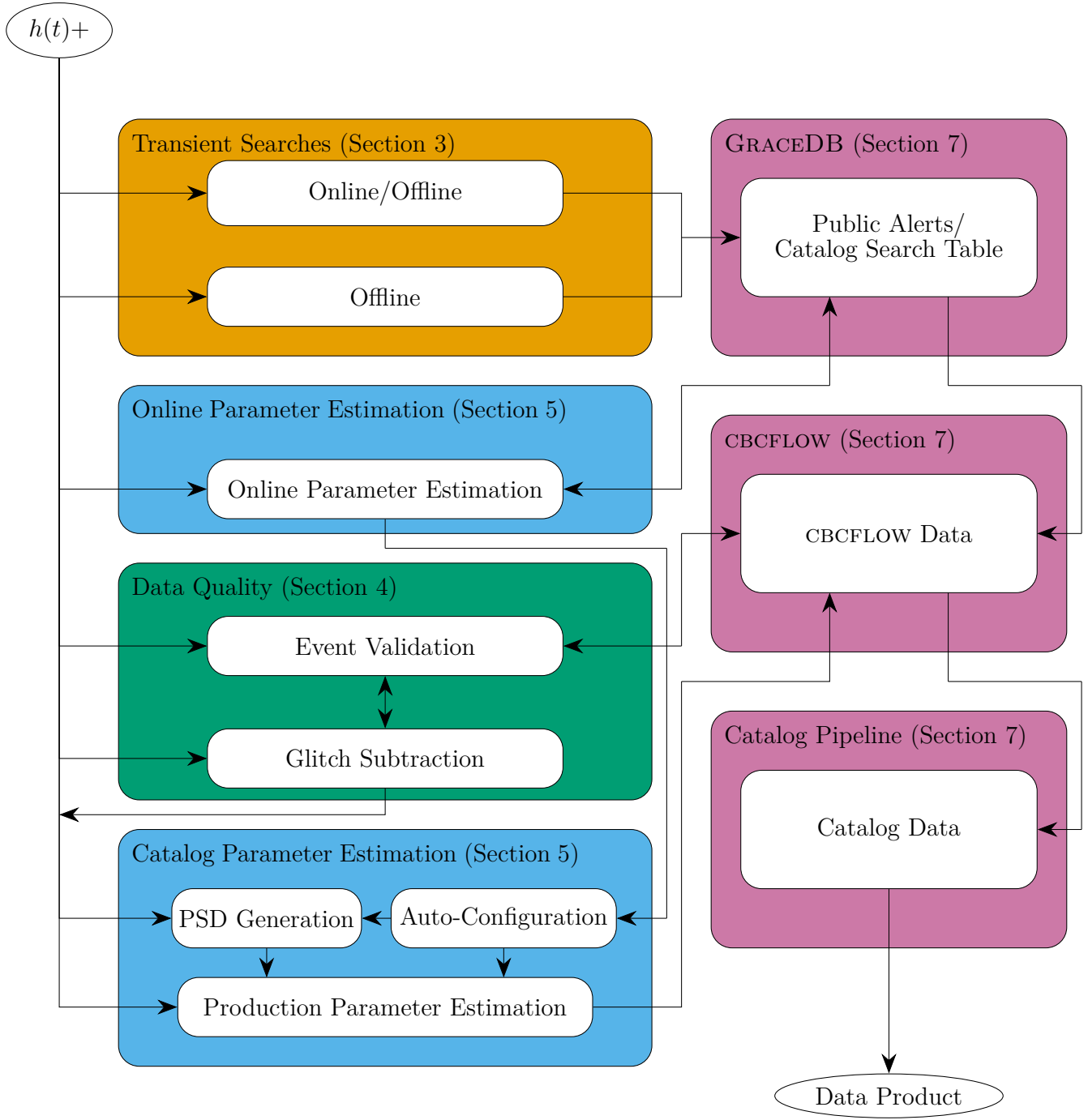

**Figure 1.** A high-level rendering of the flow of data from the strain and auxiliary data, denoted $h(t)+$, to the production of GWTC-4.0 described in this work. In this work, we use the terms *downstream* and *upstream* to refer to the flow of information in the analysis process (e.g. the parameter estimation is downstream of the compilation of search results since it depends on their outputs). This complex process enables the use of the most complete set of information for the final results while also leveraging preliminary studies to parallelise and reduce the overall analysis time. The term *online* (also referred to as low-latency in the literature) refers to analyses run on live data with a goal to upload candidates to GRACEDB immediately and enable rapid public alerts, while *offline* refers to analyses run with the goal to identify candidates for the GWTC. In each box, we provide references to the sections of this paper where the methods are explained in more detail.

neutron star (BNS), and neutron star–black hole binary (NSBH) systems. This modeling is a crucial aspect for the detection and astrophysical interpretation of these systems and is used by modeled search algorithms (described in Section 3) and parameter estimation (PE; described in Section 5). The focus of this section is on the waveform models used in GWTC-4.0, although we include models employed in analyses of previous GWTC versions that have been superseded in newer versions, in particular the PE analyses performed for GWTC-1.0 and GWTC-2.0. For an introduction to waveform modeling in general, see Section 5 of Abac et al. (2025a).

Several different modeling approaches have been followed in the development of CBC waveform models for the complete inspiral–merger–ringdown (IMR) stages of the signal. Inspiral-only models have been developed based on post-Newtonian (PN) theory (Blanchet 2014), using different PN-expansions of the balance equations for the two-body dynamics, under the TAYLOR family (Buonanno et al. 2009) for non-spinning systems and the SPINTAYLOR family (Sturani 2015; Isoyama et al. 2020) for spinning systems. The IMR-PHENOM approach (Ajith et al. 2007) focuses on the description of the GW signal, traditionally in Fourier domain for an efficient implementation in the data analysis pipelines, combining PN and numerical relativity (NR) information into closed-form expressions for describing the inspiral, merger and ringdown stages of the signal. The effective-one-body (EOB) approach (Buonanno & Damour 1999, 2000) focuses on accurately describing the dynamics and resulting waveform of the system in the time-domain, using a combination of resummed analytical information and calibration to NR. Inside this approach, two main different development effort have been followed, the SEOBNR (Buonanno et al. 2007) and TEOB (Damour & Nagar 2014; Nagar & Shah 2016) approaches. The NRSURROGATE approach (Blackman et al. 2017a) focuses on producing efficient surrogate models of the NR data, with or without hybridization with analytical waveforms, to deliver highly accurate waveform in a validity region limited by the input numerical data, both in the number of cycles and in the coverage of parameter space. These modeling approaches provide the waveform models described in this section.

The majority of waveform models that we describe in this section correspond to CBC systems on quasi-circular orbits (also called quasi-spherical orbits when spin precession is present), therefore neglecting orbital eccentricity effects, since at the moment all models employed in the analysis of GWTC candidates has this restriction. One of the main reasons for this limitation is that mature eccentric waveform models have only been developed recently, and reviewed versions were not yet ready at the time of starting the analyses described in this work. This limitation can lead to biases in the mass parameters (Martel & Poisson 1999; Lower et al. 2018; Lenon et al. 2020; O'Shea & Kumar 2023; Favata et al. 2022), and also influence the inferred spins (Lenon et al. 2020; Romero-Shaw et al. 2020a; O'Shea & Kumar 2023; Favata et al. 2022; Morras et al. 2025a,b). For high-mass systems, where few orbits are observed, neglecting eccen-tricity can also lead to incorrectly identifying spin-precessing effects due to possible degeneracies of both effects in this regime (Ramos-Buades et al. 2020a; Calderón Bustillo et al. 2021; Romero-Shaw et al. 2023). During a binary inspiral, the energy loss from GW emission rapidly circularizes an eccentric orbit, generally reducing expectation of significant eccentricity (Peters 1964) by the time the signal enters the sensitive frequency band of the interferometers (Tucker & Will 2021). Rate estimations (Wen 2003; Samsing & Ramirez-Ruiz 2017; Gupte et al. 2024) constrain the fraction of events with observable eccentricity to a small percentage. Nevertheless, recent independent analyses have shown evidence of nonzero eccentricity in a few previously detected signals (Romero-Shaw et al. 2020a; Gamba et al. 2023a; Gayathri et al. 2022; Gupte et al. 2024; Planas et al. 2025b; Morras et al. 2025b; Planas et al. 2025a). Therefore, we cannot exclude a priori that a few candidates in GWTC-4.0 may present nonzero eccentricity.

In the following subsections, we will describe chronologically the relevant waveform modeling efforts that lead to models employed in GWTC analyses. For consulting the specific set of waveform models employed in each GWTC version, we refer the reader to Table 1, while a summary of the physics of each model is displayed in Table 2. Their usage on specific pipelines will be described in the corresponding sections of this work.

### 2.1. BBH Models

Within the IMRPHENOM and SEOBNR families, the first complete IMR models calibrated to NR were produced for the dominant spin-weighted spherical harmonic multipole of nonspinning systems (Ajith et al. 2007; Buonanno et al. 2007), and then extended to aligned-spin systems (Ajith et al. 2011; Santamaria et al. 2010; Taracchini et al. 2014b). Improved versions were developed for GWTC analyses: IMR-PHENOMD (Husa et al. 2016; Khan et al. 2016) and SEOB-NRv4 (Bohé et al. 2017), increasing the amount of analytical information included in the models, the number and coverage of the calibration dataset, specific details of the model expressions and better accuracy with NR. A highly optimized version of SEOBNRv4 was produced to reduce its computational cost in the time domain, SEOBNRv4_OPT (Devine et al. 2016), and reduced-order model (ROM) techniques were applied to obtain a fast Fourier-domain version of the original model, SEOBNRv4_ROM (Bohé et al. 2017). Within the TEOB approach, a version of the model for the dominant multipole of BBH systems was developed, TEO-BRESUMS (Nagar et al. 2018), including a post-adiabatic approximation for the dynamics that increases its computational efficiency (Nagar & Rettegno 2019). Additionally, the inspiral-only models TAYLORF2 (Damour et al. 2001; Buonanno et al. 2009; Vines et al. 2011), which is an analytical Fourier-domain model, and SPINTAYLORT4 (Sturani 2015; Isoyama et al. 2020), a time-domain model, have also been employed for searching GW signals.

Modeling spin-precession effects is crucial for precise measurements of spins (Vitale et al. 2014; Pratten et al.

2020b; Johnson-McDaniel et al. 2022b; Biscoveanu et al. 2021; Steinle & Kesden 2022), providing key information about the formation channels of the observed systems (Rodriguez et al. 2016; Stevenson et al. 2017; Talbot & Thrane 2017; Zhu et al. 2018), and breaking degeneracies in parameter inference (Vecchio 2004; Lang & Hughes 2006; Chatziioannou et al. 2015; Krishnendu & Ohme 2022). Spin-precession can significantly increase the complexity of the signal, and modeling efforts focused first on applying the *twisting-up* approach (Buonanno et al. 2003; Schmidt et al. 2011; Boyle et al. 2011; O'Shaughnessy et al. 2012; Schmidt et al. 2012) to the dominant-multipole models, differing on the description of the spin dynamics of the system. The IMRPHENOM approach incorporated a closed-form solution of the next-to-next-to-leading order spin-precession evolution equations for single-spin configurations (Marsat et al. 2013; Bohe et al. 2013), mapping then the expressions effectively to double-spin systems (Schmidt et al. 2015), and employing the twisting-up technique first to IMRPHENOMC (Santamaria et al. 2010) to obtain IMRPHENOMP (Hannam et al. 2014), and then to the more accurate model IMRPHENOMD to obtain IMRPHENOMPV2 (Hannam et al. 2014; Bohé et al. 2016). On the SEOBNR approach, spin-dynamics were incorporated via evolution of the EOB equations of motion for a quasi-circular spin-precessing Hamiltonian, constructing the spin-precessing polarizations from the quadrupolar multipoles in the co-precessing frame for producing SEOBNRv3 (Pan et al. 2014).

Subdominant harmonics in the signal were shown to be important for reducing several degeneracies in the analysis of signals (Capano et al. 2014; Graff et al. 2015; Abbott et al. 2021a, 2024), in particular in inclination–distance degeneracy and modeling efforts focused on their inclusion into current models. IMRPHENOMHM (London et al. 2018) incorporated a set of the most important subdominant harmonics to the IMRPHENOMD model, although without additional calibration of these harmonics to numerical data. SEOBNRv4HM (Cotesta et al. 2018) incorporated a similar list of harmonics to SEOBNRv4, with explicit calibration of the waveform modes to NR and test-particle waveforms, and provided an efficient ROM version in Fourier-domain, SEOBNRv4HM_ROM (Cotesta et al. 2020). Spin-precessing versions of these multipolar waveform models were developed in parallel, with some improvements in the spin-dynamics description in the IMRPHENOM approach, resulting in IMRPHENOMPV3HM (Khan et al. 2020) and SEOBNRv4PHM (Ossokine et al. 2020). Additionally, the NRSURROGATE approach started producing the first surrogate models for multipolar spin-precessing signals, first with NRSUR7DQ2 (Blackman et al. 2017b) and then extending the parameter space coverage with NRSUR7DQ4 (Varma et al. 2019a).

Inside the IMRPHENOM approach, a new generation of waveform models was developed improving substantially the accuracy of the previous generation, providing a new model for the dominant harmonic of aligned-spin signal, IMRPHENOMXAS (Pratten et al. 2020a), a model for subdominant harmonics explicitly calibrated to NR simulations, IMRPHENOMXHM (García-Quirós et al. 2020) and its extension to spin-precessing signals through the twisting-up technique and the employment of the multiscale expression for the spin dynamics (Klein et al. 2013; Chatziioannou et al. 2013; Klein et al. 2014), IMRPHENOMXPHM (Pratten et al. 2021). This generation also reduced substantially the computational cost of waveform generation via an implementation of the multi-banding method (Vinciguerra et al. 2017; García-Quirós et al. 2021). On a parallel effort, a new phenomenological family of models was developed in the time domain, IMRPhenomTPHM (Estellés et al. 2021, 2022b,a), with the aim of overcoming the limitations of the stationary-phase approximation (SPA) in the modeling of precessing signals, and including for the first time in a phenomenological model accurate numerical solutions of the spin-precession equations.

Further improvements for spin-precessing signals were recently introduced in the IMRPHENOMXPHM model, introducing for the first time explicit calibration with spin-precessing NR simulations in IMRPHENOMXO4A (Hamilton et al. 2021; Thompson et al. 2024) as well as the inclusion of the dominant multipole equatorial asymmetry (Ghosh et al. 2024), a key effect for accurately describing systems with large remnant recoil (Varma et al. 2020; Borchers et al. 2024) and improving accuracy of spin-precessing models (Ramos-Buades et al. 2020b). This effect has been shown in several recent studies to be relevant for the correct inference of spin-precessing signals (Kolitsidou et al. 2024; Estellés et al. 2025). Additionally, improvements in the description of the spin dynamics during the inspiral were incorporated in IMRPHENOMXPHM_SPINTAYLOR (Colleoni et al. 2025b), performing numerically the integration of the PN spin dynamic equations and enhancing the accuracy for the inspiral stage of the signal.

The SEOBNR approach has recently been developed with a new and more accurate generation of BBH waveform models, SEOBNRv5HM (Pompili et al. 2023) for aligned-spin systems and SEOBNRv5PHM (Ramos-Buades et al. 2023) for spin-precessing systems. Improvements include more analytical PN information (Henry 2023; Khalil et al. 2023) and recent developments from second-order gravitational self-force for the flux and gravitational modes (Warburton et al. 2021; van de Meent et al. 2023), increasing the coverage of parameter space employed in the calibration with NR and test-mass limit simulations (Barausse et al. 2012; Taracchini et al. 2014a). These are also substantial improvements to the computational efficiency of the models via the implementation of the post-adiabatic technique (Nagar & Rettegno 2019; Mihaylov et al. 2021) and the release of the modular and highly optimized Python package pySEOBNR (Mihaylov et al. 2023). The dominant and multipolar spin-aligned models SEOBNRv5 and SEOBNRv5HM also provide efficient ROM versions in the Fourier domain, SEOBNRv5_ROM and SEOBNRv5HM_ROM (Pompili et al. 2023).

### 2.2. BNS Models

When modeling BNS mergers, the inclusion of tidal interactions is critical for accurately describing the two-body dynamics and the corresponding GW emission. These interactions are characterized by the dimensionless tidal deformability parameter $\Lambda$, which encodes the response of a NS to the gravitational field of its companion and is directly related to the equation of state (EoS) of dense nuclear matter (Flanagan & Hinderer 2008; Hinderer 2008). Measurements of the tidal deformability through GW observations can constrain the EoS of dense nuclear matter (Chatziioannou 2020). Tidal interactions in BNS systems are first captured at 5 PN order (Flanagan & Hinderer 2008; Vines et al. 2011) in the velocity expansion and have been extended up to 7.5 PN order (Damour et al. 2012; Henry et al. 2020). Tidal corrections have been added to the GW phase of the PN inspiral Taylor Fourier-domain model TaylorF2 (Vines et al. 2011).

Besides being tidally deformed by their companions, rapidly-spinning NSs acquire an intrinsic oblateness where the NS mass-quadrupole moment depends on the dimensionless spin and encodes information about the EoS. The interaction between this quadrupole and the monopole of the companion generates an additional conservative potential, usually termed quadrupole–monopole coupling, entering the two-body dynamics at the same 2 PN order as the spin–spin term and contributing corrections to both the binding energy and the GW flux (Poisson 1998). Neglecting the quadrupole-monopole term can bias tidal-deformability and spin measurements once spins approach millisecond-pulsar values (Agathos et al. 2015; Samajdar & Dietrich 2018).

In addition to PN-based tidal corrections and quadrupole-monopole interaction, NR simulations play an essential role in accurately modeling the merger and post-merger phases of BNS coalescence, where nonlinear effects become significant, and several catalogs of BNS simulations have been produced (Dietrich et al. 2018; Gonzalez et al. 2023b; Kiuchi et al. 2017). The NRTidal approach (Dietrich et al. 2017) has been developed to incorporate tidal effects into the frequency-domain BBH waveform models, maintaining the computational efficiency of the BBH model baseline. The tidal phase in frequency domain is modeled by employing the PN-expanded phase and calibrating it to NR BNS simulations. The original NRTidal model included PN-phase corrections augmented with calibration in the time domain using a non-spinning set of equal-mass BNS NR simulations, and then transformed to frequency-domain using the SPA. It was incorporated (Dietrich et al. 2019b) in the construction of the BNS models IMRPhenomD_NRTidal and SEOBNRv4_ROM_NRTidal, for aligned-spin systems, IMRPhenomPv2_NRTidal, for precessing systems, which was employed in the analysis of GW170817 (Abbott et al. 2017a) together with SEOBNRv4_ROM_NRTidal.

NRTidalv2 (Dietrich et al. 2019a) improved upon the previous model by the employment of improved NR data, the incorporation of spin effects, and the addition of amplitude corrections to the GW signal. It was incorporated to several BBH baselines models, including IMRPhenomPv2_NRTidalv2 (Dietrich et al. 2019a) and more recently IMRPhenomXP_NRTidalv2 (Colleoni et al. 2025a), for precessing systems, and SEOBNRv4_ROM_NRTidalv2 for aligned-spin systems.

Within the EOB framework, the SEOBNRv4T (Hinderer et al. 2016; Steinhoff et al. 2016) and TEOBResumS (Bernuzzi et al. 2015; Nagar et al. 2018; Akcay et al. 2019) models aim to provide a more accurate description of the inspiral phase, particularly in regimes where tidal interactions become significant. SEOBNRv4T includes tidally-induced multipole moments (up to $\ell = 3$), the effect of dynamical tides from f-mode resonances (Hinderer et al. 2016; Steinhoff et al. 2016), and the spin-induced quadrupole moment (Poisson 1998; Harry & Hinderer 2018). An efficient surrogate version of the model in the frequency domain, based on Gaussian process regression, was developed to reduce the cost of its employment in data-analysis applications (Lackey et al. 2019).

Similarly, TEOBResumS incorporates tidally-induced quadrupole moments using a different resummation and new information from gravitational self-force, with the lack of f-mode resonances but the addition of calibration to NR BNS simulations in a more recent version of the model (Gamba et al. 2023b). In order to improve its efficiency in data-analysis applications, a combination of the post-adiabatic technique and the SPA is employed to produce fast frequency-domain inspiral templates (Nagar & Rettegno 2019; Gamba et al. 2021).

### 2.3. NSBH Models

Modeling NSBH mergers presents distinct challenges compared to BNS systems, primarily because the NS may be fully disrupted and accreted by the BH without producing a post-merger remnant. As in BNS systems, the tidal response of the NS affects the phase evolution of the inspiral, requiring the inclusion of tidal corrections in the waveform phase (Pannarale et al. 2011). Additionally, in an EoS-dependent region of parameter space, typically involving unequal masses or highly-spinning BHs, the NS can be tidally disrupted before reaching the innermost stable circular orbit. In such cases, the signal amplitude is strongly suppressed at high frequencies (Kyutoku et al. 2011; Foucart et al. 2014; Kawaguchi et al. 2015). While tidal disruption and post-merger remnants can also occur in BNS mergers, disrupted NSBH systems may lack a distinct merger signature altogether if the NS is fully disrupted before crossing the BH's horizon.

Current NSBH models employed to analyse the NSBH candidate signals from GWTC-2.0 onwards are the SEOBNRv4_ROM_NRTidalv2_NSBH (Matas et al. 2020) model and IMRPhenomNSBH (Thompson et al. 2020) model. Both models incorporate tidal information into the phase of the respective Fourier-domain BBH baseline model using the NRTidalv2 phase model described in Section 2.2. Disruptive and non-disruptive mergers are handled by adding tidal correction to the GW amplitude, as well as establishing a parameter-space dependent cut-off frequency for suppressing the GW amplitude in the case of disruptive events. In particular, IMRPhenomNSBH employs an NR-calibrated amplitude model (Pannarale et al. 2013, 2015), while SEOBNRv4_ROM_NRTidalv2_NSBH implements NR-calibrated correction factors to the underlying BNS NRTidal model. Both the amplitude model of SEOBNRv4_ROM_NRTidalv2_NSBH and the correction factors in SEOBNRv4_ROM_NRTidalv2_NSBH incorporate disruptive events by establishing a cut-off frequency and tapering the waveform at frequencies above this cut-off.

One of the main limitations of current NSBH models is that they are restricted to the dominant multipole and assume spin-aligned configurations. Including only the dominant mode can lead to degeneracies in distance and inclination, as well as to a non-negligible loss in signal-to-noise ratio (SNR) for unequal-mass NSBH systems, since higher-order multipoles can contain a significant fraction of the signal power for asymmetric binaries. Therefore, future models will have to address this limitation, and there is active development towards this (Gonzalez et al. 2023a).

### 2.4. *Luminosity and Remnant Properties*

Besides modeling the GW waveform, accurate predictions of the mass and spin of the final remnant BH, as well as estimations of the peak luminosity, are also crucial for population studies and tests of general relativity (GR). Several analytical fits have been developed using the remnant properties from NR BBH simulations (Hofmann et al. 2016; Keitel et al. 2017; Healy & Lousto 2017; Jiménez-Forteza et al. 2017), and recently accurate Gaussian process regression surrogate models have been developed (Varma et al. 2019b; Islam et al. 2023). For NSBH systems, dependency on the tidal deformability of the NS have been included (Zappa et al. 2019). These fitting formulae generally depend on the component masses and spins, typically specified at some reference frequency in the inspiral stage of the signal. For estimating these quantities from the inferred source properties (see Section 5.9), the input spin values are evolved forward until a fiducial orbital frequency near merger is achieved, and then results from different fitting formulae are averaged to provide the final estimates. For precessing systems, these formulae are often augmented with in-plane spin contribution before the average procedure is applied (Johnson-McDaniel et al. 2016).

## 3. SIGNAL IDENTIFICATION

The GW strain time series produced by advanced-era interferometric detectors can be treated as a linear superposition of continuous non-astrophysical *noise* and occasional transient astrophysical *signals*. On timescales of tens of seconds, the noise is generally well approximated by colored stationary Gaussian processes, allowing for relatively stable modeling and analysis. However, on longer timescales, the noise becomes non-stationary, exhibiting time-dependent statistical properties. It is frequently contaminated by transient non-Gaussian artifacts, known as *glitches* (Nuttall 2018; Glanzer et al. 2023; Soni et al. 2025), and is also affected by slowly time-varying broadband disturbances (Abbott et al. 2020b).

On the other hand, the signals in the data are presently relatively rare, occurring at a rate of just a few per week. GW *searches* are thus required to perform statistical data reduction, taking in the kilohertz-sampled strain data and producing a list of astrophysical candidates. This marks the beginning of the data-analysis workflow illustrated in Figure 1.

The search for GW transient candidates is carried out through two distinct phases. Initially, *online* analyses enable prompt follow-up observations by the global astronomical community (Abbott et al. 2019b, 2023a; LIGO Scientific Collaboration et al. 2025). Later, *offline* analyses are conducted to produce a more accurate list of candidates by re-evaluating the significance of the initial online candidates and identifying new ones. Some of the factors responsible for the differences in the online and offline analyses are the use of the final strain data, improved data quality, enhanced algorithms, and a better understanding of the noise background in the data.

Two broad families of detection algorithms are used: search pipelines for minimally-modeled transient sources that do not rely on specific waveform predictions and search pipelines that, conversely, aim to maximize sensitivity to CBCs by using sets of template waveforms (a subset of the models described in Section 2). For the GWTC, we use multiple pipelines concurrently and combine their results. This approach minimizes the risk of missing astrophysical signals and allows us to cover a wider range of the CBC parameter space.

Template-based pipelines target specific regions of the CBCs parameter space, usually defined by the redshifted (detector-frame) masses (Krolak & Schutz 1987) and spins of the two compact objects, which are arranged into *template banks*. Template waveforms are currently calculated neglecting several physical effects, specifically non-dominant multipole emission, orbital precession, eccentricity and tidal deformability. The templates thus only model the dominant multipole emission from quasi-circular BBH coalescences with spins parallel to the orbital angular momentum. This simplification reduces the dimensionality of the parameter space to be covered, and therefore the number of templates. Several studies have shown that this has a limited impact on the detection rate while offering a significant reduction in computational cost (Dal Canton et al. 2015; Calderón Bustillo et al. 2016; Harry et al. 2016; Calderón Bustillo et al. 2017; Dietrich et al. 2019b; Ramos-Buades et al. 2020c; Phukon et al. 2025; Chia et al. 2024; Mehta et al. 2025).

In modeled searches, we search the data from each detector by matched filtering each template $u(f|\boldsymbol{\theta}_{\mathrm{int}})$ (where $\boldsymbol{\theta}_{\mathrm{int}}$ is a vector over a subset of intrinsic CBC parameters), to produce a SNR time series defined as (Maggiore 2007; Allen et al. 2012)

$$\rho(t) = 4 \left| \int_{f_{\mathrm{low}}}^{f_{\mathrm{high}}} \frac{h(f)u^*(f|\boldsymbol{\theta}_{\mathrm{int}})}{S_n(f)} e^{2i\pi ft} df \right|, \quad (1)$$

**Table 1.** A summary of waveform models used in each release of the GWTC. Since the catalog is cumulative, later releases (e.g., GWTC-3.0, GWTC-4.0) include all previous candidates. In most cases, PE results from earlier releases remain unchanged. An exception is GWTC-2.1, which reanalyzed O1 and O2 data using updated waveform models, replacing the original PE results from GWTC-1.0 and GWTC-2.0. As a result, the PE analyses with the waveform approximants listed for GWTC-1.0 and GWTC-2.0 are not part of GWTC-4.0. The models employed in GWTC-4.0 are discussed in Section 2, and their use within specific pipelines is described in the following sections. Special candidate analyses might employ additional waveform models not listed here, but discussed for those particular signals.

| Catalog release | Data analysed | Search templates | Sensitivity estimates | Parameter estimation |
|---|---|---|---|---|
| GWTC-1.0 (Abbott et al. 2019a) | O1, O2 | SEOBNRv4_ROM, TAYLORF2 | - | IMRPHENOMPV2, IMRPHENOMPV2_NRTIDAL, SEOBNRV3, SEOBNRV4_ROM_NRTIDAL, SEOBNRV4T, TAYLORF2, TEOBRESUMS |
| GWTC-2.0 (Abbott et al. 2021a) | O3a | SEOBNRv4_ROM, TAYLORF2 | SEOBNRV4_OPT | IMRPHENOMD, IMRPHENOMD_NRTIDAL, IMRPHENOMHM, IMRPHENOMPV2, IMRPHENOMPV2_NRTIDAL, IMRPHENOMPV3HM, IMRPHENOMNSBH, NRSUR7DQ4, SEOBNRV4_ROM_NRTIDALV2_NSBH, SEOBNRV4_ROM, SEOBNRV4HM_ROM, SEOBNRV4P, SEOBNRV4PHM, SEOBNRV4T_SURROGATE, TAYLORF2, TEOBRESUMS |
| GWTC-2.1 (Abbott et al. 2024) | O1–O3a | SEOBNRV4_OPT, SEOBNRV4_ROM, SPINTAYLORT4, TAYLORF2 | SEOBNRV4P, SEOBNRV4PHM, SPINTAYLORT4 | IMRPHENOMPV2_NRTIDAL, IMRPHENOMXPHM, SEOBNRV4PHM |
| GWTC-3.0 (Abbott et al. 2023a) | O3b | SEOBNRV4_OPT, SEOBNRV4_ROM, SPINTAYLORT4, TAYLORF2 | SEOBNRV4P, SEOBNRV4PHM, SPINTAYLORT4 | IMRPHENOMNSBH, IMRPHENOMXPHM, SEOBNRV4_ROM_NRTIDALV2_NSBH, SEOBNRV4PHM |
| GWTC-4.0 (Abac et al. 2025b) | O4a | IMRPHENOMD, SEOBNRV4_OPT, SEOBNRV4_ROM, SEOBNRV5_ROM, SPINTAYLORT4, TAYLORF2 | IMRPHENOMXPHM | IMRPHENOMNSBH, IMRPHENOMPV2_NRTIDALV2, IMRPHENOMXO4A IMRPHENOMXPHM_SPINTAYLOR, NRSUR7DQ4, SEOBNRV4_ROM_NRTIDALV2_NSBH, SEOBNRV5PHM |

where $h$ represents the strain data, and $S_n$ is the power spectral density (PSD) of the detector noise (see Appendix B of Abac et al. 2025a). The lower integration limit frequency $f_{low}$ is determined by considerations of computational cost, data quality, and calibration reliability. The upper limit $f_{high}$ is predominantly based on considerations of computational cost. Both limits may vary with search implementations and templates.

Modeled search pipelines produce triggers by looking for peaks exceeding a given threshold in the SNR time series. This same procedure is applied to the calibrated data provided by all available detectors. All search pipelines attempt to pair candidates from one detector with coincident candidates from others. In such cases, under the assumption of independent detector noise, the network SNR is computed as the root-sum-square of the individual detector SNRs (Maggiore 2007). Some pipelines also consider signals detected by a single interferometer. Each pipeline has its own approach for combining the SNR values it measures with its own signal consistency tests to rank its candidates. Based on their ranking, search pipelines assign each candidate a false alarm rate (FAR) value, which quantifies the expected rate of non-astrophysical candidates produced with a rank at least as high as the candidate under consideration.

In addition to the measure of detection significance provided by the FAR of a candidate, we have also provided the probability $p_{astro}$ that a candidate is of astrophysical rather than terrestrial (noise) origin. The calculation of $p_{astro}$ accounts for both the rate of noise events corresponding to the FAR and an estimated rate of astrophysical signals under some prior model of the CBC population (Farr et al. 2015; Kapadia et al. 2020). It is formally defined as (Abbott et al. 2016a)

$$p_{astro}(x) = \frac{\mathcal{R}_{astro}f(x)}{\mathcal{R}_{astro}f(x) + \mathcal{R}_{noise}b(x)},$$

where $\mathcal{R}_{astro}$ is the total rate of astrophysical candidates in a given pipeline, $\mathcal{R}_{noise}$ is the total rate of candidates due to noise, and $f(x)$ and $b(x)$ are the probability density function (PDF) of signal and noise events, respectively, at the candidate ranking value $x$. Each pipeline estimates these rates and calculates a $p_{astro}$ value for the candidates it identifies using its own method. Since O4a, most pipelines use a common signal model based on a distribution of simulated signals (*injections*) over source masses, spins and distance (see

**Table 2.** Waveform models used in GWTC analyses. We indicate if the model includes spin-precession, matter effects and which multipoles are included for each model. For spin-precessing models, the multipoles correspond to those available in the coprecessing frame. We indicate the most relevant reference for each model, additional references are given in the main text of Section 2. Special candidate analyses might employ additional waveform models not listed here, but discussed for those particular signals.

| Model | Precession | Multipoles $(\ell, |m|)$ | Matter | Reference |
|---|:---:|:---:|:---:|---|
| SPINTAYLORT4 | ✓ | $\ell \leq 4$ | ✓ | Isoyama et al. (2020) |
| TAYLORF2 | ✗ | $(2,2)$ | ✓ | Damour et al. (2001) |
| IMRPHENOMD | ✗ | $(2,2)$ | ✗ | Khan et al. (2016) |
| IMRPHENOMD_NRTIDAL | ✗ | $(2,2)$ | ✓ | Dietrich et al. (2019b) |
| IMRPHENOMHM | ✗ | $(2,2),(2,1),(3,3),(3,2),(4,4),(4,3)$ | ✗ | London et al. (2018) |
| IMRPHENOMNSBH | ✗ | $(2,2)$ | ✓ | Thompson et al. (2020) |
| IMRPHENOMPV2 | ✓ | $(2,2)$ | ✗ | Bohé et al. (2016) |
| IMRPHENOMPV2_NRTIDAL | ✓ | $(2,2)$ | ✓ | Dietrich et al. (2019b) |
| IMRPHENOMPV2_NRTIDALV2 | ✓ | $(2,2)$ | ✓ | Dietrich et al. (2019a) |
| IMRPHENOMPV3HM | ✓ | $(2,2),(2,1),(3,3),(3,2),(4,4),(4,3)$ | ✗ | Khan et al. (2020) |
| IMRPHENOMXPHM | ✓ | $(2,2),(2,1),(3,3),(3,2),(4,4)$ | ✗ | Pratten et al. (2021) |
| IMRPHENOMXPHM_SPINTAYLOR | ✓ | $(2,2),(2,1),(3,3),(3,2),(4,4)$ | ✗ | Colleoni et al. (2025b) |
| IMRPHENOMXO4A | ✓ | $(2,2),(2,1),(3,3),(3,2),(4,4)$ | ✗ | Thompson et al. (2024) |
| NRSUR7DQ4 | ✓ | $\ell \leq 4$ | ✗ | Varma et al. (2019a) |
| SEOBNRV3 | ✓ | $(2,2),(2,1)$ | ✗ | Pan et al. (2014) |
| SEOBNRV4_OPT | ✗ | $(2,2)$ | ✗ | Devine et al. (2016) |
| SEOBNRV4_ROM | ✗ | $(2,2)$ | ✗ | Bohé et al. (2017) |
| SEOBNRV4HM_ROM | ✗ | $(2,2),(2,1),(3,3),(4,4),(5,5)$ | ✗ | Cotesta et al. (2020) |
| SEOBNRV4_ROM_NRTIDAL | ✗ | $(2,2)$ | ✓ | Dietrich et al. (2019b) |
| SEOBNRV4_ROM_NRTIDALV2_NSBH | ✗ | $(2,2)$ | ✓ | Matas et al. (2020) |
| SEOBNRV4P | ✓ | $(2,2),(2,1)$ | ✗ | Ossokine et al. (2020) |
| SEOBNRV4PHM | ✓ | $(2,2),(2,1),(3,3),(4,4),(5,5)$ | ✗ | Ossokine et al. (2020) |
| SEOBNRV4T | ✗ | $(2,2)$ | ✓ | Steinhoff et al. (2016) |
| SEOBNRV4T_SURROGATE | ✗ | $(2,2)$ | ✓ | Lackey et al. (2019) |
| SEOBNRV5_ROM | ✗ | $(2,2)$ | ✗ | Pompili et al. (2023) |
| SEOBNRV5PHM | ✓ | $(2,2),(2,1),(3,3),(3,2),(4,4),(4,3),(5,5)$ | ✗ | Ramos-Buades et al. (2023) |
| TEOBRESUMS | ✗ | $(2,2),(2,1),(3,3),(3,2),(3,1)$ | ✓ | Akcay et al. (2019) |

Essick et al. 2025, and Section 3.7) to calculate the signal PDF $f(x)$; a slightly different model used by one analysis is described in Section 3.2.

The total astrophysical probability $p_{\mathrm{astro}}$ is then distributed between three mutually-exclusive source categories: a BNS class corresponding to both component (source-frame) masses ranging between $1\,M_\odot$ and $3\,M_\odot$; a NSBH class for one mass between $1\,M_\odot$ and $3\,M_\odot$ and the other above; and a BBH class corresponding to both masses above

$3\,M_\odot$. The associated probabilities for each category thus sum to $p_{\mathrm{astro}}$:

$$p_{\mathrm{BNS}} + p_{\mathrm{NSBH}} + p_{\mathrm{BBH}} = p_{\mathrm{astro}}. \qquad (3)$$

The complementary probability that the candidate is of terrestrial origin, i.e. caused by noise rather than a CBC, is notated as $p_{\mathrm{terr}} = 1 - p_{\mathrm{astro}}$.

Transient search methods used to compile our catalog for data from the first observing run (O1) and second observing run (O2) are described in Section 3 of Abbott et al. (2019a), and for data from third observing run (O3), in Section 4 of Abbott et al. (2021a), Section 3 of Abbott et al. (2024) and Section 4 of Abbott et al. (2023a). We describe here the methods used for data from O4a, in the context of this cumulative catalog. In the coming subsections, we present details of the different pipelines used to identify transient GW candidates, including coherent WaveBurst (cWB) for minimally-modeled sources (Klimenko et al. 2016; Mishra et al. 2022, 2025), and GStreamer LIGO Algorithm Library (GstLAL; Cannon et al. 2012a; Messick et al. 2017; Sachdev et al. 2019; Hanna et al. 2020; Cannon et al. 2020; Sakon et al. 2024; Ray et al. 2023; Tsukada et al. 2023; Ewing et al. 2024; Joshi et al. 2025a,b), Multi-Band Template Analysis (MBTA; Allené et al. 2025), PyCBC (Dal Canton et al. 2021), and Summed Parallel Infinite Impulse Response (SPIIR; Chu et al. 2022) for model-based analyses. Each pipeline analyzed strain data from the available detectors, using the data summarized in Table 3.

For offline analyses, a set of *Category 1* (CAT1) flags are determined, which flag periods of severe data-quality or technical issues affecting the interferometers during the observation time. During CAT1 periods (detailed in Section 4.1 of Soni et al. 2025), the data are considered too contaminated by noise to be analyzable within realistic computational or human resources.

During the O1 to O3 observing runs, short-duration *Category 2* (CAT2) flags were also provided for transient searches (Abbott et al. 2020b). CAT2 flags are based on statistical correlations found between excess noise transients in auxiliary channels and in the strain data. As of O4, these CAT2 flags are no longer used in CBC searches, though they continue to be used for minimally-modeled transient searches (Soni et al. 2025).

As an extra input for CBC searches, a timeseries is generated by the IDQ pipeline (Essick et al. 2020). This timeseries contains statistical data-quality information based on the activity of auxiliary channels deemed safe (i.e., channels that should not be influenced by GWs). Observation times marked with IDQ flags are likely to be contaminated by glitches. Unlike CAT1 flags, which are mandatory vetoes prior to any analysis, IDQ flags are optional and may be used either as vetoes, or as input in the ranking statistic of the candidates. CAT1 vetoes and the IDQ timeseries are created from the strain data and auxiliary channels as a pre-processing step and ingested by the offline searches (see Figure 1).

### 3.1. cWB

The cWB algorithm is designed to detect transient GW signals using networks of GW detectors without requiring a specific waveform model. It identifies coincident excess power events by analyzing multi-resolution time–frequency (TF) representations of detector strain data (Klimenko et al. 2008, 2016). Once potential GW signals or noise events are identified, cWB reconstructs the source sky location and the signal waveforms recorded by the detectors using the con-strained maximum-likelihood method (Klimenko et al. 2005, 2016). This analysis is performed on detector strain data, accommodating signal frequencies up to a few kilohertz and durations up to a few seconds.

**Table 3.** The O4a data processed by search pipelines. As described in Abac et al. (2025c), different levels of processing produce different channels of data. Full Frames refers to the `GDS-CALIB_STRAIN_CLEAN` channel, while Analysis Ready refers to the `GDS-CALIB_STRAIN_CLEAN_AR` channel; these channels are identical other than that the Analysis Ready channel contains only data for times ready to be analysed.

| Pipeline | Online | Offline |
|---|---|---|
| cWB | Full Frames | Full Frames |
| GstLAL | Full Frames | Full Frames |
| MBTA | Full Frames | Analysis Ready |
| PyCBC | Full Frames | Analysis Ready |
| SPIIR | Full Frames | no offline analysis |

The cWB detection statistic relies on coherent energy $E_c$, which is derived from cross-correlating the reconstructed signal waveforms in the detectors and normalizing them by the spectral amplitude of the detector noise. The square root of $E_c$ provides a lower bound on the signal network SNR and is used for the initial selection of candidates identified by cWB. As the production rate of cWB candidates is primarily driven by glitches, additional post-production vetoes are applied to further reduce the background rates.

One primary signal-independent veto statistics is the network correlation coefficient, defined as $c_c = E_c/(E_c + E_n)$, where $E_n$ represents the normalized residual noise energy after the reconstructed signal has been subtracted from the data. Typically, for glitches $c_c \ll 1$. Consequently, candidates with $c_c < 0.6$ are rejected as potential glitches. For authentic GW signals, the residual energy $E_n$ is expected to follow the $\chi^2$ distribution with the number of degrees of freedom $N_{DoF}$ proportional to the number of TF data samples comprising the signal. The reduced $\tilde{\chi}^2 = \chi^2/N_{DoF}$ statistic serves as a powerful signal-independent veto, effectively identifying and removing glitches where $\tilde{\chi}^2$ is significantly greater than 1. To enhance the search for CBC signals, cWB performs analysis in the frequency band below 512 Hz and also employs weak signal-dependent vetoes. These vetoes are based on the central frequency $f_c$ and the detector-frame chirp mass $(1 + z)\mathcal{M}$, both of which are estimated from the TF evolution of the signal power (Tiwari et al. 2016).

All candidates identified by cWB are ranked by the reduced coherent network SNR

$$\rho_{cWB} = \sqrt{\frac{E_c}{1 + \tilde{\chi}^2[\max(1, \tilde{\chi}^2) - 1]}}. \quad (4)$$

To estimate the statistical significance of the GW candidates, they are ranked against background events generated by the detector noise. A comprehensive background data sample, equivalent to approximately 1000 years of observation time, is obtained by repeating the cWB analysis on time-shifted data. To ensure that astrophysical signals are excluded from the background data, the time shifts are selected to be much larger than the expected signal time delay between the detectors. The statistical significance of each cWB candidate is quantified by its FAR. The FAR is defined as the rate of background events that exhibit a larger $\rho_{cWB}$ value than the GW candidate in question. To account for potential long-term variations in detector noise, the candidate significance is estimated using nearby data intervals, typically one to two weeks in length.

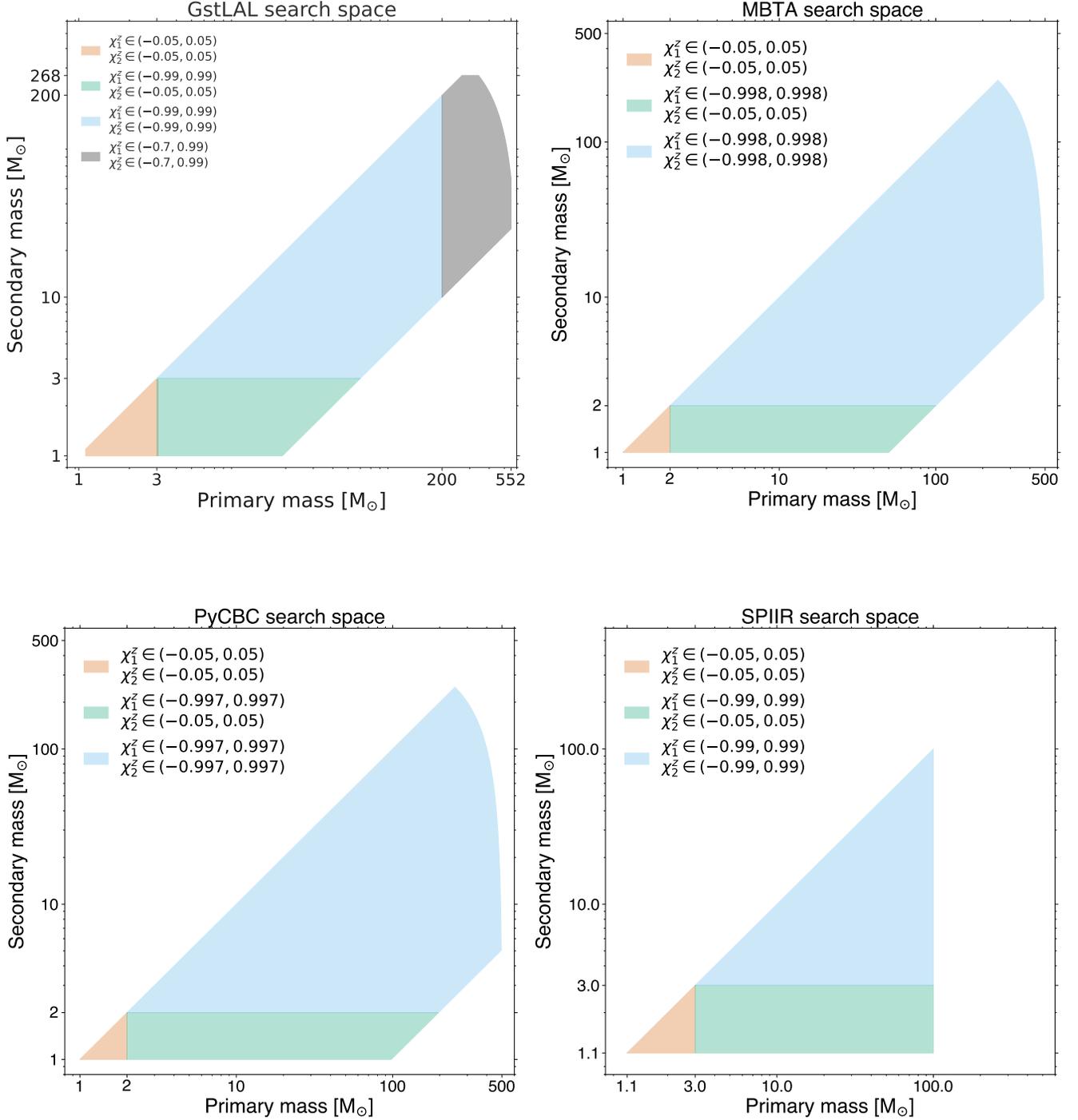

**Figure 2.** Regions of the CBC parameter space explored by our different template-based search pipelines. Individual masses are given in the detector frame, i.e. they are redshifted. The spin parameters $\chi^z_{1,2}$ are the spin projections along the orbital angular momentum.

Since the publication of GWTC-3.0, the cWB algorithm used in the previous observing runs (Klimenko et al. 2016; Drago et al. 2020), i.e., cWB-2G, has undergone significant improvements, and a new version, cWB-BBH, has been adopted for CBC searches: it incorporates two major changes compared to cWB-2G.

The first change involves replacing the Wilson–Debauchies–Meyer (WDM) wavelet transform (Necula et al. 2012) with the multi-resolution *WaveScan* transform, which is based on the Gabor wavelets (Klimenko 2022). This update effectively reduces temporal and spectral leakage in the TF data, potentially leading to more accurate signal representation.

Secondly, in addition to the traditional excess-power statistic, the algorithm now incorporates the cross-power statistic (Klimenko 2022) for identifying transient signals. The excess-power represents the total power of a TF data sample (or *WaveScan* pixel) integrated over the detector network. The cross-power represents the correlation of power in the detectors. Both statistics are maximized over possible time-of-flight delays of a GW signal across the detector network. The excess-power and the cross-power amplitudes, $a_e$ and $a_\times$, respectively, follow a predictable half-normal distribution with unity variance assuming quasi-stationary detector noise (Klimenko 2022). For analysis, TF pixels with the excess-power amplitude exceeding 2.3 standard deviations are selected. Spatially and temporally adjacent TF pixels are then clustered to form the initial cWB-BBH candidates (clusters).

The clustered excess-power $\sum_{i=1}^{I} a_e^2[i]$ and the cross-power $\sum_{i=1}^{I} a_\times^2[i]$ are used to calculate the upper bounds on the signal network SNR $\rho_p$ and the coherent network SNR $\rho_\times$ respectively (Mishra et al. 2025), where $I$ is the number of the TF pixels comprising the cluster. The $\rho_e$ is optimized for identification of clusters due to fluctuations of the quasi-stationary detector noise dominating the initial cWB-BBH rate on the order of 100 Hz. All clusters with $\rho_p > 4$ are accepted for further analysis. The accepted cWB-BBH clusters are mostly produced by glitches with a typical rate of 0.1 Hz. After selection, the initial TF clusters are aggregated if they fall within the time and frequency intervals of 0.23 s and 64 Hz respectively. It improves the energy collection for transient candidates that can be fragmented into clusters by the TF transform. The glitch rate is further reduced by requiring $\rho_\times > 7$, where the upper bound of the coherent network SNR is calculated for the defragmented candidates.

The above conditions on the $\rho_e$ and $\rho_\times$ require that the GW signal clusters with SNR > 4 and defragmented GW candidates with SNR > 7 are accepted for the analysis. The resulting reduced candidates rate, on the order of 1 mHz, makes the subsequent likelihood analysis and reconstruction of the remaining candidates computationally feasible. At this stage, the sky location of each candidate is determined and the signal waveforms are reconstructed with the inverse *WaveScan* transform (Klimenko 2022). The cWB-BBH candidates with $\rho_{cWB} > 7$ are stored for the post-production analysis.

Moreover, the post-production veto analysis has been replaced by the XGBoost machine-learning algorithm (Mishra et al. 2021, 2022, 2025), based on an ensemble of decision trees. It performs a classification of cWB-BBH candidates by using a subset of 25 summary statistics including $\rho_{cWB}$, $c_c$, $\tilde{\chi}$, $\rho_e$, $\rho_\times$, $f_c$, and $(1 + z)\mathcal{M}$. To improve the distinction between astrophysical GW signals and glitches, the cWB-BBH detection statistic is modified as

$$\rho_r = \rho_{cWB} W_{XGB}, \tag{5}$$

where $W_{XGB}$ is the XGBoost classification factor (Mishra et al. 2021) ranging from 0 (indicating a glitch) to 1 (indicating a signal). The XGBoost response is trained using a subset of background candidates and a representative set of simulated BBH signals, which are injected into the detector data and recovered with cWB-BBH. Due to a weak dependence of the cWB-BBH summary statistics on the CBC population and signal models, the algorithm is robust to a variety of CBC features, including higher multipoles, unequal component masses, misaligned spins, eccentric orbits, and possible deviations from GR (Mishra et al. 2022; Bhaumik et al. 2025).

## 3.2. GstLAL

GstLAL is a stream-based time-domain matched-filtering search pipeline (Cannon et al. 2012a; Messick et al. 2017; Sachdev et al. 2019; Hanna et al. 2020; Cannon et al. 2020; Sakon et al. 2024; Ray et al. 2023; Tsukada et al. 2023; Ewing et al. 2024; Joshi et al. 2025a,b). All previous GWTC versions included results produced by GstLAL (Abbott et al. 2019a, 2021a, 2024, 2023a). For this analysis, multiple GstLAL configurations were used: an online configuration whose results were re-evaluated using the full background, an offline configuration that processed data dropped by the online configuration, and another offline configuration that targeted intermediate-mass black hole (IMBH) mergers (Joshi et al. 2025a). The primary configuration was an online search, which identified CBC signals in near real-time, enabling rapid alerts and follow-up. The online search dropped ∼3% of the observing run data due to, e.g., computer downtime. The dropped data were processed in an offline configuration, methodologically identical to the online configuration, but executed post O4a. The online configuration did not cover the high-mass region of the parameter space, so an additional offline configuration was run to target the IMBH region, shown in grey in Figure 2. All configurations (online, dropped-data and IMBH) were used to evaluate search sensitivity by injecting simulated signals into the detector strain data and analyzing the effectiveness of signal recovery. Final candidate ranking and significance estimation incorporated the collective results from all configurations. An additional early-warning configuration, designed to identify BNS mergers prior to coalescence, was also deployed during O4a. However, results from the early-warning configuration are not part of GWTC-4.0.

The analysis begins by constructing a template bank to cover the targeted astrophysical parameter space. Strain data are then pre-processed to prepare for matched-filtering, which correlates the pre-processed strain with templates from the bank. Peaks in the matched-filter output are identified as triggers and subsequently ranked using a likelihood ratio (LR) statistic that incorporates signal consistency and detector network properties (Tsukada et al. 2023). The background LR distribution from noise triggers is used to estimate the FAR and the significance of each GW candidate. In the following paragraphs, we describe each component of the analysis in detail, beginning with the template bank.

For O4a, the template bank targeted CBCs with detector-frame total masses between $1.95\,M_\odot$ and $610\,M_\odot$, and mass ratios from 0.05 to 1 (Sakon et al. 2024; Joshi et al. 2025a). It was divided into two disjoint regions: stellar-mass (component masses up to $200\,M_\odot$) and IMBH (primary component mass between $200\,M_\odot$ and $552\,M_\odot$ and secondary component mass below $268\,M_\odot$). While no BBHs with total mass above $400\,M_\odot$ have been observed so far, cosmological redshifting of source-frame masses can shift distant stellar-mass binaries into the IMBH region in the detector frame, motivating an extended search (Abbott et al. 2023a). Templates in both regions were spin aligned, i.e. spin components in the orbital plane are zero. Aligned spin values were allowed in $[-0.99, +0.99]$ for stellar-mass templates, with additional restriction to $[-0.05, +0.05]$ for component-masses below $3\,M_\odot$, consistent with observations for BNSs (Stovall et al. 2018). In the IMBH region, spins were restricted to $[-0.70, +0.99]$ to mitigate triggers from noise transients with similar time–frequency structure as high-mass waveforms with large anti-aligned spins (Hanna et al. 2022). The search space is illustrated in Figure 2.

During template bank creation, templates were placed using a metric based on the IMRPHENOMD approximant (Husa et al. 2016; Khan et al. 2016), with a lower frequency cutoff of $10\,\mathrm{Hz}$ and a maximum duration of $128\,\mathrm{s}$, ensuring coverage of the detector's sensitive band while limiting computational cost. The duration constraint led to a template-dependent low-frequency cutoff for low-mass systems. For example, a $1.4\,M_\odot$–$1.4\,M_\odot$ binary would have a waveform duration of over $1000\,\mathrm{s}$ starting from $10\,\mathrm{Hz}$. To stay within the $128\,\mathrm{s}$ constraint, its corresponding lower frequency cutoff must be raised to $\sim$22 Hz (Peters & Mathews 1963).

Template placement was performed using MANIFOLD (Hanna et al. 2023), an algorithm based on a geometric binary tree that tiles the parameter space dictated by the minimal match constraints. A minimal match of 97% was applied in the stellar-mass region, corresponding to a 10% expected loss in detection rate (Owen 1996), while the IMBH region used 99%, resulting in a total of $\sim 1.8 \times 10^6$ templates.

To improve fault tolerance of the online configuration, the stellar-mass bank was interleaved into two complementary halves, each independently covering the stellar-mass region and analyzed independently at separate data centers (Godwin 2020; Sakon et al. 2024). Each half was subdivided into template bins and each bin was decomposed into filters using singular value decomposition (SVD; Cannon et al. 2012b; Sakon et al. 2024).

After template bank construction, the detector strain data were preprocessed to ensure compatibility with the templates and to reduce noise contamination. Data were resampled to 2048 Hz and whitened. To suppress short-duration high-amplitude noise transients that can mimic astrophysical signals, especially at high masses where template waveforms are short, amplitude-based gating was applied to the whitened data. The gating threshold was a linear function of the chirp-mass, with values computed per template bin and expressed in units of the standard deviation of the whitened strain data. This adaptive approach balances glitch suppression with sensitivity to real signals across the mass range (Sachdev et al. 2019; Ewing et al. 2024).

In parallel with gating, accurate whitening of strain data and template waveforms required careful PSD estimation to characterize frequency-dependent noise. In the online configuration, PSDs were estimated using a 4 s fast Fourier transform (FFT) computed continuously on incoming data and used to whiten the strain in real time.

To account for longer-term noise variations that could impact the SVD, template waveforms were re-whitened weekly using updated PSDs derived from recent data. For template whitening, the analysis period was divided into continuous segments of up to 8 h, with each segment producing a PSD using a 8 s FFT. The final PSD was constructed by taking the median power in each frequency bin across segments (Messick et al. 2017). The SVD basis for each template bin were then recomputed weekly using the same PSD. Weekly cadence balanced the need to track gradual detector noise evolution with the high computational cost of recomputing SVDs (Ewing et al. 2024).

After whitening the strain data using the PSD, matched-filtering was performed in the time domain, using the SVD basis, starting at 15 Hz for the stellar-mass region (10 Hz for the IMBH region). The output, an SNR time series, was scanned for peaks exceeding a threshold of 4, which were defined as triggers. A signal-consistency statistic, $\xi^2$, was computed for each trigger to quantify the deviation between observed and expected SNR values across a fixed number of points centered on the peak. Only the highest SNR trigger in each 1 s window per template bin was kept. Templates with chirp masses up to $1.73\,M_\odot$ used the TAYLORF2 approxi-mant and 701 samples to compute $\xi^2$. Higher-mass templates used the SEOBNRv4_ROM approximant and 351 samples to compute $\xi^2$ (Messick et al. 2017). The resulting triggers were used to form candidates for further processing.

Candidates were classified as either single-detector or coincident. Coincident candidates are defined as triggers from the same template in two or more detectors and occurring within the light-travel time between detectors, plus a 5 ms window to account for timing uncertainties. Single-detector candidates had no corresponding trigger in another detector within the window (Sachdev et al. 2019). Candidates from the IMBH region were excluded if they were a single-detector candidate. Each candidate was ranked using a LR quantifying the probability of astrophysical versus noise origin. The LR incorporated per-detector SNRs, the signal-consistency statistic $\xi^2$, local trigger rates, detector sensitivities, and prior assumptions about the CBC population. For coincident candidates, the relative arrival times and phases at the different detectors were included, with probabilities computed assuming isotropically-distributed sources. In contrast, single-detector candidates were down-weighted using a *singles-penalty*, empirically tuned via signal simulation campaigns, in order to suppress false positives while retaining sensitivity (Sachdev et al. 2019).

The LR was computed independently for each template bin to account for variations in how templates interact with detector noise. Background distributions for each template bin were built from single-detector triggers occurring during coincident observing time. Candidates were clustered in an 8 s window, and the candidate with the highest LR across all template bins was selected for further processing. Further details on the LR calculation can be found in Cannon et al. (2015); Messick et al. (2017); Tsukada et al. (2023).

Candidate significance was quantified using FAR, defined as the rate at which detector noise could produce a candidate with equal or higher LR than the candidate under consideration. FAR was estimated by populating the background LR distribution via Monte Carlo sampling of SNR and $\xi^2$ values drawn from the empirical background, smoothed via kernel density estimation (KDE). Samples were assigned coalescence times, phases, and templates from uniform distributions (Cannon et al. 2013; Joshi et al. 2025a). To reduce contamination of the background by astrophysical signals, a 10 s window around each GW candidate is vetted and excluded from the background, but only for candidates identified online by the online configuration (Joshi et al. 2023).

Dropped data from the online configuration was processed in an offline configuration, yielding additional candidates. Results from both configurations were merged, and FARs recomputed using the full background collected by the online configuration (Joshi et al. 2025b). The combined set of candidates and background defined the stellar-mass search analysis. To produce the final GWTC-4.0 results, candidates and background from the stellar-mass and IMBH search-analysis regions were combined into a unified analysis. Each region was assigned a weight, determined from injection campaigns (Essick et al. 2025), reflecting its relative sensitivity and the expected number of detectable astrophysical signals. The stellar-mass and IMBH search analyses were assigned weights of 0.94 and 0.06, respectively. The weights account for the fact that different analyses contribute unequally to the final set of candidates under the assumption of a given prior source population. FARs were rescaled by the inverse of these weights, and LRs were recomputed to provide a unified, unbiased ranking (Joshi et al. 2025a).

In addition to FAR, the total probability of astrophysical origin $p_{\mathrm{astro}}$, and the classification probability for the mutually-exclusive BNS, NSBH, or BBH source classes, were calculated for each candidate. Unlike other pipelines, GstLAL adopts a signal model that assumes the Salpeter primary mass function (Salpeter 1955) for each source class, with uniform distributions in mass ratio and spin. The source-specific probabilities were computed using a Bayesian framework that models the posterior probability distribution of the rates associated with each class. The framework treats GW triggers as realizations of independent Poisson processes, with rate estimates derived from previous observing runs and the population properties of each source class (Ray et al. 2023).

O3 differed from O4a in search configuration, template bank design, and candidate ranking. While GWTC-4.0 results were produced by using online results supplemented with offline analyses to avoid redundant processing, the GWTC-3.0 results were produced entirely from a post-run offline analysis (Abbott et al. 2023a). The O3 template bank extended to detector-frame total masses of 758 $M_\odot$, with aligned-spin parameters in $[-0.999, 0.999]$ for components more massive than 3 $M_\odot$, and was constructed using a stochastic placement algorithm (Abbott et al. 2021a). No duration constraint was imposed during matched filtering with the O3 template bank. Ranking of candidates identified by matched-filtering used an LR similar to O4a, but also incorporated the iDQ glitch likelihood (Godwin et al. 2020; Abbott et al. 2021a). The iDQ glitch likelihood was used only for single-detector candidates prior to GWTC-2.1, and the singles-penalty parameter, used to down-weight such triggers, was tuned differently than in O4a. Additionally, the LR term modeling the distribution of $\xi^2$ used an empirically-tuned analytical function in O3, rather than being directly informed by the statistical properties of $\xi^2$ as done in O4a. Furthermore, data near vetted GW candidates were not excluded from background estimation before O4a.

The GstLAL pipeline used during O2 differed from O3 primarily in the template bank construction and candidate ranking methodology. The O2 template bank covered detector-frame total masses up to 400 $M_\odot$ and mass ratios from 1/98 to 1, with aligned-spin parameters in $[-0.999, 0.999]$ for components above 2 $M_\odot$ (Abbott et al. 2019a; Mukherjee et al. 2021). The LR in O2 assumed uniform signal recovery across the template bank, ignoring the non-uniform template density and any astrophysical prior on the source population (Abbott et al. 2021a; Fong 2018).

The O2 GstLAL pipeline differed from O1 in its template bank construction, candidate ranking methodology and data conditioning. The O1 template bank covered detector-frame total masses from $2\,M_\odot$ to $100\,M_\odot$ (Mukherjee et al. 2021). In O1, the LR was computed only for coincident candidates and excluded time and phase differences between detectors (Hanna et al. 2020). The LR included the joint probability of observed SNRs but assumed equal horizon distances across detectors: an approximation that neglected how detector-specific noise characteristics impact the horizon distance. It was improved in O2 by pre-computing joint SNR distributions for a discrete set of horizon-distance ratios. Additionally, the O1 pipeline also lacked a template-mass-dependent glitch excision threshold (Sachdev et al. 2019).

### 3.3. MBTA

MBTA (Abadie et al. 2012a; Adams et al. 2016; Aubin et al. 2021; Alléné et al. 2025) has performed online searches for CBCs since the initial-detector era science runs (Abadie et al. 2012b). MBTA analyzes each detector independently using matched-filtering, before searching for candidates seen in coincidence. To reduce the computational cost, the pipeline splits the matched-filtering process over several frequency bands (typically two). Since O3, MBTA runs offline search analyses as well, with the aim of more accurately assessing the significance of candidates and contributing to the GWTC (Abbott et al. 2024, 2023a).

The parameter space explored by MBTA's template bank is entirely defined by the detector-frame masses and spins (assumed parallel to the orbital angular momentum) of the binary components. The O4a analysis covers individual masses greater than $1\,M_\odot$, total masses up to $500\,M_\odot$, and mass ratios ranging from $1/50$ to $1$. In line with astrophysical expectations for NSs (Stovall et al. 2018), spin magnitudes are restricted to $0.05$ for objects with masses below $2\,M_\odot$, while more massive objects are allowed to have spin magnitudes of up to $0.998$. Very short templates, which tend to generate a background excess, are removed by introducing a cut-off of $200\,\mathrm{ms}$ on the template duration, calculated starting from $18\,\mathrm{Hz}$. A visual representation of the parameter space covered by the O4a MBTA stellar-mass bank is given in the upper-right part of Figure 2.

The template banks used for O4a are generated with different algorithms. The stellar-mass search space is first divided into two regions, corresponding to BNS and symmetrical BBH (no more than a factor 3 between the two component masses). Templates in the BNS region are geometrically placed according to a local metric estimate (Brown et al. 2013), aiming for SNR recovery of at least $98\%$. Symmetrical BBH template placement uses a hybrid algorithm (Roy et al. 2017, 2019), with a similar expected SNR recovery. The rest of the bank is then completed using the same algorithm, with the objective of limiting the maximum SNR loss to $3.5\%$. As MBTA filters data in two disjoint frequency bands, a bank has been created for each of them, as well as one for the whole frequency range. Each template from the latter bank is associated with a template from other bands, maximizing their overlap in their common frequency band. MBTA uses a total of almost $825000$ different templates, which are almost entirely described by approximately $55000$ low-frequency-band templates and $20000$ high-frequency-band templates. Templates use the SPINTAYLORT4 approximant (Klein et al. 2014) for total masses below $4\,M_\odot$ and SEOBNRV4_OPT (Bohé et al. 2017) above.

Before matched filtering, MBTA applies multiple pre-processing steps to the calibrated strain data delivered by the detectors. Data are first resampled to $4096\,\mathrm{Hz}$. Next, a gating procedure is applied, designed to remove short and high-amplitude glitches. The gating used for O4 is similar to the version deployed for O3 and is triggered by drops in the detector sensitivity (Aubin et al. 2021). It is also applied to periods of poor quality data identified by data-quality vetoes (Abbott et al. 2020b). Less than $0.1\%$ of the data provided by each detector was removed by MBTA gating during O4a (Alléné et al. 2025). An ungated search analysis is also performed for part of the parameter space, with the aim of finding the massive, intense CBCs that are likely to trigger the gating (Aubin et al. 2021).

MBTA regularly re-estimates the noise PSD of each detector by taking the median over thousands of seconds of data from several FFTs, whose lengths range from seconds to hundreds of seconds, depending on the targeted region of the parameter space (Aubin et al. 2025).

For most templates, the matched-filter calculation is carried out in two disjoint frequency bands, starting at $24\,\mathrm{Hz}$ and separated at $80\,\mathrm{Hz}$. When a sufficiently high SNR is detected, bands are then coherently recombined to produce a unified single-detector trigger from the two bands. The shortest templates, however, are processed using a single band starting at $20\,\mathrm{Hz}$ due to their intrinsically narrower frequency band.

Triggers retrieved with an SNR above $4.4$ (9 for the ungated analysis) are recorded. MBTA then applies several tools to reject triggers produced by transient noise. A basic $\chi^2$ test rejects any trigger whose SNR is not distributed as expected between the filtered bands (Adams et al. 2016). Since O3, a $\chi^2$ statistic is used to measure the discrepancy between the measured SNR time series and the template autocorrelation. The test is used to reweight the SNR, defining a single-detector ranking statistic (Aubin et al. 2021).

Since O3, MBTA has been downgrading the ranking statistics of triggers occurring during periods when detectors show signs of poor data quality. In the O4a offline analysis, this task is performed by a new technique called *SNR-Excess*. It builds a $\chi^2$ statistic from the differences between the time-series of the maximum SNR observed around each trigger, and models based on a large population of simulated signals (Alléné et al. 2025). This new procedure was motivated by the desire to automate and standardize a data-quality test that was performed manually online at the beginning of O4 in case of detection. Candidates such as S230622ba, which triggered an automatic alert and was manually retracted after human verification (LIGO Scientific Collaboration et al. 2023), can now be vetoed before being released from the pipeline.

MBTA looks for coincident triggers between detectors sharing the same templates, within a limited time window. A combined ranking statistic is computed as the quadrature sum of single detector ranking statistics. Since O3, this statistic also includes a term measuring the consistency of arrival times, phases and amplitudes across the detectors (Aubin et al. 2021). In addition to the search for coincident triggers between the two LIGO detectors, since the start of O4, MBTA now also produces single-detector candidates. In order to limit the risk of false alarms, while maximizing the chances of detecting electromagnetically-bright sources, the MBTA search for single-detector candidates is restricted to chirp masses below $7\,M_\odot$.

The significance of the candidates obtained by the stellar-mass search analysis is evaluated based on their probable origin, quantified by $p_{\rm astro}$. These probabilities are obtained using a method similar to the O3 method (Andres et al. 2022). For a given astrophysical source, the ratio of expected number of foreground triggers to the total number of triggers (foreground + background) is calculated. The foreground is estimated using population models similar to those used to estimate search sensitivity (see Section 3.7) and by fitting the rate observed by MBTA. The background is obtained by fitting the distribution of the ranking statistic measured by MBTA, after eliminating confidently-detected astrophysical signals. In the case of coincidences, the statistic is increased by considering fake coincidences between noise triggers, regardless of the time of arrival. These operations are performed over several bins in the mass and spin parameter space covered by the search.

The pipeline associates a FAR to candidates using a method similar to the $p_{\rm astro}$ background estimation. In order to make the values of $p_{\rm astro}$ and FAR consistent with each other, and to include an astrophysical prior in the FAR definition, $p_{\rm astro}$ is used since O4a as a new ranking statistic.

MBTA performed several online search analyses during O4a using a pipeline version similar to the offline analysis described above, with a few differences documented in Alléné et al. (2025). All coincident candidates, as well as single-detector candidates with chirp mass below $7\,M_\odot$, from the full-bandwidth analysis with a FARs below $2\,{\rm h}^{-1}$ were submitted to GRACEDB with a median latency of $21\,{\rm s}$. In the absence of prior realistic O4 data, the online $p_{\rm astro}$ calculation performed by the pipeline during O4a used the same model as for the offline second half of the third observing run (O3b) analysis (Abbott et al. 2023a), adjusting the signal rates to account for the increased sensitivity. MBTA was also conducting *early-warning* searches for CBCs with component masses between $1\,M_\odot$ and $2.5\,M_\odot$ and aligned spins below 0.05, with the aim of rapidly identifying high-SNR signals.

MBTA also conducted offline search analyses during the O3 period, and thus participated in the production of the previous catalog versions GWTC-2.1 (Abbott et al. 2024) and GWTC-3.0 (Abbott et al. 2023a). At that time, the pipeline processing was slightly different. The O3 stellar-mass search template bank was created using a purely stochastic method (Privitera et al. 2014), guaranteeing a SNR loss below 3%, compared with 2% for parts of the O4 bank. During O3, the pipeline only searched for CBCs with (detector-frame) total mass below $200\,M_\odot$, but no constraints on waveform duration were applied. Prior to the introduction of the SNR-Excess method, re-ranking of triggers occurring during periods of poor data quality was handled by comparing the trigger rate before and after the application of various other rejection criteria. Before O4, MBTA only calculated significance for candidates seen in at least two detectors. The FAR values were based on a ranking that assumed all templates were equiprobable, which was responsible for some inconsistencies with the associated $p_{\rm astro}$ values. Trials factors were used to account for the various coincidence types and parameter space regions (Aubin et al. 2021).

### 3.4. PyCBC

The PyCBC search pipeline (Dal Canton et al. 2014; Usman et al. 2016; Nitz et al. 2017) is a descendant of the IHOPE pipeline (Allen et al. 2012; Babak et al. 2013), which was used to search LIGO–Virgo data during the initial detector era up to 2010 (Abadie et al. 2012b; Aasi et al. 2013). The input to PyCBC is the calibrated strain data from the detectors, which undergo a series of conditioning steps to prepare for analysis. After removing invalid or contaminated strain data flagged by CAT1 vetoes, we then apply high-pass filtering to suppress low-frequency noise (below $15\,{\rm Hz}$), downsampling from $16\,{\rm kHz}$ to $2\,{\rm kHz}$ to reduce data volume, and gating (windowing out) to remove loud glitches (Usman et al. 2016). The pipeline is also limited to analyze data segments with a minimum length of $500\,{\rm s}$, to ensure a sufficiently precise PSD estimate.

After data selection and input conditioning, data from each detector are filtered using a fixed template bank. The templates are placed in a four-dimensional parameter space of component detector-frame masses and orbit-aligned spins. For O4, the template bank is generated with a hybrid geometric-random placement algorithm (Roy et al. 2017), imposing a minimum template waveform duration of $70\,{\rm ms}$ which allows coverage of systems with IMBHs, and applying a dynamic minimal match criterion to ensure smooth template density across the mass range (Roy et al. 2017, 2019). The match is calculated assuming the projected LIGO O4 sensitivity that predicts a $160\,{\rm Mpc}$ BNS range (LIGO Scientific Collaboration et al. 2022). The waveform model used is SEOBNRv5_ROM (Pompili et al. 2023) for total mass above $4\,M_\odot$, and TaylorF2 (Vines et al. 2011) below. A fixed lower cutoff frequency of $15\,{\rm Hz}$ is applied for systems with total mass above $100\,M_\odot$. The bank covers a total mass range of $2\,M_\odot$ to $500\,M_\odot$, with mass ratios from $1/97.989$ to 1, and aligned spin ranges from $-0.997$ to $0.997$ for BH components and from $-0.05$ to $0.05$ for NS components, as shown in Figure 2.

Data from each detector are independently matched filtered against the template bank, producing an SNR time series for each template: triggers, corresponding to potential signals, are then identified as local SNR maxima above a given threshold within predefined time windows. A $\chi^2$ test is applied to assess whether the time–frequency distribution of power in the data matches the expected distribution from the template waveform (Allen 2005). The pipeline re-weights the matched-filter SNR using the reduced $\chi^2$ normalized such that the expected value in Gaussian noise for a signal matching the template is unity (Abadie et al. 2012b; Usman et al. 2016). Triggers with re-weighted SNR below a threshold of 4 are discarded as they are likely to correspond to noise transients. Since short-duration blip glitches (Cabero et al. 2019) may pass the time–frequency $\chi^2$ test for some templates, a high-frequency sine-Gaussian $\chi^2$ test is used to further identify such glitches (Nitz 2018). A single-detector ranking statistic $\hat{\rho}$ is then calculated, incorporating the time–frequency and sine-Gaussian $\chi^2$ tests, as well as a correction to the SNR due to short-term PSD variability (Mozzon et al. 2020).

The next step of the pipeline is to identify candidates consistent with potential GW signals by comparing the coalescence times and template parameters of triggers from multiple detectors. Coincident candidates are found if triggers from two or more detectors, associated with the same template, occur in a time window that accounts for the light travel time between detectors and a small margin for timing errors. If data from more than two detectors are analyzed, coincident candidates may be formed across multiple combinations of two and three detectors via the same process. Starting in O4, candidates consisting of triggers in only a single detector are also included (Davies & Harry 2022).

Following this, the pipeline calculates the statistical significance of candidates by estimating their FAR. A ranking statistic is assigned to each candidate, reflecting its likelihood of being a true GW signal versus background noise. The FAR is computed for a coincident candidate by comparing its ranking statistic to a set of artificially generated background candidates created by time-shifting triggers in one detector relative to another. In O4a, where only LIGO Hanford Observatory (LHO) and LIGO Livingston Observatory (LLO) data are available, background candidates are generated by repeatedly shifting triggers in one detector with respect to the other by a fixed interval. To ensure these candidates are statistically independent, clustering is performed on both coincident (un-shifted) and time-shifted analyses: only the candidate with the highest detection value within a set time window is kept. To reduce background contamination due to loud signals, any candidate detected with FAR below a threshold of 1 per 100 yr is removed from the background estimation for less significant candidates (Abbott et al. 2016b; Nitz et al. 2019).

For single-detector candidates in O4, since their FARs cannot be estimated from time-shifted analysis, the ranking statistic distribution is fitted with a falling exponential, and extrapolated with a maximum possible inverse false alarm rate (IFAR) assignment of $1000\,\mathrm{yr}$ (Davies & Harry 2022). In times when multiple detectors are active, FAR estimates from all candidate types at the ranking statistic value of the candidate are added to give the final FAR estimate.

The ranking statistic assigned to both coincident and single-detector candidates is designed to optimize the detection rate by reflecting the relative densities of signal vs. noise over the binary source parameter space. To ac-count for noise variations across different templates, the single-detector background distributions are fitted, with the fit parameters allowed to vary over the template parameter space (Nitz et al. 2017). We define an intermediate expression $R_3$, which is the basis for both the multi-detector and single-detector statistic:

$$
\begin{aligned}
R_3 &= \ln\left( \frac{\sigma_{\min,i}^3 / \bar{\sigma}_{\mathrm{HL},i}^3}{A_{\mathrm{N}\{d\}} \sum_d r_{d,i}(\hat{\rho}_d)} \frac{p(\boldsymbol{\Omega}|\mathrm{S})}{p(\boldsymbol{\Omega}|\mathrm{N})} \right) \\
&= \ln p(\boldsymbol{\Omega}|\mathrm{S}) - \ln p(\boldsymbol{\Omega}|\mathrm{N}) - \ln A_{\mathrm{N}\{d\}} \\
&\quad - \sum_d \ln r_{d,i}(\hat{\rho}_d) + 3(\ln \sigma_{\min,i} - \ln \bar{\sigma}_{\mathrm{HL},i}).
\end{aligned}
\tag{6}
$$

Here $p(\boldsymbol{\Omega}|\mathrm{S})$ and $p(\boldsymbol{\Omega}|\mathrm{N})$ are the probabilities of a signal or noise candidate, respectively, to have extrinsic parameters $\boldsymbol{\Omega}$, comprising relative phases, arrival times and amplitudes between detectors (Nitz et al. 2017). We consider the noise distribution $p(\boldsymbol{\Omega}|\mathrm{N})$ as uniform for a given combination of detectors, absorbing this constant factor into the normalization of $p(\boldsymbol{\Omega}|\mathrm{S})$. The rate of noise events is predicted from the allowed time window $A_{\mathrm{N}\{d\}}$ for coincident events involving detectors $\{d\}$ and the expected rate of noise triggers $r_{d,i}(\hat{\rho}_d)$ in template $i$ at re-weighted SNR $\hat{\rho}_d$. The last term accounts for the time-dependent rate of signals, which is proportional to the network sensitive volume for a given template, and scales with the cube of the minimum sensitive distance $\sigma_{\min,i}$ over triggering detectors, normalized to an average LIGO sensitive distance $\bar{\sigma}_{\mathrm{HL},i}$.

The ranking statistic used in O4 includes two additional terms beyond Equation (6). First, an explicit model of the signal distribution over binary masses and spins is incorporated, intended to optimize the detection rate of the known CBC population while remaining sensitive to so far unpopulated parameter regions (Kumar & Dent 2024). Second, the noise event rate model is updated to account for variations in the rate of noise artifacts, correlated with auxiliary channels that monitor the detector state or environmental disturbances, as measured by IDQ (Essick et al. 2020). We then have:

$$
\begin{aligned}
R_{\mathrm{O4}} &= R_3 + \ln d_{\mathrm{S}}(\boldsymbol{\theta}_{\mathrm{int}}) - \ln d_{\mathrm{N}}(\boldsymbol{\theta}_{\mathrm{int}}) \\
&\quad - \sum_d \ln \delta_d\left(\Delta_d(t), B_d(\boldsymbol{\theta}_{\mathrm{int}})\right),
\end{aligned}
\tag{7}
$$

where $d_{\mathrm{S}}(\boldsymbol{\theta}_{\mathrm{int}})$ and $d_{\mathrm{N}}(\boldsymbol{\theta}_{\mathrm{int}})$ represent a KDE-based model distribution of signals over template parameter space $\boldsymbol{\theta}_{\mathrm{int}}$, and the distribution of template points, respectively. The time-dependent excess rate of non-Gaussian triggers is noted as $\delta_d(\Delta_d(t), B_d(\boldsymbol{\theta}_{\mathrm{int}})) \geq 1$ for a detector $d$. Our estimate of $\delta_d$ depends on a discrete detector state variable $\Delta_d(t)$ and on an index $B_d(\boldsymbol{\theta}_{\mathrm{int}})$ labelling template bins. We consider three possible detector states, corresponding to (i) auxiliary channels indicating significant likelihood of noise artefacts; (ii) presence of loud gated glitches nearby in time, or (iii) neither. In the case of single-detector candidates, the ranking statistic omits terms that depend on multi-detector coincidence quantities (Davies & Harry 2022); specifically, the terms $A_{\mathrm{N}\{d\}}$, $p(\mathbf{\Omega}|\mathrm{S})$, and $p(\mathbf{\Omega}|\mathrm{N})$ present in Equation (6) are omitted.

The probability of astrophysical origin $p_{\mathrm{astro}}$ is estimated by a Poisson mixture inference (Abbott et al. 2016a,c), for which the distribution of signal events over the ranking statistic is estimated via histograms from a set of simulated signals and distributions of noise events are estimated via histograms of time-shifted background for coincident candidates, or via exponential fits for single-detector candidates (Davies & Harry 2022). Such histograms are generated separately for each type of (single-detector or coincident) candidate in single-detector or multi-detector observing time to allow for differences in event rates and distributions (Dent 2025). Marginalization over the unknown rate of astrophysical signals is performed by generalized Gauss–Laguerre quadrature (Creighton 2019). The resulting candidate $p_{\mathrm{astro}}$ values are apportioned between different astrophysical categories using an estimate of the source chirp mass based on search outputs (Dal Canton et al. 2021; Villa-Ortega et al. 2022).

Several features of the PyCBC online search analysis (Dal Canton et al. 2021; Nitz et al. 2018) differ from the offline search analysis described above. The main differences in O4a, and their consequences for online results, are:

- To reduce latency, multiple overlapping matched-filter analysis segments with a stride of $8\,\mathrm{s}$ are used, in contrast to a typical stride of hundreds of seconds for offline search.

- The bank construction is comparable to the offline search except that templates with duration below $0.15\,\mathrm{s}$ are not included, increasing robustness to loud, short-duration noise transients. The SEOBNRv4_ROM model is used instead of the more recent SEOB-NRv5_ROM.

- The ranking statistic for multi-detector candidates is simpler than for the offline search (Nitz et al. 2017, Equation 2): in addition to the quadrature sum of re-weighted SNRs over detectors, it includes only the network consistency term $\ln(p(\mathbf{\Omega}|\mathrm{S})/p(\mathbf{\Omega}|\mathrm{N}))$.

- The significance of multi-detector candidates is initially calculated using only the two most sensitive detectors, with other detectors (if available) incorporated via a follow-up procedure.

- To account for possible short-term changes in detector behaviour, the background for multi-detector candidates is estimated from the preceding $\sim 5$ hours of data in each of the two most sensitive detectors. The FAR estimate is limited in precision by this restricted data extent, to a minimum of $1$ per $100\,\mathrm{yr}$ in the two-detector case.

- Single-detector candidates are generated only for templates with duration above $7\,\mathrm{s}$, as these correspond to systems with potential electromagnetic emission (Foucart et al. 2018).

- The probability of astrophysical origin is estimated via simple approximations to the densities of signal and background events (Dent 2023), in contrast to the detailed histogram estimates used by the offline calculation.

- After the initial candidate upload to GRACEDB, a follow-up analysis is performed to maximize the network SNR over template masses and aligned spin components using all available detector data, and thus to optimize the resulting spatial localization, within minutes.

- An early-warning search algorithm is also operated, targeting nearby mergers only, with component masses between $1\,M_\odot$ and $3\,M_\odot$ and zero spins (Nitz et al. 2020a).

The above descriptions concern PyCBC-based analyses in the first part of O4. Here, for completeness, we summarize significant changes in search methods used for archival catalog (offline) results since the GWTC-1.0 release. The O3 template bank was constructed using the same hybrid geometric-random method as in O4 to ensure efficient coverage of the parameter space (Roy et al. 2017). Three different independent search analyses (PyCBC-BBH, PyCBC-Broad, and PyCBC-IMBH) were used, with each covering distinct regions of the parameter space (Abbott et al. 2023a; Chandra et al. 2021). The waveform model SEOBNRv4_ROM was used for systems above $4\,M_\odot$, and TAYLORF2 for lower-mass systems (Roy et al. 2019). Templates with durations below $0.15\,\mathrm{s}$ were excluded to reduce false alarms from short glitches, thereby also placing an upper limit on the mass of the systems, unlike in O4, where shorter-duration templates are included to improve coverage of high-mass systems. The template bank used in the earlier O2 and O1 analyses (Dal Canton & Harry 2017) spanned total masses from $2\,M_\odot$ to $500\,M_\odot$, with mass ratios from $1/98$ to $1$, and spin limits based on component mass: $0.05$ for masses below $2\,M_\odot$ and up to $0.998$ for higher masses (Abbott et al. 2019a). For O3, the background estimation was extended to a three-detector network by using fixed relative time shifts between detectors (Davies et al. 2020), whereas in earlier runs, a two-detector method (also used in O4a) was employed. Unlike in O4, single-detector candidates were not included in O3 or previous runs. The O3 broad search statistic is given by Equation (6), i.e., the O4 statistic without the additional KDE and data-quality terms. For the O3-BBH search, an alternate ranking statistic was adopted, incorporating an additional term in the broad search statistic dependent on the template chirp mass to enhance sensitivity to the observed BBH population (Nitz et al. 2020b). In earlier runs, only one type of event (LHO–LLO) was considered, and sensitivity variations across the parameter space, such as those described by $\sigma_i$, were neglected. As a result, many terms in the O3 statistic in Equation (6) either disappeared or became constants and were omitted (Nitz et al. 2017).

### 3.5. SPIIR

SPIIR is a coherent search pipeline developed primarily for the online detection of GW signals from CBCs (Hooper et al. 2012; Luan et al. 2012; Chu et al. 2022). It became operational during O3, during which it registered 38 out of the 56 non-retracted public alerts (Abbott et al. 2021a, 2024; Chu et al. 2022). The design of SPIIR focuses on efficiency and rapid processing, enabling it to provide detections with latencies around 11 s. This capability makes SPIIR particularly useful for issuing early warning alerts for electromagnetic follow-ups of GW candidates (Kovalam et al. 2022). While an offline version of SPIIR is under development, it has not yet participated in any of the offline searches through the end of O4a.

The SPIIR pipeline employs infinite impulse-response (IIR) filters as template banks to perform matched filtering directly in the time domain (Hooper et al. 2012). Unlike traditional methods, which can be time-consuming, the IIR filtering approach allows SPIIR to execute matched filtering rapidly. The primary advantage of IIR filters is their ability to approximate matched-filtering SNR in a way that can be efficiently parallelized and executed on graphics processing units (GPUs), which significantly accelerate high-performance computing tasks (Liu et al. 2012; Guo et al. 2018). The SPIIR template-bank generation is a two-step process. The first step is to generate the template points in the CBC parameter space using the stochastic placement algorithm (Privitera et al. 2014; Harry et al. 2009) with a minimum match of $0.97$. The approximant SPINTAYLORT4 (Klein et al. 2014) is used for chirp mass below $1.73\,M_\odot$ and SEOBNRv4_ROM for larger masses. In the second step, each corresponding waveform is decomposed into approximately 350 IIR filters (Hooper et al. 2012; Liu et al. 2012; Guo et al. 2018). These IIR filters are then utilized for matched filtering using GPU parallelization.

The real-time data analysis process begins by retrieving strain data from a shared directory at the Caltech computing cluster for all detectors in observing mode. The data are then conditioned, applying data-quality vetoes, downsampling from $16\,384\,\mathrm{Hz}$ to $2048\,\mathrm{Hz}$, and whitening, which requires real-time PSD calculation. The PSD calculation takes place using the Welch method with tens of $4\,\mathrm{s}$ overlapping data blocks. This method prevents sporadic transient signals from biasing the PSD estimate. The data are also subjected to gating to mitigate the effects of loud transient noises. Once conditioned, the whitened data from each detector is passed through a set of IIR filters that span the parameter space of CBC templates, generating corresponding SNR time series for each template.

A key distinguishing feature of the SPIIR pipeline is its coherent search algorithm (Bose et al. 2011; Harry & Fairhurst 2011). The coherent search algorithm looks for time and phase consistency across all detectors. SPIIR calculates a maximum network likelihood ratio statistic to evaluate the likelihood of a coherent GW signal being present in the data. This approach enhances the pipeline's sensitivity, particularly to weaker signals, and improves localization accuracy by combining data from multiple detectors.

The coherent analysis can be computationally demanding. To address this challenge, SPIIR uses SVD (Wen 2008), which allows for efficient calculation of the coherent statistic by focusing on the most significant signal components. This optimization reduces the complexity of the search and enables rapid scanning across different sky positions to localize the source.

After determining the coherent SNR, SPIIR applies a $\xi^2$ signal-consistency test (Messick et al. 2017) to asses the morphological consistency between data and template and combines it with the coherent SNR into a multidimensional ranking statistic. FARs are assigned by comparing candidates with background distributions generated from time-shifted analyses. To ensure robust estimation, the pipeline collects at least one million background candidates, typically accumulated over several hours. A minimum of one week of background data are used to assign FARs, which is extrapolated using a $k$-nearest-neighbor (KNN) KDE method with $k = 11$, smoothing the probability density using neighboring bins. This approach allows accurate FAR estimation even for rare, high-significance candidates. To account for non-stationary noise, SPIIR additionally maintains shorter-term background sets over $2\,\mathrm{h}$ and $1\,\mathrm{d}$. Furthermore, it stores individual single-detector SNR and $\xi^2$ values and applies the same KNN-based extrapolation method to estimate single-detector FARs used in veto stages before making the final decision about the trigger's validity.

In preparation for O4, SPIIR has undergone several upgrades, with some already implemented and others still in testing and review. These improvements include: automated signal trigger removal from the background estimation, preventing astrophysical triggers from affecting the background calculation; fine tuning of the signal consistency test $\xi^2$, aiming to improve sensitivity to BNS and NSBH; and a dynamic approach to background collection. Furthermore, SPIIR has implemented the necessary infrastructure to process KAGRA data.

### 3.6. Criteria for Inclusion in the Catalog

Search pipelines can in principle produce a rate of candidates as high as one every few seconds, depending on their configuration choices. With the sensitivity of current detectors, however, the vast majority of such candidates would have low significance and would be unlikely to have an astrophysical origin. For practical consideration, we therefore select a smaller number of candidates to report in the GWTC.

The criterion for selecting which candidates to include has evolved with the observing runs and GWTC versions. In GWTC-1.0 (Abbott et al. 2019a), we published all candidates from O1 and O2 for which at least one matched-filtering pipeline (PYCBC or GSTLAL at the time) estimated a FAR below 1 per $30\,\mathrm{d}$. GWTC-2.1 (Abbott et al. 2024) and GWTC-3.0 (Abbott et al. 2023a) respectively added candidates from first half of the third observing run (O3a) and O3b with a FAR below 1 per $2\,\mathrm{d}$ in any pipeline. Following this logic, we now add candidates from O4a whose FAR is less than 1 per $2\,\mathrm{d}$ in any of the pipelines described in Section 3, with the exception of SPIIR.

Stricter candidate selection criteria are typically applied when performing downstream analyses and for the purpose of displaying lists and properties of the candidates. Stricter criteria might, for example, be motivated by computational or person-power considerations. Furthermore, downstream analyses might not necessarily have the same requirements; for instance, a sufficiently large SNR might be necessary, as opposed to a sufficiently low FAR. Such considerations, and consequently the corresponding selection criteria, might in turn evolve as data-analysis methods become more efficient. Stricter criteria will therefore be described wherever necessary.

### 3.7. *Search sensitivity*

Apart from the candidates themselves, an important product of search pipelines is a measurement of their sensitivity, which is necessary to diagnose the behavior of the pipelines and to infer the rate density of the astrophysical sources. We quantify the sensitivity of the search pipelines described above by their estimated time–volume product or hypervolume $\langle VT \rangle$ (Abbott et al. 2023a). The hypervolume represents the sensitivity of a given search to a set of sources assumed uniformly distributed in both comoving volume and source-frame time. An estimate of the total number of signals $\hat{N}$ that a pipeline is likely to detect during a period of observation is given by

$$\hat{N} = \langle VT \rangle \mathcal{R}, \tag{8}$$

with $\mathcal{R}$ the rate of CBCs per unit comoving volume and source-frame time (Abbott et al. 2016a). In practice, hypervolumes are obtained through weighted Monte Carlo simulations, which use simulated GW signals referred to as *injections*.

At this time, the population of injected signals focuses on CBCs only, as no transient GW signal so far detected is inconsistent with a CBC source at high confidence. The parameter ranges and distribution models were chosen to represent CBC signals the detectors were possibly sensitive to, based on previous releases of GWTC (Abbott et al. 2023b). The rate of injections was chosen to be much higher than the astrophysical rate in order to reach a sufficient precision in $\langle VT \rangle$. As the production of injections required detector PSDs to be measured over each month of observation, injection analysis was performed offline. To ensure consistency in the $\langle VT \rangle$ measurements, all pipelines contributing GWTC candidates (GSTLAL, MBTA, PYCBC, and CWB-BBH) analyzed the same sets of injections. The pipelines processed data with injections using methods equivalent to those used to produce candidates from injection-free data, described earlier in this section. All pipelines used background data collected by their injection-free analyses to assign significances to the recovered injections. The choice of population priors and the injection generation process are detailed in Essick et al. (2025), while results from the injection analysis are described in Abac et al. (2025b).

The sensitive $\langle VT \rangle$ differs between pipelines, reflecting their abilities to detect CBCs within a given range of parameters as well as their different effective live times. Furthermore, as CWB-BBH is designed to search for BBH signals as described in Section 3.1, we calculate $\langle VT \rangle$ for the corresponding masses for this pipeline.

There are differences in how pipelines process injections compared to injection-free data, which arise from computational efficiency or technical considerations and do not affect the sensitivity estimation. The CWB-BBH pipeline performed injection generation in a $5\,\mathrm{s}$ time window around the time of the injections. The GSTLAL and PYCBC pipelines processed data with injection using subsets of their template banks based on the known chirp mass of the injections, in order to limit the computational cost. In addition, GSTLAL had two other major differences for its injection processing compared to its injection-free data processing. First, injection processing was done entirely in an offline configuration, as opposed to reassigning significances to triggers obtained in online analysis, due to injection data not being available at the time of the online analysis. Meanwhile, other pipelines processed both injection-free and injection data in their offline configurations to produce the GWTC-4.0 results and $\langle VT \rangle$ estimations. Second, the SNR time series were only computed for a short time window around the time of the injections to limit the computational cost, as opposed to computing the SNR time series for all available data in O4a.

## 4. DATA QUALITY AND CANDIDATE VETTING

A subset of the candidates produced by the search algorithms described in Section 3 undergo studies to understand the quality of the interferometric data surrounding the candidates with a goal to characterise and, if required, mitigate against any departures from stationary Gaussian noise such as glitches. This is vital to ensure that downstream analyses, particularly PE (see Figure 1 and Section 5), which typically assume stationary Gaussian noise, produce unbiased results (Pankow et al. 2018; Powell 2018; Ghonge et al. 2024)

Event validation is the process of identifying and examining potential data-quality issues in the vicinity of a GW candidate. We apply the process to all candidates reported by online search pipelines and labeled as significant, with an internationally-coordinated team who can provide rapid vetting and aided by an automated data-quality report framework (Soni et al. 2025).

In contrast to O3, where LIGO and Virgo conducted separate validation processes (Davis et al. 2022; Acernese et al. 2023), O4 employs a unified infrastructure across all active detectors. This uniformity in validation improves efficiency, streamlines information flow between different analysis groups, and reduces the demand on human resources. Once the initial online vetting is completed, all significant candidates undergo further scrutiny to determine a final assessment of the data quality status.

If data-quality issues such as glitches are identified around a candidate and these issues are not severe enough to warrant a retraction, we quantify if further action is required by comparing the PSD variance in the relevant region with expected Gaussian noise and computing a $p$-value (Mozzon et al. 2020). Until November 2023, a conservative threshold of 0.1 was adopted. This threshold was then relaxed to 0.05. For $p$-values greater than this threshold, we conclude the data are consistent with Gaussian noise, and the candidates are ready for downstream analyses. On the other hand, if $p \leq 0.05$, we attempt to subtract the noise using either a modeled Bayesian inference approach implemented in BAYESWAVE (Pankow et al. 2018; Cornish et al. 2021; Chatziioannou et al. 2021; Hourihane et al. 2022), or a linear noise subtraction based on auxiliary witness channels (Davis et al. 2022). For all candidates in O4a, BAYESWAVE was used and applied to the GDS-CALIB_STRAIN_CLEAN_AR channel as described in Abac et al. (2025c). For past observing runs, BAYESWAVE has predominantly been applied, except in cases where a witness channel was available (Abbott et al. 2023a). However, if the noise is extended in time or frequency, as is often the case for certain glitch classes (e.g., Soni et al. 2020, 2024), making subtraction difficult, or if the data remains insufficiently stationary after subtraction, we instead restrict the time and frequency analysis window to avoid the afflicted region. For badly afflicted candidates, preliminary parameter-estimation studies are routinely performed to validate the mitigation technique and understand the impact on parameter estimates. Once a satisfactory mitigation approach has been identified, the candidates are ready for downstream analyses.

# 5. PARAMETER ESTIMATION FOR CBC

For a selection of candidates identified by the search algorithms (see Section 3) and either having passed validation or with appropriate mitigation (see Section 4), we use Bayesian inference to estimate the source parameters $\boldsymbol{\theta}$ of the GW signal (see Figure 1), which includes both the intrinsic parameters of the CBC source and the extrinsic parameters that localize and orient the source in spacetime (Abac et al. 2025a). These inferences come in the form of posterior distributions $p(\boldsymbol{\theta}|d)$ over the source parameters. The posteriors represent our best understanding of the properties of the source and its location, including all correlations and degeneracies in the measured parameters. The posteriors for the selected GW candidates from O4a are reported for the first time in (Abac et al. 2025b). We have not updated our inferences for candidates identified in previous observing runs. Thus, our preferred parameter inferences for candidates identified before O4a remain the same as previously reported in Abbott et al. (2024, 2023a), with the exception of the BNS candidate GW170817 whose preferred parameter inferences are reported in Abbott et al. (2019a).

In Sections 5.2 to 5.8, we describe the broad elements of GW PE and discuss the particular choices made in each version of the catalog in Section 5.9, noting that configuration files to reproduce the analysis of individual candidates are provided with the data release.

## 5.1. Bayesian Formalism

To carry out Bayesian PE, we assume the data $d$ contains colored Gaussian noise and an astrophysical GW signal which is well-approximated by a quasi-circular CBC waveform model (see Section 2), consistent with GR and absent of environmental effects. These assumptions determine the form of our likelihood $p(d|\boldsymbol{\theta})$, which for a single detector is Gaussian in the residuals between data and waveform (Finn 1992; Cutler & Flanagan 1994; Abac et al. 2025a). We use the standard formulation for this likelihood and the associated frequency-domain noise-weighted inner product (e.g., Veitch et al. 2015b; Thrane & Talbot 2019). Given the likelihood and a prior distribution $\pi(\boldsymbol{\theta})$, we compute the posterior distribution from Bayes' theorem

$$p(\boldsymbol{\theta}|d) = \frac{p(d|\boldsymbol{\theta})\pi(\boldsymbol{\theta})}{p(d)}, \qquad (9)$$

where $p(d)$ is the normalizing evidence:

$$p(d) = \int p(d|\boldsymbol{\theta})\pi(\boldsymbol{\theta}) \, d\boldsymbol{\theta}. \qquad (10)$$

The evidence is not required in most approaches to PE, but can be used for model comparison.

In our baseline analyses the joint likelihood also includes additional parameters to account for the calibration state of the detector and the uncertainty in this calibration (Section 5.4). The full likelihood is evaluated coherently across the detector network by multiplying the individual interferometer likelihoods, under the assumption that the noise is independent in each.

## 5.2. Preliminary PE

After initial identification of a candidate by a search algorithm (Section 3), we carry out preliminary PE in order to understand the properties of the source and guide our analysis settings. We choose our initial settings and waveform model using the output of the search pipelines, e.g., the point estimates of the source masses. As needed, we carry out further analyses with modified priors or configuration settings, in some cases utilizing multiple waveform models to understand systematic modeling errors. If there are data-quality issues, this preliminary estimation is iterated alongside glitch-mitigation studies, as described in Section 4.

Preliminary PE also guides our choice of waveform models for inference (Section 2). Specifically, since $3 \, M_\odot$ provides a stringent upper limit on the maximum mass of a NS (Rhoades & Ruffini 1974; Kalogera & Baym 1996), candidates where both source masses are $> 3 \, M_\odot$ are classified as unambiguous BBH candidates. Otherwise the source potentially contains a NS, and may be a NSBH or BNS. For sources which may include a NS we carry out PE with waveforms that include the imprint of matter onto the GW signal in order to constrain such effects.

### 5.3. *Likelihood Calculation: Elements of the Inner Product*

The integration of the inner product that defines the likelihood is carried out in the frequency domain (Abac et al. 2025a; Veitch et al. 2015b; Thrane & Talbot 2019). Following our standard conventions (Abac et al. 2025a), given a discrete raw time-series data $d(t)$ with a duration $T$ and sampling frequency $f_s$, we first apply a FFT and divide by the sampling frequency to produce the discrete frequency series $\tilde{d}$. Similarly, time-domain waveforms are transformed into frequency series, while frequency-domain models can be evaluated directly for use in the inner product.

The amount of data analyzed for each candidate depends on the estimated mass of the system as determined by preliminary PE analyses, with lower mass signals in the sensitive band for longer periods of time. The segment of data used in PE is selected with the estimated time of coalescence set 2 s before the end of the segment, with the total duration set to powers of 2 ranging from 4 s up to 128 s such that the evolution of the signal starting from $f_{low}$ is included. This choice sets the discrete frequency spacing $\Delta f = 1/T$ in the inner product. Before transforming time-domain data to the frequency domain to evaluate the likelihood, we apply a symmetric Tukey window function (Harris 1978) with a roll-on time that balances how much data in the segment is affected by the windowing and the impact of spectral leakage of the noise, especially at the lowest frequencies analyzed where the noise is steeply varying.

We compute the required integral from a lower frequency $f_{low}$ to $f_{high}$ and discard the data outside this range. Generally, $f_{low}$ is set to 20 Hz, below which the detector noise steeply increases. The upper frequency $f_{high}$ is chosen to be the Nyquist frequency, half the discrete sampling rate $f_s$ of the data, with further modifications to account for power lost due to applying a low-pass Butterworth filter when downsampling the data to the desired sampling rate (Veitch et al. 2015a; Romero-Shaw et al. 2020b). The result is $f_{high} = \alpha^{\text{roll-off}} f_s/2$, with the particular choice of $\alpha^{\text{roll-off}} = 0.875$ selected to limit power loss to 1% (Abbott et al. 2023a). The data are downsampled both to limit the computational cost of the likelihood evaluation, and $f_s$ is chosen for each candidate to ensure that selected higher multipolar modes are resolved for candidates which coalesce in the sensitive frequency band of the detectors, or else set to a high enough value where noise dominates the signal power for low-mass candidates that coalesce out of the sensitive band. This upper limit is generally $f_s = 4096$ Hz or $f_s = 8192$ Hz.

A key ingredient in the likelihood evaluation is an estimate of the noise PSD for the analyzed data. In order to mitigate the effects of the nonstationary nature of the instrumental noise, for PE results presented in the GWTC, we use an on-source estimate for the PSD generated with the BAYESWAVE algorithm (Cornish & Littenberg 2015; Littenberg & Cornish 2015; Cornish et al. 2021). This approach uses Bayesian inference to construct a posterior distribution over PSD realizations, and we take the median PSD value at each frequency (Chatziioannou et al. 2019) for the point-estimate used in the inner product.

### 5.4. *Calibration Marginalization*

In order to account for the uncertain calibration of the strain data, we allow for independent frequency-dependent shifts of the amplitude and phase of the waveform in each interferometer (Farr et al. 2014). The frequency-dependent shifts are modeled using additional free parameters in our waveform model, which we call calibration parameters. We marginalize over the calibration parameters when reporting PE results (Abbott et al. 2016d).

The strain data are produced in near-real time using an initial calibration model, which may be subsequently updated in order to recalibrate the data (Abac et al. 2025c). For the LIGO detectors the methods used to calibrate the strain data provide a measured distribution of frequency-dependent correction factors $\tilde{\eta}_R(f)$ (Cahillane et al. 2017; Sun et al. 2020, 2021; Dartez et al. 2025). These factors correct the calibrated strain data $d$ in each detector to the strain that would be measured with perfect calibration $d_\star$,

$$\tilde{d}_\star = \tilde{\eta}_R \tilde{d}. \tag{11}$$

For the Virgo detector the methods used for calibration give an estimate of the inverse of the correction factor $1/\tilde{\eta}_R$ (which has been unity through O3) and the corresponding frequency-dependent uncertainties (Accadia et al. 2014; Acernese et al. 2018, 2022).

The likelihood used in PE relies on knowledge of the noise properties of the imperfectly calibrated data $d$ through the estimated PSD. The likelihood $p(d|\boldsymbol{\theta})$ is evaluated in the frequency domain using $\tilde{d}$,

$$\tilde{d} = \frac{1}{\tilde{\eta}_R}[\tilde{h}(\boldsymbol{\theta}) + \tilde{n}_\star] = \frac{\tilde{h}(\boldsymbol{\theta})}{\tilde{\eta}_R} + \tilde{n}, \tag{12}$$

where $\tilde{n}_\star$ is the true noise in the detector, assumed to be Gaussian, and $\tilde{n} = \tilde{n}_\star/\tilde{\eta}_R$ is the noise in the calibrated strain, also assumed to be Gaussian. The PSD describes the properties of $\tilde{n}$, so the correction factors $\tilde{\eta}_R$ must be applied to correct the GW model $\tilde{h}$ to account for imperfect calibration. The corrections $1/\tilde{\eta}_R$ are approximated by small frequency-dependent amplitude and phase corrections, which are modeled using splines (Farr et al. 2014). The values of these splines at fixed frequency nodes are the calibration parameters, and for PE we use Gaussian priors on these additional parameters. For LIGO data we use the median and standard deviation of the measured distribution of $\tilde{\eta}_R$ (Cahillane et al. 2017; Sun et al. 2020, 2021; Dartez et al. 2025) to set the mean and standard deviation of the priors over the calibration parameters describing $1/\tilde{\eta}_R$. For Virgo data, the priors on the calibration parameters through O3 are zero-mean Gaussians with standard deviations corresponding to the directly measured uncertainties in $1/\tilde{\eta}_R$.

Up to O3, the data from LHO and LLO were recalibrated prior to final analysis in order to approximately remove the systematic miscalibration arising from the initial model. For PE of these candidates, we used final calibrated data as described in Table 4 of Abac et al. (2025c). Because of this recalibration, the errors and uncertainty on the calibration parameters are small, such that the priors for the calibration parameters are nearly centered at zero (corresponding to no calibration correction). However, in O4a the data in LHO and LLO are no longer recalibrated (Abac et al. 2025c). Nevertheless, recalibration during PE can be effectively carried out using calibration priors centered away from zero. For this reason the priors associated with the new GW candidates presented in GWTC-4.0 may lie more than a standard deviation away from zero for GWTC-4.0 at various frequency points.

### 5.5. *Priors*

To estimate the posterior distribution, Equation (9), we must select an appropriate prior distribution over the binary parameters $\theta$. Since our inferences are made over a catalog of CBC systems with a wide range of properties and using a variety of waveform models, our prior ranges are selected to be appropriate for each candidate. The priors chosen are agnostic and wide enough to cover the region of parameter space where the posteriors have support, while ensuring a reasonable amount of analysis time and accounting for the parameter ranges over which our waveform models are calibrated. Prior ranges (e.g., on the component masses) are selected during preliminary PE analysis and adjusted afterward as needed, for example to ensure that an arbitrary prior boundary does not affect the posterior.

For all candidates we use priors that are uniform over the (redshifted) detector-frame component masses $(1 + z)m_i$, with boundaries in detector-frame total mass, detector-frame chirp mass, and mass ratio appropriate for each candidate and applied model (see Section 2). The priors are uniform on the spin magnitudes and isotropic in spin orientations. We use isotropic priors on the binary orientation, and priors that are uniform in comoving volume and comoving time so that the priors on the sky location are isotropic. For this we carry out sampling in a reference cosmology (Abac et al. 2025a) and reweight our final samples to priors appropriate for the cosmological model of Ade et al. (2016), which is a flat-$\Lambda$CDM model with $H_0 = 67.9\,\mathrm{km\,s^{-1}\,Mpc^{-1}}$ and $\Omega_\mathrm{m} = 0.3065$, following our approach in past versions of the GWTC.

### 5.6. *Sampling from the Posterior*

Due to the high dimensionality of the source parameters $\theta$, brute-force evaluation of the posteriors is intractable. Instead we represent the posteriors for each candidate with discrete samples $\theta_i$ which represent fair draws from the posterior distribution. The challenge then reduces to the problem of stochastically sampling from the posterior distribution $p(\theta|d)$.

A number of algorithms exist to tackle this sampling challenge. These include nested sampling (Skilling 2006), a Monte Carlo technique to estimate the evidence and which produces samples as a byproduct, and Markov chain Monte Carlo (MCMC; Hastings 1970), which samples through random walks and forms the core of many sampling algorithms. The evidence can be computed with MCMC methods by e.g., thermodynamic integration (Goggans & Chi 2004; Littenberg & Cornish 2009) via parallel tempering (Earl & Deem 2005; van der Sluys et al. 2008; Veitch et al. 2015b). For the new candidates presented in GWTC-4.0, we use the DYNESTY (Speagle 2020) nested-sampling algorithm, and accessed through the BILBY package (Ashton et al. 2019; Romero-Shaw et al. 2020b) which includes a custom stepping method optimized for use on parallelized computing resources, specifically large-core-count CPUs.

For PE results presented with previous catalog releases, we have made use of additional, highly parallelized sampling techniques to tackle computationally expensive analyses. These include PARALLELBILBY (Smith et al. 2020) and the RIFT algorithm (Pankow et al. 2015; Lange et al. 2017; Wysocki et al. 2019). PARALLELBILBY is optimized to use distributed computing to perform nested sampling. RIFT is a highly parallel, iterative sampling algorithm that uses adaptive, grid-based explorations of the likelihood while marginalizing over the extrinsic parameters, followed by a final stage of Monte Carlo sampling of the intrinsic and extrinsic parameters. For a small number of candidates, parameter inference was carried out using LALINFERENCE (Veitch et al. 2015b), which provides an implementation of MCMC and nested sampling, as well as tailored methods for proposing new sample points.

To improve sampling performance, certain parameters in the waveform model can be marginalized out of the posterior during sampling. Depending on the waveform model employed and the GW candidate, any of the coalescence phase (Veitch et al. 2015b; Farr 2014), coalescence time (Farr 2014; Romero-Shaw et al. 2020b), or luminosity distance (Singer & Price 2016) have been marginalized over, as discussed in Section 5.9. This can be done analytically or numerically depending on the parameter and the details of the waveform model. When marginalization is carried out, the full posterior distributions are reconstructed in postprocessing (Thrane & Talbot 2019).

The method of marginalizing over calibration uncertainties varies depending on the sampling method used. For inferences carried out using RIFT, the marginalization over the calibration model parameters is carried out in post-processing, by reweighting the likelihood of samples produced first without utilizing the spline calibration model (Payne et al. 2020; Abbott et al. 2023a). Meanwhile, for PE using BILBY, PARALLELBILBY, and LALINFERENCE, the parameters of the calibration model are inferred alongside the intrinsic and extrinsic binary parameters during sampling.

A large number of PE analyses and steps within each analysis is required to produce the inferences included in GWTC-4.0. These analyses were managed using the ASIMOV software library (Williams et al. 2023). Postprocessing of the samples is carried out with the PESUMMARY software library (Hoy & Raymond 2021). This includes the computation of derived parameters such as the remnant properties following BBH merger and the evolution of the spin components to large binary separations as described in Section 2.4. These are implemented with routines in LALSuite (LIGO Scientific Collaboration et al. 2018; Wette 2020) and used by PESummary. For reweighting of samples, e.g., to our preferred cosmological model or to incorporate calibration marginalization with RIFT, either routines in PESummary or simple custom routines (as in the case for new candidates in GWTC-4.0) are used and provided with our data release.

### 5.7. *Posterior Samples*

Following sampling, our measurements of the binary parameters $\boldsymbol{\theta}$ for each candidate are represented by a discrete set of parameter values $\boldsymbol{\theta}_i$ sampled from the posterior distribution. These samples can be used to compute expectation values over quantities of interest, e.g., through Monte Carlo integration. Marginalization is achieved by only considering the dependence of the samples on the quantities that are not marginalized out. Each sample includes the fifteen intrinsic and extrinsic parameters required to describe the GW strain from quasi-circular BBHs (systems which include NSs require an additional tidal parameter per NS) as well as the parameters required to describe the inferred calibration state. In addition to these a number of quantities can be derived from the intrinsic parameters of the binary using the methods described in Section 2.4. These include the final mass $M_f$ and final dimensionless spin $\chi_f$ of the remnant compact object following coalescence, the peak luminosity $\ell_{peak}$ and total energy radiated $E_{rad}$, as well as the final kick velocity of the remnant $v_k$ when the relevant waveform model supports such estimates.

Some intrinsic properties of the binaries vary over the course of the binary coalescence, i.e., the orientations of the dimensionless spins $\boldsymbol{\chi}_1$ and $\boldsymbol{\chi}_2$ in precessing systems. Our initial inferences report these quantities at a reference frequency $f_{ref}$, taken to be 20 Hz, corresponding to a variable reference time before coalescence. The spin tilt angles with respect to the orbital angular momentum, $\theta_1$ and $\theta_2$, in particular carry valuable information about the formation mechanism of the binary. These angles approach well-defined limits at infinite binary separation (Gerosa et al. 2015), where they can be used in population inferences (e.g., Mould & Gerosa 2022), and so we report the spin tilts and derived quantities such as $\chi_{eff}$ and $\chi_p$ in this limit. The spins are evolved from their reference values (Johnson-McDaniel et al. 2022b) using PN expressions for the precession-average dynamics (Gerosa et al. 2015; Chatziioannou et al. 2017) to formally infinite separations (cf. Gerosa et al. 2023). Wherever possible, the more accurate hybrid evolution using a combination of orbit-averaged precession followed by precession-average evolution is employed, but for a small number candidates only the computationally faster precession-averaging approach is used. We report these evolved spin quantities in our final samples for BBH candidates, and for NSBH candidates when analyzing them with waveform models that neglect the imprint of matter onto the GW signal.

In reporting and plotting the inferred parameters for each candidate, we marginalize out all but one or two parameters at a time, and report the median value and credible intervals (CIs) of the marginalized posteriors for those parameters. Unless otherwise noted, we report the median and 90% CI of each binary parameter, marginalizing over the others. Generally we use symmetric CIs for single parameters, so that 5% of the posterior density lies below the lower bound of the 90% CI and 5% of the density lies above the upper bound. In some instances where the posteriors have support near a physical prior boundary, the symmetric CI gives the appearance of excluding the boundary value even if there is high probability there. In such cases we may report the 90% highest posterior density (HPD), the smallest interval containing 90% of the posterior density. When plotting marginalized densities we make use of one- and two-dimensional KDE to produce continuous densities from samples. Our two-dimensional credible regions are constructed using the HPD method.

Our inferences on the location of GW candidates are available in two formats. The samples themselves represent our full posterior over the source parameters, including the sky location and distance of the detected GW sources. In addition, we provide three-dimensional localizations in the same format as the localizations included in our public GW alerts, by applying KDE to the samples. These localizations are created using LIGO.SKYMAP (Singer & Price 2016; Singer et al. 2016a,b), which includes the BAYESTAR package used to localize GW candidates from modeled online and offline searches.

### 5.8. *Approximate Spatial Localization*

We do not carry out full PE for all candidates which meet the criteria for inclusion in GWTC-4.0 (Section 3.6). Similar to the criteria for inclusion in our past catalog releases, the criteria for full PE analysis has evolved with GWTC versions (Section 5.9). However, we provide approximate spatial localization for all candidates in GWTC-4.0 to enable multimessenger analyses of large samples of weak GW candidates using the same methods as for our public GW alerts. Such methods are computationally much cheaper than full PE, and already integrated into the candidate management system. In particular, for candidates produced by modeled CBC searches, the approximate localization is carried out with BAYESTAR (Singer & Price 2016; Singer et al. 2016a,b). For candidates produced by cWB-BBH, the localization method is directly part of cWB and described in Section 3.1.

### 5.9. *Analysis Settings and Details*

Here we describe the particular analysis settings which vary across the inferences carried out for each candidate in GWTC-4.0.

#### 5.9.1. *New Candidates Found in O4a*

We perform full Bayesian PE on a high-purity subset of the candidates (Section 3.6), namely those with FARs $< 1 \, yr^{-1}$

and $p_{astro} > 0.9$, which is the least strict threshold we set for use in further downstream analyses. For this subset, the analysis settings depend on the nature of the binary as inferred using preliminary PE and confirmed in our final analysis. As described in Abac et al. (2025b) these candidates include a large number of BBHs as well as new NSBHs, but no significant BNS candidates.

Based on preliminary PE and any input from data validation (cf. Section 4), we categorize the signals and determine an appropriate prior, waveform model, and data segment to analyze (in all cases, the data product is based on the `GDS-CALIB_STRAIN_CLEAN_AR` channel, as described in Abac et al. (2025c)). For nearly all cases, the durations are selected so that the $\ell = 3$, $|m| = 3$ higher harmonic is fully captured within the frequency range integrated over in our likelihood. In each case it is possible that at the starting frequency of the waveform, even higher harmonics begin in band, which may cause aliasing in the case of time-domain waveform models. Due to the small amplitude of such higher harmonics relative to the lower-frequency multipole moments, the effect is expected to be negligible. Further, we adopt the additional prior during sampling such that the $\ell = 3$ mode is resolved in band, given the chosen sampling rate. This only modifies the prior for a small number of candidates, and we have checked that in these cases, the prior makes no discernible difference in the final posterior samples.

For all candidates the Tukey window applied before transforming time-domain data to the frequency domain has a roll-on of 1 s. This is longer than the 0.4 s roll-on used in prior PE analyses. The longer window is chosen to reduce spectral leakage, because for the first time the instrumental noise near $f_{low}$ is sufficiently low that the small amount of leakage from even lower frequencies into frequency range integrated over in our likelihood made noticeable impact on our analyses.

The analysis of the initial PE results to determine our final settings is carried out using the PECONFIGURATOR software package. The resulting information is also used to determine the appropriate waveform models to use to capture systematic uncertainties across waveform models.

For BBH candidates, we use both the phenomenological IMRPHENOMXPHM_SPINTAYLOR (García-Quirós et al. 2020; Pratten et al. 2020a; Colleoni et al. 2025b) model and the effective one body model SEOBNRV5PHM (Khalil et al. 2023; Pompili et al. 2023; Ramos-Buades et al. 2023) for all candidates. In addition, we use NRSUR7DQ4 (Varma et al. 2019a) for those candidates whose parameters lie within the range of total mass and mass ratio values supported by the model, as determined by preliminary PE. In the case of NRSUR7DQ4 we use a fixed duration of $10000(G/c^3)M(1+z)$ for the waveform model when evaluating the likelihood, with $M(1 + z)$ the total detector-frame mass of the source. For a subset of the candidates that are inferred to have unequal masses or orbital precession, we additionally use IMRPHENOMXO4A (Hamilton et al. 2021; Thompson et al. 2024), since these are cases where systematic differences between waveform models are largest. In the case of the exceptional

BBH candidate GW231123_135430 (Abac et al. 2025d), we additionally employed the time-domain phenomenological model IMRPHENOMTPHM (Estellés et al. 2025a).

For potential NSBHs (where the secondary has posterior support in the range $m_2 < 3M_\odot$ as expected for NSs, and the primary has posterior support in the range $m_1 > 3M_\odot$ as expected for BHs), we perform additional analyses with and without matter effects. For our baseline results we employ our BBH models IMRPHENOMXPHM_SPINTAYLOR and SEOBNRV5PHM in the case of GW230529_181500; Abac et al. 2024 which include higher-multipole emission and precession but neglect matter effects. We additionally use IMRPHENOMNSBH (Thompson et al. 2020) and SEOBNRV4_ROM_NRTIDALV2_NSBH (Matas et al. 2020) which include matter effects tuned to NSBH systems but which neglect precession. We also apply IMRPHENOMPV2_NRTIDALV2 (Dietrich et al. 2019a), which includes precession and tidal effects. Since these latter three models do not incorporate higher multipole emission, we analytically marginalize over the coalescence phase. These three models also differ in the ranges over which the primary spin has been calibrated, and thus we enforce $\chi_1 < 0.5$ for IMRPHENOMNSBH, $\chi_1 < 0.9$ for SEOBNRV4_ROM_NRTIDALV2_NSBH, and allow nearly maximal spins $\chi_1 < 0.99$ for IMRPHENOMPV2_NRTIDALV2. In all three cases we adopt the bound $\chi_2 < 0.05$, corresponding to the largest projected spins near merger for Galactic BNSs which will merge in a Hubble time (Burgay et al. 2003; Stovall et al. 2018). Depending on the model, multiple methods to accelerate the likelihood were employed, including multibanding for IMRPHENOMXPHM (García-Quirós et al. 2021; Morisaki 2021), heterodyning (also called relative binning; Cornish 2010, 2021; Zackay et al. 2018; Krishna et al. 2023) for models including matter effects, and reduced-order quadrature (Canizares et al. 2015; Smith et al. 2016; Morisaki et al. 2023) for some analyses involving IMRPHENOMPV2_NRTIDALV2. Finally, for GW230529_181500, we employ additional models with reduced physical content in order to test for the presence of effects such as precession (Abac et al. 2024).

For all new candidates presented in GWTC-4.0, the luminosity distance is marginalized over during sampling. For sampling we use a distance prior uniform in comoving volume and comoving time with the default cosmological model of the ASTROPY software package, a flat-$\Lambda$CDM cosmology with $H_0 = 67.66 \text{ km s}^{-1} \text{ Mpc}^{-1}$ and $\Omega_m = 0.30966$. The cosmology assumed during sampling is not the default cosmology we present our results in. Instead, the final samples are reweighted to our preferred cosmology (Ade et al. 2016) as described in Section 5.5.

### 5.9.2. Candidates found in O1, O2 and O3

As a cumulative catalog, GWTC-4.0 includes candidates detected during the first three observing runs of the advanced detector network. As discussed in Section 3.6, GWTC-3.0 included GW candidates identified during these runs with FARs less than a threshold value, which is 1 per 30 d for O1 and O2 and 1 per $2\,\mathrm{d}$ for O3. We have previously performed full PE for the subset of these candidates having $p_{\mathrm{astro}} > 0.5$ plus the NSBH candidate GW200105_162426 (Abbott et al. 2021b), and our PE results for these candidates remain the same in GWTC-4.0. These inferences were first presented in GWTC-2.1 (Abbott et al. 2024, for candidates found in O1, O2, and O3a) and GWTC-3.0 (Abbott et al. 2023a, for candidates found in O3b). The exception is the BNS candidate GW170817, for which the PE results were not updated in GWTC-2.1, and so remain the same as those presented in GWTC-1.0 (Abbott et al. 2019a). Both sets of analyses from GWTC-2.1 and GWTC-3.0 used the same methods and settings which we now summarize.

For each candidate, PSDs were generated using BAYESWAVE (Cornish & Littenberg 2015; Littenberg & Cornish 2015; Cornish et al. 2021; Hourihane et al. 2022) as described in Section 5.3. For these candidates the roll-on of Tukey window applied to the time domain-data before transforming to the frequency domain was $0.4\,\mathrm{s}$. Multiple waveform models were used in each case in order to understand systematic uncertainties in the inferences.

For BBH candidates, PE was carried out using two waveform models which include higher harmonics and the effects of orbital precession. The phenomenological model IMRPHENOMXPHM (García-Quirós et al. 2020; Pratten et al. 2020a) was used with a multiscale prescription for the precession dynamics (Chatziioannou et al. 2017) and sampling was carried out with BILBY. The EOB model SEOBNRv4PHM (Ossokine et al. 2020) was also used for each BBH candidate, using RIFT to carry out the PE.

For the NSBH candidates, multiple waveform models were also used for PE. Our baseline results are drawn from the same models as for BBH candidates, namely IMRPHENOMXPHM and SEOBNRv4PHM. These models include higher harmonics and precession but neglect matter effects such as the tidal deformation or disruption of the NS on the waveform. This is because such effects are negligible for the NSBH candidates identified in O3b (Abbott et al. 2021b). In order to assess the importance of matter effects on the GW signal, models which include the effects to tidal deformation and models which are specifically tailored for NSBH systems were also used for inference. These were IMRPHENOMNSBH and SEOBNRv4_ROM_NRTIDALv2_NSBH. As with the new NSBH candidates from O4a, we analytically marginalize over the coalescence phase of these latter models which neglect higher multipolar emission. For each model two analyses were performed, one restricting the dimensionless spin magnitude on the secondary to $\chi_2 \leq 0.05$, and the second allowing it to range to $\chi_2 \leq 0.99$ or the largest value allowed by the model. These inferences are computationally expensive, and a number of samplers were employed depending on the waveform model, including DYNESTY as accessed through BILBY, RIFT, PARALLELBILBY, and LALINFERENCE.

Up to GWTC-4.0, two BNS candidates have been identified, GW170817 and GW190425. Multiple models are used for PE of GW170817 (Abbott et al. 2019a). GW170817 was analysed with three frequency-domain models using LALINFERENCE: TAYLORF2 including tidal effects (Sturani 2015; Isoyama et al. 2020; Flanagan & Hinderer 2008; Vines et al. 2011), IMRPHENOMPv2_NRTIDAL, and SEOBNRv4_ROM_NRTIDAL, and with two time-domain models using RIFT: SEOBNRv4T (Hinderer et al. 2016; Steinhoff et al. 2016), and TEOBRESUMS (Nagar et al. 2018). For this candidate we marginalize over the phase analytically for models that assume spins aligned with the orbital plane. As with our other analyses on candidates including NSs, multiple analyses are carried out allowing only for relatively small spin magnitudes $\chi_i \leq 0.05$ and allowing for spin magnitudes up to the maximum allowed for a given model, as large as $\chi_i \leq 0.99$. The second BNS candidate, GW190425 (Abbott et al. 2020c), was detected during O3a and its PE results updated in GWTC-2.1. For this candidate we present results using the precessing, tidal approximant IMRPHENOMPv2_NRTIDAL using DYNESTY through BILBY, and both relatively low- and high-spin prior limits on the component spins. To accelerate inference we employ reduced-order quadrature.

When using BILBY or PARALLELBILBY to analyze candidates from the first three observing runs, the posteriors are marginalized over luminosity distance and geocenter time, with the exception of the BILBY analysis of GW190425 which only used distance marginalization. For these candidates, initial sampling was carried out with a distance prior uniform in Euclidean volume. During postprocessing, the posterior samples were then reweighted to the cosmological model described in Section 5.5 (Ade et al. 2016).

### 5.9.3. Calibration Prior Settings for Candidates from O1, O2, and O3

We marginalized over the calibration uncertainties when producing the PE results from GWTC-1.0, GWTC-2.0, GWTC-2.1 and GWTC-3.0, as discussed in Section 5.4. Due to an error in implementation, an incorrect prior on the calibration parameters for LHO and LLO was used for these results. The priors were set using the median and CIs of $\tilde{\eta}_R$ from Equation (11) for data from the LIGO detectors, rather than those of $1/\tilde{\eta}_R$ as required by the method. In the limit of small calibration uncertainties, this amounts to a sign error in the means of the Gaussian priors for the calibration parameters. In the case that the means of the priors are zero the error has no effect. However, the calibration uncertainties for LHO and LLO have nonzero means. Meanwhile, the priors on the calibration parameters were set correctly for Virgo data.

For candidates detected in O1, O2, and O3 the means are generally small relative to the standard deviation of the priors on the calibration parameters. Further the absolute sizes of the standard deviations of these parameters are small in the sensitive band of the detectors, of the order of a few percent in amplitude and a few degrees in phase for the two LIGO interferometers (Abbott et al. 2019a, 2021a, 2024, 2023a), and so *a priori* the impact of this error on our inferences is expected to be small. We have verified this expectation through preliminary re-analysis of the potentially impacted candidates, carried out by repeating PE with corrected priors and by reweighting the likelihood values of existing PE samples in order to correct for the erroneous calibration priors. None of the scientific conclusions reported in previous studies is affected by the error. In addition, the error does not impact the significance of any of our candidates, since the GW searches described in Section 3 do not incorporate the effect of uncertain strain calibration.

The typical change in the posteriors following preliminary re-analysis is within statistical sampling error, as quantified by the Jensen–Shannon divergence (Lin 1991) between one-dimensional marginal distributions before and after correction (cf. Romero-Shaw et al. 2020b; Abbott et al. 2021a, Appendix A). As a particular case, the localization of the source of GW170817 receives only a small correction when reanalyzed, and its association with the electromagnetic counterpart emission from AT 2017gfo (Abbott et al. 2017b) is unaffected by the error.

Another case is GW150914, which displays visible differences in the sky location posteriors when re-analyzed (although the bounds of the 90% CI of the right ascension and declination remain nearly unchanged). Meanwhile, our inferences of the intrinsic parameters of GW150914 remain unchanged to within sampling errors.

While the impact on individual candidates is small, a possible concern is that this error can bias analyses that aggregate data from multiple candidates, such as population studies and cosmological inferences. We are investigating the impact of the calibration marginalization error on these analyses, but the effects are expected to be negligible compared to other sources of systematic error. This error does not impact the PE of new candidates observed in O4a (Abac et al. 2025b), and we have updated the PE results for GW230529_181500 (Abac et al. 2024) to correct for the error.

### 5.10. On the Likelihood used for Inference

Late in the preparation of this manuscript, we discovered a normalization error in the likelihood used for the inference codes BayesWave, Bilby, LALInference, Parallel-Bilby, and RIFT. The error arises due to the incorrect application of a window factor that was intended to account for the power lost in the noise residuals due to the Tukey window applied to the data before transforming them to the frequency domain. This error causes the likelihood to be overly constrained by a factor depending on the window function used to mitigate spectral leakage. The incorrect likelihood $\hat{p}(d|\boldsymbol{\theta})$ is related to the correct likelihood $p(d|\boldsymbol{\theta})$ via the average power in the Tukey window $\hat{p}(d|\boldsymbol{\theta}) = p(d|\boldsymbol{\theta})^{\beta}$ with

$$\beta = \left(1 - \frac{5T_{\mathrm{w}}}{4T}\right)^{-1}, \qquad (13)$$

where $T_{\mathrm{w}}$ is the roll-off time of the Tukey window to one side, and $T$ is the segment duration. The window factor is applied when computing the PSD via standard methods, but should not be applied to the data when the signal has support only where the window function is unity, as is the case in our analyses. Although accounting for the windowing using a single window factor in PSD estimation is still an approximation when computing the likelihood, multiple investigations and the use of probability–probability tests (Veitch et al. 2015b; Romero-Shaw et al. 2020b) have confirmed its accuracy. Further, these tests have confirmed the normalization error in our inference codes and the correctness of our updated likelihood. More details on the error and validation of the updated likelihood can be found in Talbot et al. (2025).

We have reanalysed the O4a candidates presented in GWTC-4.0 using the correct likelihood and we find that the differences in the posteriors are small, but systematically widen the posterior distributions. For the reanalyses, we use rejection sampling to reweight the samples from the original posteriors, which were produced using the incorrect likelihood. The acceptance ratio is the ratio of the likelihoods using the correct and incorrect likelihoods, $p(d|\boldsymbol{\theta})/\hat{p}(d|\boldsymbol{\theta})$. In some cases the rejection efficiency is poor, and so for all candidates we resample with replacement until we achieve the original number of samples. This procedure produces a new set of unbiased samples, but the samples are not independent. For the worst cases, $T_{\mathrm{w}} = 1\,\mathrm{s}$ and $T = 4\,\mathrm{s}$, and so $\beta = 1.45$, which also corresponds to overestimated SNRs from PE by a factor 1.21. The impact is less for lower-mass candidates whose durations are larger, so that for BBH candidates with $T_{\mathrm{w}} = 1\,\mathrm{s}$ and $T = 8\,\mathrm{s}$, $\beta = 1.19$ and the SNRs are overestimated by a factor of 1.09. The error also impacts previous analyses from O1 through O3, but the impact is reduced because of the smaller Tukey window applied when performing PE for candidates from those observing runs, and we have not corrected these past inferences. For results from these previous runs, the worst cases have $T_{\mathrm{w}} = 0.4\,\mathrm{s}$ and $T = 4\,\mathrm{s}$, and so $\beta = 1.14$ and the SNRs are overestimated by a factor of 1.07. However, these new posteriors were not created in time to be included in the downstream analyses. Therefore, for O4a candidates, we release both the original posteriors using the incorrect likelihood and the new reweighted posteriors using the correct likelihood.

## 6. WAVEFORM CONSISTENCY TESTS

As we have seen in the earlier Sections, a common assumption made so far for GWTC candidates is that the signal source is a quasi-circular CBC, in vacuum, as predicted by GR. Although no candidates have yet been proven to violate this assumption, GW190521 (Abbott et al. 2020d), GW200105_042309 (Abbott et al. 2021b), and GW231123_135430 (Abac et al. 2025d) highlight the importance of continuously checking its validity, in order to ensure the reliability of astrophysical interpretations. One way of checking this assumption is to perform *waveform consistency tests*. These tests compare different waveform reconstruction techniques to assess their agreement and identify any unexpected features in the reconstructed signals. Waveform reconstruction techniques can be minimally modeled (as seen for cWB in Section 3.1) or template based (Sections 3 and 5).

Minimally-modeled techniques use time–frequency wavelets to identify coherent features in the data of a network of multiple detectors. This generic approach enables the discovery of unexpected phenomena (which might be present if the signal source violates the assumption of a quasi-circular CBC) but does not provide a direct mapping between the reconstructed waveform and the source's physical properties, such as the masses and spins of a binary system and the distance to the source.

To evaluate the consistency between template-based and minimally-modeled reconstructions, we implement a systematic injection study (Abbott et al. 2019a, 2021a, 2023a; Salemi et al. 2019; Ghonge et al. 2020; Johnson-McDaniel et al. 2022a). This involves injecting CBC waveform samples from the posterior parameter distributions (Section 5) into detector data at times near but distinct from the candidate (*off-source* injections). These injections are then reconstructed using minimally-modeled methods. By comparing the reconstructed waveforms from off-source injections ($w_i$) with the reconstruction of the candidate ($\hat{w}$), we can assess how well the CBC PE posteriors align with the minimally-modeled reconstruction. We quantify the agreement between waveform reconstructions using the *overlap*, defined as

$$\mathcal{O}(h_1, h_2) = \frac{\langle h_1 | h_2 \rangle}{\sqrt{\langle h_1 | h_1 \rangle \langle h_2 | h_2 \rangle}}, \quad (14)$$

where $h_1$ and $h_2$ are the waveforms being compared and $\langle \cdot | \cdot \rangle$ denotes the noise-weighted inner product (Abac et al. 2025a, Appendix B). The overlap, $\mathcal{O}(h_1, h_2)$, is bounded between $[-1, +1]$.

For each candidate, we compute two types of overlap measurements and compare them to assess waveform consistency in terms of $p$-values:

1. The *off-source* overlaps $\mathcal{O}(w_i, h_i)$ between injected waveforms and their minimally-modeled reconstructions, forming a reference distribution $\mathcal{O}$;

2. The *on-source* overlap $\mathcal{O}(\hat{w}, h_{\text{maxL}})$ between the maximum-likelihood posterior sample and the actual minimally-modeled reconstruction.

This comparison has an inherent asymmetry: for off-source cases, we calculate the overlap between known waveforms and their minimally-modeled reconstructions, while for on-source cases, we compare the maximum-likelihood template from PE with the minimally-modeled reconstruction. This asymmetry typically results in off-source matches being systematically lower than what would be expected from a true null distribution (Abbott et al. 2023a). Consequently, the derived $p$-values are conservative by construction. Despite this limitation, the on-source $p$-value remains a useful indicator for identifying unexpected signal features, though with reduced statistical power compared to an unbiased test.

To ensure reliable waveform consistency tests, we selected all O4a candidates that meet four key criteria: (i) data availability from both LIGO detectors, (ii) presence of PE results using the NRSur7dq4 waveform family, (iii) detector-frame chirp mass $(1 + z)\mathcal{M} > 15 M_\odot$, and (iv) a network SNR $> 10$. These criteria reflect that minimally-modeled methods require multi-detector data, and that they are particularly effective for high-mass CBCs with high SNR signals.

The candidates satisfying these criteria are analyzed in a companion paper (Abac et al. 2025b) using three minimally-modeled waveform reconstruction methods: BAYESWAVE (Cornish & Littenberg 2015; Cornish et al. 2021; Ghonge et al. 2020), and two configurations of the CWB pipeline described in Section 3.1. Specifically, CWB-2G (Klimenko et al. 2016; Drago et al. 2020) which employs the WDM wavelet transform and an excess power statistic to identify coherent features and CWB-BBH (Klimenko 2022) which uses the WaveScan transform and a cross-power statistic. Both BAYESWAVE and CWB-2G are designed for generic GW transients, whereas CWB-BBH is optimized for CBC signals, also adopting specialized frequency bands and time–frequency resolutions.

For each selected candidate, our injection campaign used for

- BAYESWAVE approximately 400 random posterior samples injected within $\pm 8192\,\text{s}$ of the candidate time,

- CWB-2G and CWB-BBH several thousand random posterior samples injected across a 2–3 week period surrounding the candidate time.

This analysis provides a robust statistical baseline to quantify the degree of agreement between template-based and minimally-modeled reconstructions for a relatively large subset of the candidates from GWTC-4.0. By comparing the on-source overlap to the distribution from off-source injections, we can verify their consistency with the quasi-circular CBC hypothesis, as well as identify those that may exhibit unexpected features.

## 7. DATA MANAGEMENT

The workflow of data analyses described in this paper outlines a complex chain of disparate analyses required to find and characterize GW transients (see Figure 1). In addition to the internal complexity of each analysis, coordinating each stage of this analysis and effectively tracking that input and output data is a significant challenge. This challenge grows with the number of GW candidates observed, necessitating the development of tools to manage these tasks with little to no human intervention. For O4a, this development included the augmentation of existing infrastructure and the development of new software packages including CBCFLOW and the catalog data-product pipeline.

The online and offline analysis results from the search pipelines described in Section 3 are stored in GRACEDB (Moe et al. 2014). For offline results, GRACEDB is also used to generate the final candidate list for the catalog. To facilitate the tracking of these offline results, GRACEDB has been augmented to provide version-controlled snapshots of the state of the catalog during the progression of the offline analyses.

As in GWTC-2.0, GWTC-2.1, and GWTC-3.0, the PE analyses described in Section 5 are managed with the ASIMOV software package (Williams et al. 2023). ASIMOV ingests data-quality recommendations and preliminary PE analyses, and uses this to automatically determine appropriate configuration for PE analyses, as well as automating the production of PSD estimates. Once all desired PE is complete, ASIMOV packages results into a standard format including all inputs and configuration required for reproduction using PESUMMARY (Hoy & Raymond 2021) .

The CBCFLOW (Ashton et al. 2022) software package is used to manage the flow of data between the various analyses and to track metadata about each stage of the analysis. A monitor process fetches search metadata from GRACEDB and preliminary PE results from a shared directory on the Caltech computing cluster. Data-quality information about candidates is updated following the studies described in Section 4. ASIMOV reads CBCFLOW for search, data-quality, and preliminary-PE metadata, which is used to configure production analyses. Upon completion of those analyses, metadata about them are written back to CBCFLOW. All other downstream analyses such as searches for lensed pairs of candidates (Abac et al. 2025e) and tests of GR (Abac et al. 2025f,g,h) also utilize CBCFLOW to track their progress and results.

Finally, the release data product is assembled by a collection of scripts. These read CBCFLOW to identify the preferred search and PE results, as well as tracking their finalization status. These are then collated into the data product itself, and tables of summary information are generated to facilitate downstream use.

## 8. CONCLUSION

Leading on from the introduction presented in Abac et al. (2025a), this article describes the analysis methods used to transform the interferometric strain data from the LVK detectors into version 4.0 of the GWTC; the results of these analyses are presented in Abac et al. (2025b). We began in Section 2 with a description of the waveform models used to describe the GW signals from CBCs involving BHs and NSs. In Section 3, we described the search methods we use to filter the strain data to identify candidate transient GW signals (sensitive to both CBC sources as well as minimally-modeled bursts of GW radiation), and how these candidates are then ranked to identify the most significant ones. Likely GW candidates are then studied to understand the quality of data and identify any transient non-Gaussian noise that may bias later analyses (see Section 4). Section 5 described how a subset of likely GW candidates are then characterized using computational Bayesian inference methods to estimate the parameters of the GW source, such as the masses and spins of the compact objects involved. The PE results however assume that the GW source is a quasi-circular CBC in vacuum as modeled in Section 2. To check this assumption, we described in Section 6 how waveform consistency tests compare different waveform reconstruction techniques to assess their agreement and identify any unexpected features in the reconstructed signals. Finally, in Section 7, we described the data management and workflow tools used to coordinate the various analyses and track the input and output data.

The methods described in this work are continually developed to improve the capabilities of the LVK network to detect and characterize GW transients. Specifically, as the detectors evolve towards higher sensitivity, and more detectors are added to the network, we continue to see a greater number of signals and more high-fidelity signals at larger SNR (Abbott et al. 2020a). This necessitates efficiency improvements across all the methods to avoid computational bottlenecks and to ensure that the GW transient candidates can be processed in a timely manner. Moreover, the increased SNR requires refinements to the waveform models to reduce systematic biases as much as possible. The evolution of the detectors also necessitates continual improvements to the data analysis methods to ensure optimal data processing (e.g., as the noise floor of the detectors lowers, new glitch classes may become relevant, requiring development to searches and data-quality studies). Finally, as the time–volume explored increases, we hope to further expand the range of astrophysical sources that can be detected and characterized, which in turn requires the development of new models, search, and PE methods.

*Data Availability:* The data products generated by the methods described within this work are openly available in the GWTC-4 online catalog, which is hosted at https://gwosc.org/GWTC-4.0 and documented further in Abac et al. (2025c).

## ACKNOWLEDGEMENTS

This material is based upon work supported by NSF's LIGO Laboratory, which is a major facility fully funded by the National Science Foundation. The authors also gratefully acknowledge the support of the Science and Technology Facilities Council (STFC) of the United Kingdom, the Max-Planck-Society (MPS), and the State of Niedersachsen/Germany for support of the construction of Advanced LIGO and construction and operation of the GEO 600 detector. Additional support for Advanced LIGO was provided by the Australian Research Council. The authors gratefully acknowledge the Italian Istituto Nazionale di Fisica Nucleare (INFN), the French Centre National de la Recherche Scientifique (CNRS) and the Netherlands Organization for Scientific Research (NWO) for the construction and operation of the Virgo detector and the creation and support of the EGO consortium. The authors also gratefully acknowledge research support from these agencies as well as by the Council of Scientific and Industrial Research of India, the Department of Science and Technology, India, the Science & Engineering Research Board (SERB), India, the Ministry of Human Resource Development, India, the Spanish Agencia Estatal de Investigación (AEI), the Spanish Ministerio de Ciencia, Innovación y Universidades, the European Union NextGenerationEU/PRTR (PRTR-C17.I1), the ICSC - Cen-troNazionale di Ricerca in High Performance Computing, Big Data and Quantum Computing, funded by the European Union NextGenerationEU, the Consellaria de les Illes Balears through the Conselleria d'Educació i Universitats, the Conselleria d'Innovació, Universitats, Ciència i Societat Digital de la Generalitat Valenciana and the CERCA Programme Generalitat de Catalunya, Spain, the Polish National Agency for Academic Exchange, the National Science Centre of Poland and the European Union - European Regional Development Fund; the Foundation for Polish Science (FNP), the Polish Ministry of Science and Higher Education, the Swiss National Science Foundation (SNSF), the Russian Science Foundation, the European Commission, the European Social Funds (ESF), the European Regional Development Funds (ERDF), the Royal Society, the Scottish Funding Council, the Scottish Universities Physics Alliance, the Hungarian Scientific Research Fund (OTKA), the French Lyon Institute of Origins (LIO), the Belgian Fonds de la Recherche Scientifique (FRS-FNRS), Actions de Recherche Concertées (ARC) and Fonds Wetenschappelijk Onderzoek - Vlaanderen (FWO), Belgium, the Paris Île-de-France Region, the National Research, Development and Innovation Office of Hungary (NKFIH), the National Research Foundation of Korea, the Natural Sciences and Engineering Research Council of Canada (NSERC), the Canadian Foundation for Innovation (CFI), the Brazilian Ministry of Science, Technology, and Innovations, the International Center for Theoretical Physics South American Institute for Fundamental Research (ICTP-SAIFR), the Research Grants Council of Hong Kong, the National Natural Science Foundation of China (NSFC), the Israel Science Foundation (ISF), the US-Israel Binational Science Fund (BSF), the Leverhulme Trust, the Research Corporation, the National Science and Technology Council (NSTC), Taiwan, the United States Department of Energy, and the Kavli Foundation. The authors gratefully acknowledge the support of the NSF, STFC, INFN and CNRS for provision of computational resources.

This work was supported by MEXT, the JSPS Leading-edge Research Infrastructure Program, JSPS Grant-in-Aid for Specially Promoted Research 26000005, JSPS Grant-in-Aid for Scientific Research on Innovative Areas 2402: 24103006, 24103005, and 2905: JP17H06358, JP17H06361 and JP17H06364, JSPS Core-to-Core Program A. Advanced Research Networks, JSPS Grants-in-Aid for Scientific Research (S) 17H06133 and 20H05639, JSPS Grant-in-Aid for Transformative Research Areas (A) 20A203: JP20H05854, the joint research program of the Institute for Cosmic Ray Research, University of Tokyo, the National Research Foundation (NRF), the Computing Infrastructure Project of the Global Science experimental Data hub Center (GSDC) at KISTI, the Korea Astronomy and Space Science Institute (KASI), the Ministry of Science and ICT (MSIT) in Korea, Academia Sinica (AS), the AS Grid Center (ASGC) and the National Science and Technology Council (NSTC) in Taiwan under grants including the Science Vanguard Research Program, the Advanced Technology Center (ATC) of NAOJ, and the Mechanical Engineering Center of KEK.

Additional acknowledgements for support of individual authors may be found in the following document: https://dcc.ligo.org/LIGO-M2300033/public. For the purpose of open access, the authors have applied a Creative Commons Attribution (CC BY) license to any Author Accepted Manuscript version arising. We request that citations to this article use 'A. G. Abac et al. (LIGO-Virgo-KAGRA Collaboration), ...' or similar phrasing, depending on journal convention.

*Software:* Plots were prepared with MATPLOTLIB (Hunter 2007), NUMPY (Harris et al. 2020), and TIKZ (Tantau 2023).

# All Authors and Affiliations

A. G. Abac,[1] I. Abouelfettouh,[2] F. Acernese,[3,4] K. Ackley,[5] S. Adhicary,[6] D. Adhikari,[7,8]
N. Adhikari,[9] R. X. Adhikari,[10] V. K. Adkins,[11] S. Afroz,[12] D. Agarwal,[13,14] M. Agathos,[15]
M. Aghaei Abchouyeh,[16] O. D. Aguiar,[17] S. Ahmadzadeh,[18] L. Aiello,[19,20] A. Ain,[21] P. Ajith,[22]
S. Akcay,[23] T. Akutsu,[24,25] S. Albanesi,[26,27] R. A. Alfaidi,[28] A. Al-Jodah,[29] C. Alléné,[30]
A. Allocca,[31,4] S. Al-Shammari,[32] P. A. Altin,[33] S. Alvarez-Lopez,[34] O. Amarasinghe,[32] A. Amato,[35,36]
C. Amra,[37] A. Ananyeva,[10] S. B. Anderson,[10] W. G. Anderson,[10] M. Andia,[38] M. Ando,[39,40] T. Andrade,[41]
M. Andrés-Carcasona,[42] T. Andrić,[7,8,43] J. Anglin,[44] S. Ansoldi,[45,46] J. M. Antelis,[47] S. Antier,[48]
M. Aoumi,[49] E. Z. Appavuravther,[50,51] S. Appert,[10] S. K. Apple,[52] K. Arai,[10] A. Araya,[53] M. C. Araya,[10]
M. Arca Sedda,[43] J. S. Areeda,[54] L. Arginas,[55] N. Aritomi,[2] F. Armato,[56,57] S. Armstrong,[58]
N. Arnaud,[38,59] M. Arogeti,[60] S. M. Aronson,[11] G. Ashton,[61] Y. Aso,[24,62] M. Assiduo,[63,64]
S. Assis de Souza Melo,[59] S. M. Aston,[65] P. Astone,[66] F. Attadio,[67,66] F. Aubin,[68] K. AultONeal,[69]
G. Avallone,[70] S. Babak,[71] F. Badaracco,[56] C. Badger,[72] S. Bae,[73] S. Bagnasco,[27] E. Bagui,[74]
L. Baiotti,[75] R. Bajpai,[24] T. Baka,[76] T. Baker,[77] M. Ball,[78] G. Ballardin,[59] S. W. Ballmer,[79]
S. Banagiri,[80] B. Banerjee,[43] D. Bankar,[10] T. M. Baptiste,[11] P. Baral,[9] J. C. Barayoga,[10]
B. C. Barish,[10] D. Barker,[2] P. Barneo,[41,81] F. Barone,[82,4] B. Barr,[28] L. Barsotti,[34]
M. Barsuglia,[71] D. Barta,[84] A. M. Bartoletti,[84] M. A. Barton,[28] I. Bartos,[44] S. Basak,[22]
A. Basalaev,[85] R. Bassiri,[86] A. Basti,[87,88] D. E. Bates,[32] M. Bawaj,[89,50] P. Baxi,[90] J. C. Bayley,[28]
A. C. Baylor,[9] P. A. Baynard II,[60] M. Bazzan,[91,92] V. M. Bedakihale,[93] F. Beirnaert,[94] M. Bejger,[71]
D. Belardinelli,[20] A. S. Bell,[28] D. S. Bellie,[80] L. Bellizzi,[88,87] W. Benoit,[96] I. Bentara,[97]
J. D. Bentley,[85] M. Ben Yaala,[58] S. Bera,[98] F. Bergamin,[7,8] B. K. Berger,[86] S. Bernuzzi,[26]
M. Beroiz,[10] C. P. L. Berry,[28] D. Bersanetti,[56] A. Bertolini,[36] J. Betzwieser,[65] D. Beveridge,[29]
G. Bevilacqua,[99] N. Bevins,[55] R. Bhandare,[100] R. Bhatt,[10] D. Bhattacharjee,[101,102] S. Bhaumik,[44]
S. Bhowmick,[103] V. Biancalana,[99] A. Bianchi,[36,104] I. A. Bilenko,[105] G. Billingsley,[10] A. Binetti,[106]
S. Bini,[107,108] C. Binu,[109] O. Birnholtz,[110] S. Bisceveanu,[80] A. Bisht,[8] M. Bitossi,[59,88]
M.-A. Bizouard,[48] S. Blair,[111] J. K. Blackburn,[10] L. A. Blagg,[112] C. D. Blair,[29,65] D. G. Blair,[29]
F. Bobba,[70,112] N. Bode,[7,8] G. Boileau,[48] M. Boldrini,[66,67] G. N. Bolingbroke,[113] A. Bolliand,[114,37]
L. D. Bonavena,[44,91] R. Bondarescu,[41] F. Bondu,[115] E. Bonilla,[86] M. S. Bonilla,[54] A. Bonino,[116]
R. Bonnand,[30,114] P. Booker,[7,8] A. Borchers,[7,8] S. Borhanian,[6] V. Boschi,[88] S. Bose,[117] V. Bossilkov,[65]
A. Boudon,[97] A. Bozzi,[59] C. Bradaschia,[88] P. R. Brady,[9] A. Branch,[65] M. Branchesi,[43,118] I. Braun,[101]
T. Briant,[119] A. Brillet,[48] M. Brinkmann,[7,8] P. Brockill,[9] E. Brockmueller,[7,8] A. F. Brooks,[10]
B. C. Brown,[44] D. D. Brown,[113] M. L. Brozzetti,[89,50] S. Brunett,[10] G. Bruno,[114] R. Bruntz,[2] J. Bryant,[116]
Y. Bu,[121] F. Bucci,[64] J. Buchanan,[120] O. Bulashenko,[41,81] T. Bulik,[122] H. J. Bulten,[36] A. Buonanno,[123,1]
K. Burtnyk,[2] R. Buscicchio,[124,125] D. Buskulic,[30] C. Buy,[10] R. L. Byer,[86] G. S. Cabourn Davies,[77]
G. Cabras,[45,46] R. Cabrita,[13] V. Cáceres-Barbosa,[47] L. Cadonati,[60] G. Cagnoli,[127] C. Cahillane,[10]
A. Calafat,[98] J. Calderón Bustillo,[128] T. A. Callister,[129] E. Calloni,[31,4] G. Caneva Santoro,[42]
K. C. Cannon,[40] H. Cao,[34] L. A. Capistran,[130] E. Capocasa,[71] E. Capote,[79] G. Capurri,[88,87]
G. Carapella,[70,112] F. Carbognani,[59] M. Carlassara,[7,8] J. B. Carlin,[121] T. K. Carlson,[131] M. F. Carney,[101]
M. Carpinelli,[124,132,59] G. Carrillo,[78] J. J. Carter,[7,8] G. Carullo,[133] J. Casanueva Diaz,[59]
C. Casentini,[134,19,20] S. Y. Castro-Lucas,[103] S. Caudill,[131,36,76] M. Cavaglià,[102] R. Cavalieri,[59]
G. Cella,[88] P. Cerdá-Durán,[135,136] E. Cesarini,[20] W. Chaibi,[48] P. Chakraborty,[7,8] S. Chakraborty,[100]
S. Chalathadka Subrahmanya,[85] J. C. L. Chan,[137] M. Chan,[111] R.-J. Chang,[132] S. Chao,[139,140]
E. L. Charlton,[120] P. Charlton,[141] E. Chassande-Mottin,[71] C. Chatterjee,[142] Debarati Chatterjee,[14]
Deep Chatterjee,[34] M. Chaturvedi,[100] S. Chaty,[71] K. Chatziioannou,[10] C. Checchia,[29] A. Chen,[15]
A. H.-Y. Chen,[143] D. Chen,[144] H. Chen,[143] H. Y. Chen,[145] J. Chen,[139] Yanbei Chen,[10]
Yitian Chen,[147] H. P. Cheng,[148] P. Chessa,[89,50] H. T. Cheung,[90] S. Y. Cheung,[149] F. Chiadini,[150,112]
G. Chiarini,[92] R. Chierici,[92] A. Chincarini,[56] M. L. Chiofalo,[87,88] A. Chiummo,[59] C. Chou,[143]
S. Choudhary,[29] N. Christensen,[48] S. S. Y. Chua,[33] P. Chugh,[149] G. Ciani,[107,108] P. Ciecielag,[95]
M. Cieślar,[122] M. Cifaldi,[20] R. Ciolfi,[151,92] F. Clara,[2] J. A. Clark,[10,60] J. Clarke,[32] T. A. Clarke,[149]
P. Clearwater,[152] S. Clesse,[74] S. M. Clyne,[153] E. Coccia,[43,118,42] E. Codazzo,[154] P.-F. Cohadon,[119]
S. Colace,[57] E. Colangeli,[2] M. Colleoni,[98] C. G. Collette,[155] J. Collins,[65] S. Colloms,[28]
A. Colombo,[156,125] C. M. Compton,[2] G. Connolly,[28] L. Conti,[92] T. R. Corbitt,[11] I. Cordero-Carrión,[157]
S. Corezzi,[89,50] N. J. Cornish,[158] A. Corsi,[159] S. Cortese,[59] R. Cottingham,[65] M. W. Coughlin,[96]
A. Couineaux,[66] J.-P. Coulon,[48] J.-F. Coupechoux,[97] P. Couvares,[10,60] D. M. Coward,[29] R. Coyne,[153]
K. Craig,[58] J. D. E. Creighton,[9] T. D. Creighton,[160] P. Cremonese,[98] A. W. Criswell,[96] S. Crouch,[65]
R. Crouch,[2] J. Csizmazia,[2] J. R. Cudell,[161] T. J. Cullen,[10] A. Cumming,[28] E. Cuoco,[162,163]
M. Cusinato,[135] P. Dabadie,[127] L. V. Da Conceição,[164] T. Dal Canton,[38] S. Dall'Osso,[66] S. Dal Pra,[165]

G. Dálya,[126] B. D'Angelo,[56] S. Danilishin,[35,36] S. D'Antonio,[20] K. Danzmann,[8,7,8] K. E. Darroch,[120]
L. P. Dartez,[65] A. Dasgupta,[93] S. Datta,[166] V. Dattilo,[59] A. Daumas,[71] N. Davari,[167,132] I. Dave,[100]
A. Davenport,[103] M. Davier,[38] T. F. Davies,[29] D. Davis,[10] L. Davis,[29] M. C. Davis,[96] P. Davis,[168,169]
M. Dax,[1] J. De Bolle,[94] M. Deenadayalan,[14] J. Degallaix,[170] U. Deka,[171] M. De Laurentis,[31,4]
S. Deléglise,[119] F. De Lillo,[21] D. Dell'Aquila,[167,132] F. Della Valle,[99] W. Del Pozzo,[87,88]
F. De Marco,[67,66] G. Demasi,[172,64] F. De Matteis,[19,20] V. D'Emilio,[10] N. Demos,[34] T. Dent,[128]
A. Depasse,[13] N. DePergola,[55] R. De Pietri,[173,174] R. De Rosa,[31,4] C. De Rossi,[59] M. Desai,[34]
R. DeSalvo,[175] A. DeSimone,[176] R. De Simone,[160] A. Dhani,[3] R. Diab,[44] M. C. Díaz,[160] M. Di Cesare,[31,4]
G. Dideron,[177] N. A. Didio,[79] T. Dietrich,[3,1] L. Di Fiore,[4] C. Di Fronzo,[29] M. Di Giovanni,[67,66]
T. Di Girolamo,[31,4] D. Diksma,[36,35] A. Di Michele,[89] J. Ding,[34,71,178] S. Di Pace,[67,66] I. Di Palma,[67,66]
F. Di Renzo,[97] Divyajyoti,[179] A. Dmitriev,[116] Z. Doctor,[80] N. Doerksen,[164] E. Dohmen,[2]
D. Dominguez,[180] L. D'Onofrio,[66] F. Donovan,[32] K. L. Dooley,[32] T. Dooney,[76] S. Doravari,[14]
O. Dorosh,[181] M. Drago,[67,66] J. C. Driggers,[32] J.-G. Ducoin,[182,71] L. Dunn,[121] U. Dupletsa,[43]
D. D'Urso,[167,154] H. Duval,[183] S. E. Dwyer,[2] C. Eassa,[2] M. Ebersold,[30] T. Eckhardt,[2] G. Eddolls,[10,5]
B. Edelman,[78] T. B. Edo,[10] O. Edy,[77] A. Effler,[65] J. Eichholz,[33] H. Einsle,[48] M. Eisenmann,[24]
R. A. Eisenstein,[34] A. Ejlli,[32] M. Emma,[91] K. Endo,[184] R. Enficiaud,[3] A. J. Engl,[86] L. Errico,[31,4]
R. Espinosa,[160] M. Esposito,[4,31] R. C. Essick,[185] H. Estellés,[1] T. Etzel,[10] M. Evans,[34] T. Evstafyeva,[186]
B. E. Ewing,[3] J. M. Ezquiaga,[137] F. Fabrizi,[63,64] F. Faedi,[64,63] V. Fafone,[19,20] S. Fairhurst,[10]
A. M. Farah,[129] B. Farr,[78] W. M. Farr,[187,188] G. Favaro,[91] M. Favata,[189] M. Fays,[161] M. Fazio,[58]
J. Feicht,[10] M. M. Fejer,[86] R. Felicetti,[190] E. Fenyvesi,[83,191] D. L. Ferguson,[145] T. Fernandes,[192,135]
D. Fernando,[109] S. Ferraiuolo,[193,67,66] I. Ferrante,[87,88] T. A. Ferreira,[11] F. Fidecaro,[87,88] P. Figura,[95]
A. Fiori,[88,87] I. Fiori,[59] M. Fishbach,[185] R. P. Fisher,[120] R. Fittipaldi,[194,112] V. Fiumara,[195,112]
R. Flaminio,[30] S. M. Fleischer,[196] L. S. Fleming,[18] E. Floden,[96] H. Fong,[111] J. A. Font,[135,136] C. Foo,[1]
B. Fornal,[197] P. W. F. Forsyth,[33] K. Franceschetti,[173] N. Franchini,[198] S. Frasca,[67,66] F. Frasconi,[88]
A. Frattale Mascioli,[67,66] Z. Frei,[199] A. Freise,[36,104] O. Freitas,[192,135] R. Frey,[78] W. Frischhertz,[65]
P. Fritschel,[34] V. V. Frolov,[65] G. G. Fronzé,[27] M. Fuentes-Garcia,[160] S. Fujii,[200] T. Fujimori,[201] P. Fulda,[44]
M. Fyffe,[76] H. A. Gabbard,[10] B. U. Gadre,[3] J. R. Gair,[11] S. Galaudage,[202] V. Galdi,[175] H. Gallagher,[109] B. Gallego,[203]
R. Gamba,[6,26] A. Gamboa,[1] D. Ganapathy,[34] A. Ganguly,[14] B. Garaventa,[56,57]
J. García-Bellido,[204] C. García Núñez,[78] I. Ferrante,[205] J. W. Gardner,[33] K. A. Gardner,[111]
J. Gargiulo,[59] A. Garron,[98] F. Garufi,[31,4] P. A. Garver,[98] C. Gasbarra,[19,20] B. Gateley,[2]
F. Gautier,[206] V. Gayathri,[9] T. Gayer,[79] G. Gemme,[56] A. Gennai,[88] V. Gennari,[126] J. George,[100]
R. George,[145] O. Gerberding,[85] L. Gergely,[207] Archisman Ghosh,[94] Sayantan Ghosh,[208]
Shaon Ghosh,[189] Shrobana Ghosh,[7,8] Suprovo Ghosh,[14] Tathagata Ghosh,[14] J. A. Giaime,[11,65]
K. D. Giardina,[65] D. R. Gibson,[78] D. T. Gibson,[186] C. Gier,[58] S. Gkaitatzis,[87,88] J. Glanzer,[10] F. Glotin,[38]
J. Godfrey,[78] P. Godwin,[10] A. S. Goettel,[32] E. Goetz,[111] J. Golomb,[10] S. Gomez Lopez,[131]
B. Goncharov,[43] Y. Gong,[209] G. González,[11] P. Goodarzi,[210] S. Goode,[149] A. W. Goodwin-Jones,[10,29]
M. Gosselin,[59] R. Gouaty,[30] D. W. Gould,[3] K. Govorkova,[34] S. Goyal,[14] B. Grace,[33] A. Grado,[89,50]
V. Graham,[28] A. E. Granados,[96] M. Granata,[170] V. Granata,[70] S. Gras,[34] P. Grassia,[10] A. Gray,[96]
C. Gray,[2] R. Gray,[28] G. Greco,[50] A. C. Green,[36,104] S. M. Green,[77] S. R. Green,[211] A. M. Gretarsson,[69]
E. M. Gretarsson,[69] D. Griffith,[10] W. L. Griffiths,[32] H. L. Griggs,[32] G. Grignani,[89,50] C. Grimaud,[30]
H. Grote,[32] S. Grunewald,[3] D. Guerra,[135] D. Guetta,[212] G. M. Guidi,[63,64] A. R. Guimaraes,[11]
H. K. Gulati,[93] F. Gulminelli,[168,169] A. M. Gunny,[34] H. Guo,[213] W. Guo,[207] Y. Guo,[36,35]
Anchal Gupta,[10] Anuradha Gupta,[214] I. Gupta,[3] N. C. Gupta,[93] P. Gupta,[36,76] S. K. Gupta,[44] T. Gupta,[158]
V. Gupta,[96] N. Gupte,[1] J. Gurs,[85] N. Gutierrez,[170] F. Guzman,[130] D. Haba,[180] M. Haberland,[10] S. Haino,[215]
E. D. Hall,[34] E. Z. Hamilton,[98] G. Hammond,[28] W.-B. Han,[217] M. Haney,[36,205]
J. Hanks,[2] C. Hanna,[6] M. D. Hannam,[32] O. A. Hannuksela,[218] A. G. Hanselman,[129] H. Hansen,[2] J. Hanson,[65]
R. Harada,[40] A. R. Hardison,[176] S. Harikumar,[181] K. Haris,[36,76] T. Harmark,[133] J. Harms,[43,118]
G. M. Harry,[219] I. W. Harry,[77] J. Hart,[101] B. Haskell,[95] C.-J. Haster,[28] K. Haughian,[28] H. Hayakawa,[49]
K. Hayama,[221] R. Hayes,[32] M. C. Heintze,[65] J. Heinze,[116] J. Heinzel,[34] H. Heitmann,[48] A. Heffernan,[222]
F. Hellman,[223] A. F. Helmling-Cornell,[78] G. Hemming,[59] O. Henderson-Sapir,[113] M. Hendry,[28]
I. S. Heng,[28] M. H. Hennig,[2] C. Henshaw,[60] M. Heurs,[7,8] A. L. Hewitt,[186,224] J. Heyns,[34]
S. Higginbotham,[32] S. Hild,[35,36] S. Hill,[28] Y. Himemoto,[225] N. Hirata,[24] C. Hirose,[226] S. Hochheim,[7,8]
D. Hofman,[170] N. A. Holland,[36,104] D. E. Holz,[129] L. Honet,[74] C. Hong,[86] S. Hoshino,[220] J. Hough,[28]
S. Hourihane,[10] N. Howard,[142] E. J. Howell,[29] C. G. Hoy,[77] C. A. Hrishikesh,[19] H.-F. Hsieh,[139]
H.-Y. Hsieh,[139] C. Hsiung,[227] W.-F. Hsu,[106] Q. Hu,[28] H. Y. Huang,[140] Y. Huang,[79] Y. T. Huang,[79]
A. D. Huddart,[228] B. Hughey,[69] D. C. Y. Hui,[229] V. Hui,[30] S. Husa,[207] K. Huxford,[6] L. Iampieri,[67,66]
G. A. Iandolo,[35] M. Ianni,[20,19] A. Ierardi,[43] A. Iess,[230,88] H. Imafuku,[40] K. Inayoshi,[231] Y. Inoue,[140]
G. Iorio,[91] P. Iosif,[190,46] M. H. Iqbal,[33] J. Irwin,[28] R. Ishikawa,[232] M. Isi,[187,188] Y. Itoh,[233,201]

H. Iwanaga,[233] M. Iwaya,[200] B. R. Iyer [22] C. Jacquet,[126] P.-E. Jacquet [119] S. J. Jadhav,[234] S. P. Jadhav,[152] T. Jain,[186] A. L. James [10] P. A. James,[120] R. Jamshidi,[155] K. Jani [142] J. Janquart [13] K. Janssens [21,48] N. N. Janthalur,[234] S. Jaraba [204] P. Jaranowski [235] R. Jaume [98] W. Javed,[32] A. Jennings,[2] W. Jia,[34] J. Jiang [148] S. J. Jin [29] C. Johanson,[131] G. R. Johns,[120] N. A. Johnson,[44] N. K. Johnson-McDaniel [214] M. C. Johnston [220] R. Johnston,[28] N. Johny,[7,8] D. H. Jones [35] D. I. Jones,[236] E. J. Jones,[11] R. Jones,[28] S. Jose,[179] P. Joshi [6] S. K. Joshi,[14] J. Ju,[237] L. Ju,[29] K. Jung [238] J. Junker [33] V. Juste,[74] H. B. Kabagoz [65] T. Kajita [239] I. Kaku,[233] V. Kalogera [80] M. Kalomenopoulos [220] M. Kamiizumi [49] N. Kanda [201,233] S. Kandhasamy [14] G. Kang [240] N. C. Kannachel,[149] J. B. Kanner,[10] S. J. Kapadia [14] D. P. Kapasi,[33] S. Karat,[10] R. Kashyap [6] M. Kasprzack [10] W. Kastaun,[7,8] T. Kato,[200] E. Katsavounidis,[34] W. Katzman,[65] R. Kaushik [100] K. Kawabe,[2] R. Kawamoto,[233] A. Kazemi,[96] D. Keitel [98] J. Kennington [6] R. Kesharwani,[14] J. S. Key [241] R. Khadela,[7,8] S. Khadka,[86] F. Y. Khalili [105] F. Khan [7,8] I. Khan,[242,37] T. Khanam,[159] M. Khursheed,[100] N. M. Khusid,[187,188] N. Kijbunchoo [113] C. Kim,[244] J. C. Kim,[245] K. Kim [246] M. H. Kim [238] S. Kim [229] Y.-M. Kim [246] C. Kimball [80] M. Kinley-Hanlon [28] M. Kinnear,[32] J. S. Kissel [2] S. Klimenko,[37] A. M. Knee [111] N. Knust [7,8] K. Kobayashi,[200] P. Koch,[7,8] S. M. Koehlenbeck [86] G. Koekoek,[36,35] K. Kohri [247,248] K. Kokeyama [32] S. Koley [43] P. Kolitsidou [116] K. Komori [40,39] A. K. H. Kong [249] M. Korobko [85] R. V. Kossak,[7,8] X. Kou,[96] A. Koushik [21] N. Kouvatsos [72] M. Kovalam,[29] D. B. Kozak,[10] S. L. Kranzhoff,[35,36] V. Kringel,[7,8] N. V. Krishnendu [116] A. Królak [250,181] K. Kruska,[7,8] J. Kubisz [251] G. Kuehn,[7,8] S. Kulkarni [214] A. Kulur Ramamohan [33] A. Kumar,[234] Praveen Kumar [10] Prayush Kumar [22] Rahul Kumar,[2] Rakesh Kumar,[93] J. Kume [252,253,40] K. Kuns [34] N. Kuntimaddi,[32] S. Kuroyanagi [204,254] S. Kuwahara [40] K. Kwak,[238] K. Kwan,[33] J. Kwok,[186] G. Lacaille,[28] P. Lagabbe [30,107] D. Laghi [126] S. Lai,[143] E. Lalande,[255] M. Lalleman [21] P. C. Lalremruati,[256] M. Landry,[2] B. B. Lane,[34] R. N. Lang [34] J. Lange,[145] R. Langgin [220] B. Lantz [86] A. La Rana [66] I. La Rosa [98] J. Larsen,[196] A. Lartaux-Vollard [38] P. D. Lasky [149] J. Lawrence [160,257] M. N. Lawrence,[11] C. Lazarte [10] C. Lazzaro [258,154] P. Leaci [67,66] L. Leali,[96] Y. K. Lecoeuche [111] H. M. Lee [245] H. W. Lee [259] J. Lee,[79] K. Lee [237] R.-K. Lee [139] R. Lee,[34] Sunghoon Lee [260] Sunjae Lee,[237] Y. Lee,[140] I. N. Legred,[10] J. Lehmann,[7,8] L. Lehner,[177] M. Le Jean [170] A. Lemaître,[261] M. Lenti [64,172] M. Leonardi [107,108,24] M. Lequime,[37] N. Leroy [38] M. Lesovsky,[10] N. Letendre,[30] M. Lethuillier [97] Y. Levin,[149] K. Leyde [71,77] A. K. Y. Li,[10] K. L. Li [138] T. G. F. Li,[106] X. Li,[146] Y. Li,[80] Z. Li,[28] C-Y. Lin,[262] E. T. Lin [139] C-Y. Lin [138] Y.-C. Lin,[139] C. Lindsay,[18] S. D. Linker,[203] T. B. Littenberg,[263] A. Liu [218] G. C. Liu [227] Jian Liu [29] F. Llamas Villarreal,[160] J. Llobera-Querol [98] R. K. L. Lo [137] J.-P. Locquet,[106] M. R. Loizou,[131] L. T. London,[72,34] A. Longo [63,64] D. Lopez,[69] H. N. Lopez,[161,205] M. Lopez Portilla,[76] A. Lorenzo-Medina [128] V. Loriette,[38] M. Lormand,[65] G. Losurdo [230,88] E. Lotti,[131] T. P. Lott IV [60] J. D. Lough [7,8] H. A. Loughlin,[34] C. O. Lousto [109] N. Low,[121] M. J. Lowry,[120] N. Lu,[12] S. Lucchesi [88] H. Lück,[8,7,8] D. Lumaca [20] A. P. Lundgren,[77] A. W. Lussier [255] L.-T. Ma [139] S. Ma,[177] R. Macas [72] A. Macedo [54] M. MacInnis,[34] R. R. Maciy,[7,8] D. M. Macleod [32] I. A. O. MacMillan [10] A. Macquet [38] D. Macri,[34] S. Maenaut [106] S. S. Magare,[14] R. M. Magee [10] E. Maggio [88] M. Magnozzi [56,57] M. Mahesh,[85] M. Maini,[153] S. Majhi,[14] E. Majorana,[67,66] C. N. Makarem,[10] D. Malakar [14,102] J. A. Malaquias-Reis,[17] U. Mali [185] S. Maliakal,[10] A. Malik,[100] L. Mallick [164,185] A. Malz,[61] N. Man,[48] V. Mandic [96] V. Mangano [66,67] B. Mannix,[78] G. L. Mansell [2] G. Mansingh,[219] M. Manske [9] M. Mantovani [59] M. Mapelli [91,92,264] F. Marchesoni,[51,50,265] C. Marinelli,[99] D. Marín Pina [41,81,266] F. Marion [30] S. Márka [29] Z. Márka [29] A. S. Markosyan,[86] A. Markowitz,[10] E. Maros,[10] S. Marsat [126] F. Martelli [63,64] I. W. Martin [28] R. M. Martin,[189] B. B. Martinez,[130] M. Martinez,[42,268] V. Martinez [127] A. Martini,[107,108] J. C. Martins [17] D. V. Martynov,[116] E. J. Marx,[34] L. Massaro,[35,36] A. Masserot,[30] M. Masso-Reid [28] M. Mastrodicasa,[66,67] S. Mastrogiovanni [66] T. Matcovich [50] M. Matiushechkina [7,8] M. Matsuyama,[233] N. Mavalvala [34] N. Maxwell,[12] G. McCarrol,[65] R. McCarthy,[2] D. E. McClelland [33] S. McCormick,[65] L. McCuller [10] S. McEachin,[120] C. McElhenny,[120] G. I. McGhee [28] J. McGinn,[28] K. B. M. McGowan,[142] J. McIver [111] A. McLeod [29] T. McRae,[33] D. Meacher [9] Q. Meijer,[76] A. Melatos,[7,269,121] M. Melching [7,269,121] S. Mellaerts,[106] C. S. Menoni [86] F. Mera,[4] R. A. Mercer [9] L. Mereni,[170] K. Merfeld,[159] E. L. Merilh,[65] J. R. Mérou [99] J. D. Merritt,[78] M. Merzougui,[48] C. Messenger [28] C. Messick [9] B. Mestichelli,[43] M. Meyer-Conde [270] F. Meylahn [7,8] A. Mhaske,[14] A. Miani,[107,108] H. Miao,[271] I. Michaloliakos [44] C. Michel [170] Y. Michimura,[10,40] H. Middleton [116] S. J. Miller [10] M. Millhouse [190,46] V. Milotti [91] Y. Minenkov,[20] N. Mio,[272] Ll. M. Mir [42] L. Mirasola [154,258] M. Miravet-Tenés [135] C.-A. Miritescu [42] A. K. Mishra,[22] A. Mishra,[20] C. Mishra [179] T. Mishra,[44] A. L. Mitchell,[36,104] J. G. Mitchell,[69] S. Mitra [14] V. P. Mitrofanov [105] R. Mittleman,[34] O. Miyakawa [49] S. Miyamoto,[200] S. Miyoki [49] G. Mo [34] S. R. Mohapatra,[10] L. Mobilia,[63,64] S. R. Mohite [9] M. Molina-Ruiz [223] C. Mondal [168] M. Mondin,[203] M. Montani,[63,64] C. J. Moore,[186] D. Moraru,[2] A. More [14] S. More [14] E. A. Moreno [34] G. Moreno,[2] S. Morisaki [40,200] Y. Moriwaki [184]

G. Morras [204], A. Moscatello [91], M. Mould [34], P. Mourier [222,273], B. Mours [68], C. M. Mow-Lowry [36,104]
F. Muciaccia [67,66], D. Mukherjee [263], Samanwaya Mukherjee [14], Soma Mukherjee [160], Subroto Mukherjee [93]
Suvodip Mukherjee [34], N. Mukund [34], A. Mullavey [65], H. Mulluck [111], J. Munch [113], J. Mundi [219]
C. L. Mungioli [29], Y. Murakami [200], M. Murakoshi [232], P. G. Murray [33], S. Muusse [33], D. Nabari [107,108]
S. L. Nadji [7,8], A. Nagar [27,275], N. Nagarajan [28], K. Nakagaki [49], K. Nakamura [24], H. Nakano [276]
M. Nakano [10], D. Nanadoumgar-Lacroze [42], D. Nandi [11], V. Napolano [59], P. Narayan [214], I. Nardecchia [20]
T. Narikawa [200], H. Narola [76], L. Naticchioni [66], R. K. Nayak [256], A. Nela [28], A. Nelson [130], T. J. N. Nelson [65]
M. Nery [7,8], A. Neunzert [21], S. Ng [54], L. Nguyen Quynh [277,278], S. A. Nichols [11], A. B. Nielsen [28]
G. Nieradka [95], Y. Nishino [24,280], A. Nishizawa [281], S. Nissanke [274,36], E. Nitoglia [97], W. Niu [6], F. Nocera [59]
M. Norman [32], C. North [32], J. Novak [114,282,283], J. F. Nuño Siles [204], L. K. Nuttall [77], K. Obayashi [232]
J. Oberling [2], J. O'Dell [228], M. Oertel [282,114,284,283], A. Offermans [106], G. Oganesyan [43,118], J. J. Oh [285]
K. Oh [229], T. O'Hanlon [65], M. Ohashi [49], M. Ohkawa [226], F. Ohme [7,8], R. Oliveri [114,284,283], R. Omer [96]
B. O'Neal [120], B. O'Reilly [65], R. Oram [10], N. D. Ormsby [120], M. Orselli [50,89]
R. O'Shaughnessy [109], S. O'Shea [39], S. Oshino [49], C. Osthelder [10], I. Ota [11]
D. J. Ottaway [113], A. Ouzriat [97], H. Overmier [65], B. J. Owen [288], A. E. Pace [6], R. Pagano [11], M. A. Page [24]
A. Pai [208], L. Paiella [43], A. Pal [289], S. Pal [256], M. A. Palaia [88,87], M. Pálfi [199], P. P. Palma [67,19], P. Palud [71]
C. Palomba [66], P. Palud [71], J. Pan [29], K. C. Pan [139], R. Panai [154,91], P. K. Panda [234], Shiksha Pandey [6]
Swadha Pandey [34], P. T. H. Pang [36,76], F. Pannarale [67,66], K. A. Pannone [54], B. C. Pant [100], F. H. Panther [4]
F. Paoletti [88], A. Paolone [66,290], A. Papadopoulos [28], E. E. Papalexakis [210], L. Papalini [88,87], G. Papigkiotis [291]
A. Paquis [38], A. Parisi [89,50], B.-J. Park [260], J. Park [292], W. Parker [65], G. Pascale [7,8], D. Pascucci [94]
A. Pasqualetti [59], R. Passaquieti [87,88], L. Passenger [149], D. Passuello [88], O. Patane [2], D. Pathak [14]
L. Pathak [14], A. Patra [32], B. Patricelli [87,88], A. S. Patron [11], B. G. Patterson [32], K. Paul [179], S. Paul [78]
E. Payne [10], T. Pearce [32], M. Pedraza [1], A. Pele [10], F. E. Peña Arellano [293], S. Penn [294], M. D. Penuliar [54]
A. Perego [107,108], Z. Pereira [131], J. J. Perez [44], C. Périgois [25], G. Perna [151,92,91], A. Perreca [107,108]
J. Perret [71], S. Perriès [97], J. W. Perry [36,104], D. Pesios [291], S. Petracca [89], C. Petrillo [89], H. P. Pfeiffer [1]
H. Pham [65], K. A. Pham [54], K. S. Phukon [116], H. Phurailatpam [14], M. Piarulli [126], L. Piccari [67,66]
O. J. Piccinni [33], M. Pichot [48], M. Piendibene [87,88], F. Piergiovanni [63,64], L. Pierini [97], G. Pierra [97]
V. Pierro [295,112], M. Pietrzak [95], M. Pillas [161], F. Pilo [88], L. Pinard [170], I. M. Pinto [295,112,296,31], M. Pinto [59]
B. J. Piotrzkowski [9], M. Pirello [2], M. D. Pitkin [186,224], A. Placidi [50], E. Placidi [97], M. L. Planas [98]
W. Plastino [297,20], C. Plunkett [34], R. Poggiani [87,88], E. Polini [34], L. Pompili [1], J. Poon [218], E. Porcelli [34]
E. K. Porter [71], C. Posnansky [6], R. Poulton [59], J. Powell [152], M. Pracchia [161], B. K. Pradhan [14]
T. Pradier [68], A. K. Prajapati [93], K. Prasai [86], R. Prasanna [234], P. Prasia [14], G. Pratten [4], G. Principe [190,46]
M. Principe [175], G. A. Prodi [107,108], L. Prokhorov [116], P. Prosperi [88], P. Prosposito [19,20], A. Providence [69]
A. Puecher [36,76], J. Pullin [11], M. Punturo [59], P. Puppo [66], M. Pürrer [153], H. Qi [15], J. Qin [33]
G. Quéméner [169,169,114], V. Quetschke [160], P. J. Quinonez [69], F. J. Raab [33], I. Rainho [135], S. Raja [100], C. Rajan [100]
B. Rajbhandari [109], K. E. Ramirez [65], F. A. Ramis Vidal [98], A. Ramos-Buades [36,1], D. Rana [14], S. Ranjan [100,69]
K. Ransom [65], P. Rapagnani [67,66], B. Ratto [69], A. Ray [9], V. Raymond [32], M. Razzano [87,88], J. Read [54]
M. Recaman Payo [106], T. Regimbau [30], L. Rei [56], S. Reid [58], D. H. Reitze [10], P. Relton [32], A. I. Renzini [1]
A. Renzini [124], B. Revenu [298,38], R. Reyes [203], A. S. Rezaei [66,67], F. Ricci [67,66], M. Ricci [66,67]
A. Ricciardone [87,88], J. W. Richardson [210], A. Rijal [69], K. Riles [10], H. K. Riley [32]
S. Rinaldi [264,91], J. Rittmeyer [85], C. Robertson [228], F. Robinet [8], M. Robinson [2], A. Rocchi [20], L. Rolland [25]
J. G. Rollins [10], A. E. Romano [299], R. Romano [3,4], A. Romero [33], I. M. Romero-Shaw [186], J. H. Romie [65]
S. Ronchini [6,43,118], T. J. Roocke [113], L. Rosa [4,31], T. J. Rosauer [210], C. A. Rose [60], D. Rosińska [122]
M. P. Ross [52], M. Rossello-Sastre [98], S. Rowan [28], S. Roy [13], S. K. Roy [187,188], D. Rozza [124,125]
P. Ruggi [59], N. Ruhama [238], E. Ruiz Morales [300,204], K. Ruiz-Rocha [142], S. Sachdev [60], T. Sadecki [2], J. Sadiq [128]
P. Saffarieh [36,104], S. Safi-Harb [164], M. R. Sah [12], S. Saha [139], T. Sainrat [68], S. Sajith Menon [212,67,66]
K. Sakai [301], M. Sakellariadou [72], S. Sakon [6], O. S. Salafia [156,125,124], F. Salces-Carcoba [10], L. Salconi [59]
M. Saleem [96], F. Salemi [67,66], M. Sallé [36], S. U. Salunkhe [14], S. Salvador [10], A. Samajdar [76,36]
A. Sanchez [2], E. J. Sanchez [10], J. H. Sanchez [80], L. E. Sanchez [10], N. Sanchis-Gual [135], J. R. Sanders [176]
E. M. Sänger [1], A. Santoliquido [43], F. Sarandrea [27], T. R. Saravanan [14], N. Sarin [149], P. Sarkar [7,8]
S. Sasaoka [180], A. Sasli [291], P. Sassi [50,89], B. Sassolas [170], B. S. Sathyaprakash [6,32], R. Sato [226]
Y. Sato [184], O. Sauter [44], R. L. Savage [2], T. Sawada [49], H. L. Sawant [14], S. Sayah [30], V. Scacco [19,20]
D. Schaetzl [10], M. Scheel [146], A. Schiebelbein [185], M. G. Schiworski [7], P. Schmidt [4], S. Schmidt [76]
R. Schnabel [85], M. Schneewind [7,8], R. M. S. Schofield [78], K. Schouteden [106], B. W. Schulte [7,8], B. F. Schutz [32,7,8]
E. Schwartz [86], M. Scialpi [302], J. Scott [28], S. M. Scott [33], R. M. Sedas [65], T. C. Seetharamu [28]
M. Seglar-Arroyo [42], Y. Sekiguchi [303], D. Sellers [65], A. S. Sengupta [93], D. Sentenac [59], E. G. Seo [28]
J. W. Seo [106], V. Sequino [31,4], M. Serra [66], G. Servignat [71,284], A. Sevrin [183], T. Shaffer [2], U. S. Shah [60]
M. S. Shahriar [80], M. A. Shaikh [245], L. Shao [231], A. Sharma [305], A. K. Sharma [22], P. Sharma, [100]

S. Sharma Chaudhary,[102] M. R. Shaw,[32] P. Shawhan [iD],[123] N. S. Shcheblanov [iD],[306,261] Y. Shikano [iD],[307,308]
M. Shikauchi,[40] K. Shimode [iD],[49] H. Shinkai [iD],[309] J. Shiota,[232] S. Shirke,[14] D. H. Shoemaker [iD],[34]
D. M. Shoemaker [iD],[145] R. W. Short,[2] S. ShyamSundar,[100] A. Sider,[155] H. Siegel [iD],[187,188] D. Sigg [iD],[57]
L. Silenzi [iD],[50,51] M. Simmonds,[113] L. P. Singer [iD],[310] A. Singh,[214] D. Singh [iD],[6] M. K. Singh [iD],[98] N. Singh [iD],[98]
S. Singh,[180,62] A. Singha [iD],[35,36] A. M. Sintes [iD],[98] V. Sipala,[167,154] V. Skliris [iD],[32] B. J. J. Slagmolen [iD],[33]
D. A. Slater,[196] T. J. Slaven-Blair,[29] J. Smetana,[116] J. R. Smith [iD],[54] L. Smith [iD],[28,190] R. J. E. Smith [iD],[149]
W. J. Smith [iD],[142] K. Somiya [iD],[180] I. Song [iD],[139] K. Soni,[14] S. Soni [iD],[34] V. Sordini,[97] F. Sorrentino,[56]
H. Sotani [iD],[311] A. Southgate,[32] F. Spada [iD],[35,36] A. P. Spencer [iD],[28] M. Spera [iD],[46,312]
P. Spinicelli,[59] C. A. Sprague,[277] A. K. Srivastava,[93] F. Stachurski [iD],[28] D. A. Steer [iD],[313] N. Steinle [iD],[164]
J. Steinlechner,[35,36] S. Steinlechner [iD],[35,36] N. Stergioulas [iD],[291] P. Stevens,[38] S. P. Stevenson,[152] F. Stolzi [iD],[99]
M. StPierre,[153] G. Stratta [iD],[314,134,66,315] M. D. Strong,[11] A. Strunk,[2] R. Sturani,[316] A. L. Stuver,[55,*]
M. Suchenek,[95] A. Sudhagar [iD],[95] N. Sueltmann,[85] L. Suleiman [iD],[54] J.M. Sullivan [iD],[317] K. D. Sullivan,[11]
J. Sun,[240] L. Sun [iD],[33] S. Sunil,[93] J. Suresh [iD],[48] B. J. Sutton,[72] P. J. Sutton [iD],[32] T. Suzuki [iD],[226] Y. Suzuki,[232]
B. L. Swinkels [iD],[36] A. Syx,[68] M. J. Szczepańczyk [iD],[318,44] P. Szewczyk [iD],[122] M. Tacca [iD],[36] H. Tagoshi [iD],[200]
S. C. Tait [iD],[28] H. Takahashi [iD],[270] R. Takahashi [iD],[24] A. Takamori [iD],[53] T. Takase,[49] K. Takatani,[233]
H. Takeda [iD],[319] K. Takeshita,[180] C. Talbot,[129] M. Tamaki,[200] N. Tamanini [iD],[126] D. Tanabe,[140] K. Tanaka,[49]
S. J. Tanaka [iD],[232] T. Tanaka,[319] D. Tang,[29] S. Tanioka [iD],[79] D. B. Tanner,[44] W. Tanner,[7,8] L. Tao [iD],[210]
R. D. Tapia,[6] E. N. Tapia San Martín [iD],[36] R. Tarafder,[10] C. Taranto,[19,20] A. Taruya [iD],[320] J. D. Tasson [iD],[321]
J. G. Tau [iD],[109] R. Tenorio [iD],[98] H. Themann,[203] A. Theodoropoulos [iD],[135] M. P. Thirugnanasambandam,[14]
L. M. Thomas [iD],[10] M. Thomas,[65] P. Thomas,[2] J. E. Thompson [iD],[236] S. R. Thondapu,[100] K. A. Thorne,[65]
E. Thrane,[149] S. Tibrewal [iD],[145] J. Tissino [iD],[46] A. Tiwari,[14] S. Tiwari,[43] S. Tiwari [iD],[55] V. Tiwari [iD],[32]
M. R. Todd,[79] A. M. Toivonen [iD],[96] K. Toland [iD],[28] A. E. Tolley [iD],[77] T. Tomaru [iD],[24] K. Tomita,[233]
V. Tommasini,[10] T. Tomura [iD],[49] H. Tong [iD],[149] C. Tong-Yu,[140] A. Toriyama,[232] N. Toropov [iD],[116]
A. Torres-Forné [iD],[135,136] C. I. Torrie,[10] M. Toscani [iD],[126] I. Tosta e Melo [iD],[322] E. Tournefier [iD],[30]
M. Trad Nery,[48] A. Trapananti [iD],[51,50] F. Travasso [iD],[51,50] G. Traylor,[65] C. Trejo,[10] M. Trevor,[123]
M. C. Tringali [iD],[59] A. Tripathee [iD],[90] G. Troian [iD],[190,46] A. Trovato [iD],[190,46] L. Trozzo,[4] R. J. Trudeau,[10]
T. T. L. Tsang [iD],[32] S. Tsuchida [iD],[323] L. Tsukada [iD],[220] K. Turbang [iD],[183,21] M. Turconi [iD],[48] C. Turski,[94]
H. Ubach [iD],[41,81] N. Uchikata [iD],[200] T. Uchiyama [iD],[49] R. P. Udall [iD],[10] T. Uehara [iD],[324] M. Uematsu,[233] S. Ueno,[232]
V. Undheim [iD],[279] T. Ushiba [iD],[49] M. Vacatello [iD],[88,87] H. Vahlbruch [iD],[7,8] G. Vajente [iD],[10] A. Vajpeyi,[149]
G. Valdes [iD],[325] J. Valencia [iD],[98] A. F. Valentini,[11] M. Valentini [iD],[104,36] S. A. Vallejo-Peña [iD],[299] S. Vallero,[27]
V. Valsan [iD],[9] N. van Bakel,[36] M. van Beuzekom [iD],[36] M. van Dael,[36,326] J. F. J. van den Brand [iD],[35,104,36]
C. Van Den Broeck,[76,36] D. C. Vander-Hyde,[79] M. van der Sluys [iD],[36,76] A. Van de Walle,[38] J. van Dongen [iD],[36,104]
K. Vandra,[55] H. van Haevermaet [iD],[21] J. V. van Heijningen [iD],[36,104] P. Van Hove [iD],[68] J. Vanier,[255]
M. VanKeuren,[101] J. Vanosky,[2] M. H. P. M. van Putten [iD],[16] Z. van Ranst [iD],[35,36] N. van Remortel [iD],[21]
M. Vardaro,[35,36] A. F. Vargas,[121] J. J. Varghese,[69] V. Varma [iD],[131] A. N. Vazquez,[86] A. Vecchio [iD],[116]
G. Vedovato,[92] J. Veitch [iD],[28] P. J. Veitch [iD],[113] S. Venikoudis,[13] J. Venneberg [iD],[7,8] P. Verdier [iD],[97]
M. Vereecken,[13] D. Verkindt [iD],[30] B. Verma,[131] P. Verma,[181] Y. Verma [iD],[100] S. M. Vermeulen [iD],[10] F. Vetrano,[63]
A. Veutro [iD],[66,67] A. M. Vibhute [iD],[2] A. Viceré [iD],[63,64] S. Vidyant,[79] A. D. Viets [iD],[284] A. Vijaykumar [iD],[185]
A. Vilkha,[109] V. Villa-Ortega [iD],[128] E. T. Vincent [iD],[60] J.-Y. Vinet,[48] S. Viret,[97] A. Virtuoso [iD],[46] S. Vitale [iD],[34]
A. Vives,[78] H. Vocca [iD],[89,50] D. Voigt [iD],[85] E. R. G. von Reis,[2] J. S. A. von Wrangel,[7,8] L. Vujeva,[137]
S. P. Vyatchanin [iD],[105] J. Wack,[10] L. E. Wade,[101] M. Wade [iD],[101] K. J. Wagner [iD],[109] A. Wajid,[56,57] M. Walker,[120]
G. S. Wallace,[58] S. L. Wallace,[10] J. E. Wang,[86] H. Wang [iD],[39] J. Z. Wang,[140] W. H. Wang,[160] Y. F. Wang [iD],[9]
Z. Wang,[140] G. Waratkar [iD],[208] J. Warner,[2] M. Was [iD],[30] T. Washimi [iD],[24] N. Y. Washington,[10] D. Watarai,[40]
K. E. Wayt,[101] B. R. Weaver,[32] B. Weaver,[2] C. R. Weaving,[77] S. A. Webster,[28] N. L. Weickhardt [iD],[85]
M. Weinert,[7,8] A. J. Weinstein [iD],[10] R. Weiss,[34] F. Wellmann,[7,8] L. Wen,[29] P. Weßels [iD],[7,8] K. Wette [iD],[33]
J. T. Whelan [iD],[109] B. F. Whiting [iD],[44] C. Whittle [iD],[10] E. G. Wickens,[77] J. B. Wildberger,[1] D. Wilken [iD],[7,8]
D. J. Willadsen,[84] K. Willetts,[32] D. Williams [iD],[28] M. J. Williams [iD],[28] N. S. Williams,[116] J. L. Willis [iD],[10]
B. Willke [iD],[8,7,8] M. Wils [iD],[106] C. W. Winborn,[102] J. Winterflood,[29] C. C. Wipf,[10] G. Woan [iD],[28] J. Woehler,[35,36]
N. E. Wolfe,[34] H. T. Wong [iD],[140] I. C. F. Wong [iD],[218,106] J. L. Wright,[33] M. Wright [iD],[28] C. Wu [iD],[139] D. S. Wu [iD],[7,8]
H. Wu [iD],[139] E. Wuchner,[54] D. M. Wysocki [iD],[9] A. Xu [iD],[34] Y. Xu [iD],[205] N. Yadav [iD],[95] H. Yamamoto [iD],[10]
K. Yamamoto [iD],[184] T. S. Yamamoto [iD],[40] T. Yamamoto [iD],[49] S. Yamamura,[200] R. Yamazaki [iD],[232] T. Yan,[116]
F. W. Yang [iD],[327] F. Yang,[267] K. Z. Yang [iD],[96] Y. Yang [iD],[143] Z. Yarbrough [iD],[11] H. Yasui,[49] S.-W. Yeh,[139]
A. B. Yelikar [iD],[109] X. Yin,[34] J. Yokoyama [iD],[328,40,39] T. Yokozawa,[49] J. Yoo [iD],[147] J. Yoo [iD],[146] S. Yuan,[29]
H. Yuzurihara [iD],[49] A. Zadrożny,[181] M. Zanolin,[69] M. Zeeshan [iD],[2] T. Zelenova,[59] J.-P. Zendri,[92] M. Zeoli [iD],[13]
M. Zerrad,[37] M. Zevin [iD],[80] A. C. Zhang,[267] L. Zhang,[10] R. Zhang [iD],[148] T. Zhang,[116] Y. Zhang [iD],[33] C. Zhao [iD],[29]
Yue Zhao,[327] Yuhang Zhao [iD],[71] Y. Zheng [iD],[102] H. Zhong [iD],[96] R. Zhou,[273] X.-J. Zhu [iD],[329] Z.-H. Zhu [iD],[329,209]
A. B. Zimmerman [iD],[145] M. E. Zucker,[34,10] And J. Zweizig [iD],[10]

THE LIGO SCIENTIFIC COLLABORATION, THE VIRGO COLLABORATION, AND THE KAGRA COLLABORATION

[1]*Max Planck Institute for Gravitational Physics (Albert Einstein Institute), D-14476 Potsdam, Germany*
[2]*LIGO Hanford Observatory, Richland, WA 99352, USA*
[3]*Dipartimento di Farmacia, Università di Salerno, I-84084 Fisciano, Salerno, Italy*
[4]*INFN, Sezione di Napoli, I-80126 Napoli, Italy*
[5]*University of Warwick, Coventry CV4 7AL, United Kingdom*
[6]*The Pennsylvania State University, University Park, PA 16802, USA*
[7]*Max Planck Institute for Gravitational Physics (Albert Einstein Institute), D-30167 Hannover, Germany*
[8]*Leibniz Universität Hannover, D-30167 Hannover, Germany*
[9]*University of Wisconsin-Milwaukee, Milwaukee, WI 53201, USA*
[10]*LIGO Laboratory, California Institute of Technology, Pasadena, CA 91125, USA*
[11]*Louisiana State University, Baton Rouge, LA 70803, USA*
[12]*Tata Institute of Fundamental Research, Mumbai 400005, India*
[13]*Université catholique de Louvain, B-1348 Louvain-la-Neuve, Belgium*
[14]*Inter-University Centre for Astronomy and Astrophysics, Pune 411007, India*
[15]*Queen Mary University of London, London E1 4NS, United Kingdom*
[16]*Department of Physics and Astronomy, Sejong University, 209 Neungdong-ro, Gwangjin-gu, Seoul 143-747, Republic of Korea*
[17]*Instituto Nacional de Pesquisas Espaciais, 12227-010 São José dos Campos, São Paulo, Brazil*
[18]*SUPA, University of the West of Scotland, Paisley PA1 2BE, United Kingdom*
[19]*Università di Roma Tor Vergata, I-00133 Roma, Italy*
[20]*INFN, Sezione di Roma Tor Vergata, I-00133 Roma, Italy*
[21]*Universiteit Antwerpen, 2000 Antwerpen, Belgium*
[22]*International Centre for Theoretical Sciences, Tata Institute of Fundamental Research, Bengaluru 560089, India*
[23]*University College Dublin, Belfield, Dublin 4, Ireland*
[24]*Gravitational Wave Science Project, National Astronomical Observatory of Japan, 2-21-1 Osawa, Mitaka City, Tokyo 181-8588, Japan*
[25]*Advanced Technology Center, National Astronomical Observatory of Japan, 2-21-1 Osawa, Mitaka City, Tokyo 181-8588, Japan*
[26]*Theoretisch-Physikalisches Institut, Friedrich-Schiller-Universität Jena, D-07743 Jena, Germany*
[27]*INFN Sezione di Torino, I-10125 Torino, Italy*
[28]*SUPA, University of Glasgow, Glasgow G12 8QQ, United Kingdom*
[29]*OzGrav, University of Western Australia, Crawley, Western Australia 6009, Australia*
[30]*Univ. Savoie Mont Blanc, CNRS, Laboratoire d'Annecy de Physique des Particules - IN2P3, F-74000 Annecy, France*
[31]*Università di Napoli "Federico II", I-80126 Napoli, Italy*
[32]*Cardiff University, Cardiff CF24 3AA, United Kingdom*
[33]*OzGrav, Australian National University, Canberra, Australian Capital Territory 0200, Australia*
[34]*LIGO Laboratory, Massachusetts Institute of Technology, Cambridge, MA 02139, USA*
[35]*Maastricht University, 6200 MD Maastricht, Netherlands*
[36]*Nikhef, 1098 XG Amsterdam, Netherlands*
[37]*Aix Marseille Univ, CNRS, Centrale Med, Institut Fresnel, F-13013 Marseille, France*
[38]*Université Paris-Saclay, CNRS/IN2P3, IJCLab, 91405 Orsay, France*
[39]*Department of Physics, The University of Tokyo, 7-3-1 Hongo, Bunkyo-ku, Tokyo 113-0033, Japan*
[40]*University of Tokyo, Tokyo, 113-0033, Japan.*
[41]*Institut de Ciències del Cosmos (ICCUB), Universitat de Barcelona (UB), c. Martí i Franquès, 1, 08028 Barcelona, Spain*
[42]*Institut de Física d'Altes Energies (IFAE), The Barcelona Institute of Science and Technology, Campus UAB, E-08193 Bellaterra (Barcelona), Spain*
[43]*Gran Sasso Science Institute (GSSI), I-67100 L'Aquila, Italy*
[44]*University of Florida, Gainesville, FL 32611, USA*
[45]*Dipartimento di Scienze Matematiche, Informatiche e Fisiche, Università di Udine, I-33100 Udine, Italy*
[46]*INFN, Sezione di Trieste, I-34127 Trieste, Italy*
[47]*Tecnologico de Monterrey, Escuela de Ingeniería y Ciencias, Monterrey 64849, Mexico*
[48]*Université Côte d'Azur, Observatoire de la Côte d'Azur, CNRS, Artemis, F-06304 Nice, France*
[49]*Institute for Cosmic Ray Research, KAGRA Observatory, The University of Tokyo, 238 Higashi-Mozumi, Kamioka-cho, Hida City, Gifu 506-1205, Japan*
[50]*INFN, Sezione di Perugia, I-06123 Perugia, Italy*
[51]*Università di Camerino, I-62032 Camerino, Italy*
[52]*University of Washington, Seattle, WA 98195, USA*
[53]*Earthquake Research Institute, The University of Tokyo, 1-1-1 Yayoi, Bunkyo-ku, Tokyo 113-0032, Japan*
[54]*California State University Fullerton, Fullerton, CA 92831, USA*
[55]*Villanova University, Villanova, PA 19085, USA*

[56]INFN, Sezione di Genova, I-16146 Genova, Italy

[57]Dipartimento di Fisica, Università degli Studi di Genova, I-16146 Genova, Italy

[58]SUPA, University of Strathclyde, Glasgow G1 1XQ, United Kingdom

[59]European Gravitational Observatory (EGO), I-56021 Cascina, Pisa, Italy

[60]Georgia Institute of Technology, Atlanta, GA 30332, USA

[61]Royal Holloway, University of London, London TW20 0EX, United Kingdom

[62]Astronomical course, The Graduate University for Advanced Studies (SOKENDAI), 2-21-1 Osawa, Mitaka City, Tokyo 181-8588, Japan

[63]Università degli Studi di Urbino "Carlo Bo", I-61029 Urbino, Italy

[64]INFN, Sezione di Firenze, I-50019 Sesto Fiorentino, Firenze, Italy

[65]LIGO Livingston Observatory, Livingston, LA 70754, USA

[66]INFN, Sezione di Roma, I-00185 Roma, Italy

[67]Università di Roma "La Sapienza", I-00185 Roma, Italy

[68]Université de Strasbourg, CNRS, IPHC UMR 7178, F-67000 Strasbourg, France

[69]Embry-Riddle Aeronautical University, Prescott, AZ 86301, USA

[70]Dipartimento di Fisica "E.R. Caianiello", Università di Salerno, I-84084 Fisciano, Salerno, Italy

[71]Université Paris Cité, CNRS, Astroparticule et Cosmologie, F-75013 Paris, France

[72]King's College London, University of London, London WC2R 2LS, United Kingdom

[73]Korea Institute of Science and Technology Information, Daejeon 34141, Republic of Korea

[74]Université libre de Bruxelles, 1050 Bruxelles, Belgium

[75]International College, Osaka University, 1-1 Machikaneyama-cho, Toyonaka City, Osaka 560-0043, Japan

[76]Institute for Gravitational and Subatomic Physics (GRASP), Utrecht University, 3584 CC Utrecht, Netherlands

[77]University of Portsmouth, Portsmouth, PO1 3FX, United Kingdom

[78]University of Oregon, Eugene, OR 97403, USA

[79]Syracuse University, Syracuse, NY 13244, USA

[80]Northwestern University, Evanston, IL 60208, USA

[81]Departament de Física Quàntica i Astrofísica (FQA), Universitat de Barcelona (UB), c. Martí i Franquès, 1, 08028 Barcelona, Spain

[82]Dipartimento di Medicina, Chirurgia e Odontoiatria "Scuola Medica Salernitana", Università di Salerno, I-84081 Baronissi, Salerno, Italy

[83]HUN-REN Wigner Research Centre for Physics, H-1121 Budapest, Hungary

[84]Concordia University Wisconsin, Mequon, WI 53097, USA

[85]Universität Hamburg, D-22761 Hamburg, Germany

[86]Stanford University, Stanford, CA 94305, USA

[87]Università di Pisa, I-56127 Pisa, Italy

[88]INFN, Sezione di Pisa, I-56127 Pisa, Italy

[89]Università di Perugia, I-06123 Perugia, Italy

[90]University of Michigan, Ann Arbor, MI 48109, USA

[91]Università di Padova, Dipartimento di Fisica e Astronomia, I-35131 Padova, Italy

[92]INFN, Sezione di Padova, I-35131 Padova, Italy

[93]Institute for Plasma Research, Bhat, Gandhinagar 382428, India

[94]Universiteit Gent, B-9000 Gent, Belgium

[95]Nicolaus Copernicus Astronomical Center, Polish Academy of Sciences, 00-716, Warsaw, Poland

[96]University of Minnesota, Minneapolis, MN 55455, USA

[97]Université Claude Bernard Lyon 1, CNRS, IP2I Lyon / IN2P3, UMR 5822, F-69622 Villeurbanne, France

[98]IAC3–IEEC, Universitat de les Illes Balears, E-07122 Palma de Mallorca, Spain

[99]Università di Siena, I-53100 Siena, Italy

[100]RRCAT, Indore, Madhya Pradesh 452013, India

[101]Kenyon College, Gambier, OH 43022, USA

[102]Missouri University of Science and Technology, Rolla, MO 65409, USA

[103]Colorado State University, Fort Collins, CO 80523, USA

[104]Department of Physics and Astronomy, Vrije Universiteit Amsterdam, 1081 HV Amsterdam, Netherlands

[105]Lomonosov Moscow State University, Moscow 119991, Russia

[106]Katholieke Universiteit Leuven, Oude Markt 13, 3000 Leuven, Belgium

[107]Università di Trento, Dipartimento di Fisica, I-38123 Povo, Trento, Italy

[108]INFN, Trento Institute for Fundamental Physics and Applications, I-38123 Povo, Trento, Italy

[109]Rochester Institute of Technology, Rochester, NY 14623, USA

[110]Bar-Ilan University, Ramat Gan, 5290002, Israel

[111]University of British Columbia, Vancouver, BC V6T 1Z4, Canada

[112]INFN, Sezione di Napoli, Gruppo Collegato di Salerno, I-80126 Napoli, Italy

[113]*O$_z$Grav, University of Adelaide, Adelaide, South Australia 5005, Australia*

[114]*Centre national de la recherche scientifique, 75016 Paris, France*

[115]*Univ Rennes, CNRS, Institut FOTON - UMR 6082, F-35000 Rennes, France*

[116]*University of Birmingham, Birmingham B15 2TT, United Kingdom*

[117]*Washington State University, Pullman, WA 99164, USA*

[118]*INFN, Laboratori Nazionali del Gran Sasso, I-67100 Assergi, Italy*

[119]*Laboratoire Kastler Brossel, Sorbonne Université, CNRS, ENS-Université PSL, Collège de France, F-75005 Paris, France*

[120]*Christopher Newport University, Newport News, VA 23606, USA*

[121]*O$_z$Grav, University of Melbourne, Parkville, Victoria 3010, Australia*

[122]*Astronomical Observatory Warsaw University, 00-478 Warsaw, Poland*

[123]*University of Maryland, College Park, MD 20742, USA*

[124]*Università degli Studi di Milano-Bicocca, I-20126 Milano, Italy*

[125]*INFN, Sezione di Milano-Bicocca, I-20126 Milano, Italy*

[126]*L2IT, Laboratoire des 2 Infinis - Toulouse, Université de Toulouse, CNRS/IN2P3, UPS, F-31062 Toulouse Cedex 9, France*

[127]*Université de Lyon, Université Claude Bernard Lyon 1, CNRS, Institut Lumière Matière, F-69622 Villeurbanne, France*

[128]*IGFAE, Universidade de Santiago de Compostela, 15782 Spain*

[129]*University of Chicago, Chicago, IL 60637, USA*

[130]*University of Arizona, Tucson, AZ 85721, USA*

[131]*University of Massachusetts Dartmouth, North Dartmouth, MA 02747, USA*

[132]*INFN, Laboratori Nazionali del Sud, I-95125 Catania, Italy*

[133]*Niels Bohr Institute, Copenhagen University, 2100 København, Denmark*

[134]*Istituto di Astrofisica e Planetologia Spaziali di Roma, 00133 Roma, Italy*

[135]*Departamento de Astronomía y Astrofísica, Universitat de València, E-46100 Burjassot, València, Spain*

[136]*Observatori Astronòmic, Universitat de València, E-46980 Paterna, València, Spain*

[137]*Niels Bohr Institute, University of Copenhagen, 2100 Kóbenhavn, Denmark*

[138]*Department of Physics, National Cheng Kung University, No.1, University Road, Tainan City 701, Taiwan*

[139]*National Tsing Hua University, Hsinchu City 30013, Taiwan*

[140]*National Central University, Taoyuan City 320317, Taiwan*

[141]*O$_z$Grav, Charles Sturt University, Wagga Wagga, New South Wales 2678, Australia*

[142]*Vanderbilt University, Nashville, TN 37235, USA*

[143]*Department of Electrophysics, National Yang Ming Chiao Tung University, 101 Univ. Street, Hsinchu, Taiwan*

[144]*Kamioka Branch, National Astronomical Observatory of Japan, 238 Higashi-Mozumi, Kamioka-cho, Hida City, Gifu 506-1205, Japan*

[145]*University of Texas, Austin, TX 78712, USA*

[146]*CaRT, California Institute of Technology, Pasadena, CA 91125, USA*

[147]*Cornell University, Ithaca, NY 14850, USA*

[148]*Northeastern University, Boston, MA 02115, USA*

[149]*O$_z$Grav, School of Physics & Astronomy, Monash University, Clayton 3800, Victoria, Australia*

[150]*Dipartimento di Ingegneria Industriale (DIIN), Università di Salerno, I-84084 Fisciano, Salerno, Italy*

[151]*INAF, Osservatorio Astronomico di Padova, I-35122 Padova, Italy*

[152]*O$_z$Grav, Swinburne University of Technology, Hawthorn VIC 3122, Australia*

[153]*University of Rhode Island, Kingston, RI 02881, USA*

[154]*INFN Cagliari, Physics Department, Università degli Studi di Cagliari, Cagliari 09042, Italy*

[155]*Université Libre de Bruxelles, Brussels 1050, Belgium*

[156]*INAF, Osservatorio Astronomico di Brera sede di Merate, I-23807 Merate, Lecco, Italy*

[157]*Departamento de Matemáticas, Universitat de València, E-46100 Burjassot, València, Spain*

[158]*Montana State University, Bozeman, MT 59717, USA*

[159]*Johns Hopkins University, Baltimore, MD 21218, USA*

[160]*The University of Texas Rio Grande Valley, Brownsville, TX 78520, USA*

[161]*Université de Liège, B-4000 Liège, Belgium*

[162]*DIFA- Alma Mater Studiorum Università di Bologna, Via Zamboni, 33 - 40126 Bologna, Italy*

[163]*Istituto Nazionale Di Fisica Nucleare - Sezione di Bologna, viale Carlo Berti Pichat 6/2, Bologna, Italy*

[164]*University of Manitoba, Winnipeg, MB R3T 2N2, Canada*

[165]*INFN-CNAF - Bologna, Viale Carlo Berti Pichat, 6/2, 40127 Bologna BO, Italy*

[166]*Chennai Mathematical Institute, Chennai 603103, India*

[167]*Università degli Studi di Sassari, I-07100 Sassari, Italy*

[168]*Université de Normandie, ENSICAEN, UNICAEN, CNRS/IN2P3, LPC Caen, F-14000 Caen, France*

[169]*Laboratoire de Physique Corpusculaire Caen, 6 boulevard du maréchal Juin, F-14050 Caen, France*

[170]*Université Claude Bernard Lyon 1, CNRS, Laboratoire des Matériaux Avancés (LMA), IP2I Lyon / IN2P3, UMR 5822, F-69622 Villeurbanne, France*

[171]*International Centre for Theoretical Sciences, Tata Institute of Fundamental Research, Bangalore 560089, India*

[172]*Università di Firenze, Sesto Fiorentino I-50019, Italy*

[173]*Dipartimento di Scienze Matematiche, Fisiche e Informatiche, Università di Parma, I-43124 Parma, Italy*

[174]*INFN, Sezione di Milano Bicocca, Gruppo Collegato di Parma, I-43124 Parma, Italy*

[175]*University of Sannio at Benevento, I-82100 Benevento, Italy and INFN, Sezione di Napoli, I-80100 Napoli, Italy*

[176]*Marquette University, Milwaukee, WI 53233, USA*

[177]*Perimeter Institute, Waterloo, ON N2L 2Y5, Canada*

[178]*Corps des Mines, Mines Paris, Université PSL, 60 Bd Saint-Michel, 75272 Paris, France*

[179]*Indian Institute of Technology Madras, Chennai 600036, India*

[180]*Graduate School of Science, Tokyo Institute of Technology, 2-12-1 Ookayama, Meguro-ku, Tokyo 152-8551, Japan*

[181]*National Center for Nuclear Research, 05-400 Świerk-Otwock, Poland*

[182]*Institut d'Astrophysique de Paris, Sorbonne Université, CNRS, UMR 7095, 75014 Paris, France*

[183]*Vrije Universiteit Brussel, 1050 Brussel, Belgium*

[184]*Faculty of Science, University of Toyama, 3190 Gofuku, Toyama City, Toyama 930-8555, Japan*

[185]*Canadian Institute for Theoretical Astrophysics, University of Toronto, Toronto, ON M5S 3H8, Canada*

[186]*University of Cambridge, Cambridge CB2 1TN, United Kingdom*

[187]*Stony Brook University, Stony Brook, NY 11794, USA*

[188]*Center for Computational Astrophysics, Flatiron Institute, New York, NY 10010, USA*

[189]*Montclair State University, Montclair, NJ 07043, USA*

[190]*Dipartimento di Fisica, Università di Trieste, I-34127 Trieste, Italy*

[191]*HUN-REN Institute for Nuclear Research, H-4026 Debrecen, Hungary*

[192]*Centro de Física das Universidades do Minho e do Porto, Universidade do Minho, PT-4710-057 Braga, Portugal*

[193]*Centre de Physique des Particules de Marseille, 163, avenue de Luminy, 13288 Marseille cedex 09, France*

[194]*CNR-SPIN, I-84084 Fisciano, Salerno, Italy*

[195]*Scuola di Ingegneria, Università della Basilicata, I-85100 Potenza, Italy*

[196]*Western Washington University, Bellingham, WA 98225, USA*

[197]*Barry University, Miami Shores, FL 33168, USA*

[198]*Centro de Astrofísica e Gravitação, Departamento de Física, Instituto Superior Técnico - IST, Universidade de Lisboa - UL, Av. Rovisco Pais 1, 1049-001 Lisboa, Portugal*

[199]*Eötvös University, Budapest 1117, Hungary*

[200]*Institute for Cosmic Ray Research, KAGRA Observatory, The University of Tokyo, 5-1-5 Kashiwa-no-Ha, Kashiwa City, Chiba 277-8582, Japan*

[201]*Nambu Yoichiro Institute of Theoretical and Experimental Physics (NITEP), Osaka Metropolitan University, 3-3-138 Sugimoto-cho, Sumiyoshi-ku, Osaka City, Osaka 558-8585, Japan*

[202]*Université Côte d'Azur, Observatoire de la Côte d'Azur, CNRS, Lagrange, F-06304 Nice, France*

[203]*California State University, Los Angeles, Los Angeles, CA 90032, USA*

[204]*Instituto de Fisica Teorica UAM-CSIC, Universidad Autonoma de Madrid, 28049 Madrid, Spain*

[205]*University of Zurich, Winterthurerstrasse 190, 8057 Zurich, Switzerland*

[206]*Laboratoire d'Acoustique de l'Université du Mans, UMR CNRS 6613, F-72085 Le Mans, France*

[207]*University of Szeged, Dóm tér 9, Szeged 6720, Hungary*

[208]*Indian Institute of Technology Bombay, Powai, Mumbai 400 076, India*

[209]*School of Physics and Technology, Wuhan University, Bayi Road 299, Wuchang District, Wuhan, Hubei, 430072, China*

[210]*University of California, Riverside, Riverside, CA 92521, USA*

[211]*University of Nottingham NG7 2RD, UK*

[212]*Ariel University, Ramat HaGolan St 65, Ari'el, Israel*

[213]*University of the Chinese Academy of Sciences / International Centre for Theoretical Physics Asia-Pacific, Bejing 100049, China*

[214]*The University of Mississippi, University, MS 38677, USA*

[215]*Institute of Physics, Academia Sinica, 128 Sec. 2, Academia Rd., Nankang, Taipei 11529, Taiwan*

[216]*Science and Technology Institute, Universities Space Research Association, Huntsville, AL 35805, USA*

[217]*Shanghai Astronomical Observatory, Chinese Academy of Sciences, 80 Nandan Road, Shanghai 200030, China*

[218]*The Chinese University of Hong Kong, Shatin, NT, Hong Kong*

[219]*American University, Washington, DC 20016, USA*

[220]*University of Nevada, Las Vegas, Las Vegas, NV 89154, USA*

[221]*Department of Applied Physics, Fukuoka University, 8-19-1 Nanakuma, Jonan, Fukuoka City, Fukuoka 814-0180, Japan*

[222]*IAC3–IEEC, Universitat de les Illes Balears, E-07122 Palma de Mallorca, Spain*

[223]*University of California, Berkeley, CA 94720, USA*

[224]*University of Lancaster, Lancaster LA1 4YW, United Kingdom*

[225] *College of Industrial Technology, Nihon University, 1-2-1 Izumi, Narashino City, Chiba 275-8575, Japan*

[226] *Faculty of Engineering, Niigata University, 8050 Ikarashi-2-no-cho, Nishi-ku, Niigata City, Niigata 950-2181, Japan*

[227] *Department of Physics, Tamkang University, No. 151, Yingzhuan Rd., Danshui Dist., New Taipei City 25137, Taiwan*

[228] *Rutherford Appleton Laboratory, Didcot OX11 0DE, United Kingdom*

[229] *Department of Astronomy and Space Science, Chungnam National University, 9 Daehak-ro, Yuseong-gu, Daejeon 34134, Republic of Korea*

[230] *Scuola Normale Superiore, I-56126 Pisa, Italy*

[231] *Kavli Institute for Astronomy and Astrophysics, Peking University, Yiheyuan Road 5, Haidian District, Beijing 100871, China*

[232] *Department of Physical Sciences, Aoyama Gakuin University, 5-10-1 Fuchinobe, Sagamihara City, Kanagawa 252-5258, Japan*

[233] *Department of Physics, Graduate School of Science, Osaka Metropolitan University, 3-3-138 Sugimoto-cho, Sumiyoshi-ku, Osaka City, Osaka 558-8585, Japan*

[234] *Directorate of Construction, Services & Estate Management, Mumbai 400094, India*

[235] *Faculty of Physics, University of Białystok, 15-245 Białystok, Poland*

[236] *University of Southampton, Southampton SO17 1BJ, United Kingdom*

[237] *Sungkyunkwan University, Seoul 03063, Republic of Korea*

[238] *Department of Physics, Ulsan National Institute of Science and Technology (UNIST), 50 UNIST-gil, Ulju-gun, Ulsan 44919, Republic of Korea*

[239] *Institute for Cosmic Ray Research, The University of Tokyo, 5-1-5 Kashiwa-no-Ha, Kashiwa City, Chiba 277-8582, Japan*

[240] *Chung-Ang University, Seoul 06974, Republic of Korea*

[241] *University of Washington Bothell, Bothell, WA 98011, USA*

[242] *Aix Marseille Université, Jardin du Pharo, 58 Boulevard Charles Livon, 13007 Marseille, France*

[243] *Laboratoire de Physique et de Chimie de l'Environnement, Université Joseph KI-ZERBO, 9GH2+3V5, Ouagadougou, Burkina Faso*

[244] *Ewha Womans University, Seoul 03760, Republic of Korea*

[245] *Seoul National University, Seoul 08826, Republic of Korea*

[246] *Korea Astronomy and Space Science Institute, Daejeon 34055, Republic of Korea*

[247] *Institute of Particle and Nuclear Studies (IPNS), High Energy Accelerator Research Organization (KEK), 1-1 Oho, Tsukuba City, Ibaraki 305-0801, Japan*

[248] *Division of Science, National Astronomical Observatory of Japan, 2-21-1 Osawa, Mitaka City, Tokyo 181-8588, Japan*

[249] *Bard College, Annandale-On-Hudson, NY 12504, USA*

[250] *Institute of Mathematics, Polish Academy of Sciences, 00656 Warsaw, Poland*

[251] *Astronomical Observatory, Jagiellonian University, 31-007 Cracow, Poland*

[252] *Department of Physics and Astronomy, University of Padova, Via Marzolo, 8-35151 Padova, Italy*

[253] *Sezione di Padova, Istituto Nazionale di Fisica Nucleare (INFN), Via Marzolo, 8-35131 Padova, Italy*

[254] *Department of Physics, Nagoya University, ES building, Furocho, Chikusa-ku, Nagoya, Aichi 464-8602, Japan*

[255] *Université de Montréal/Polytechnique, Montreal, Quebec H3T 1J4, Canada*

[256] *Indian Institute of Science Education and Research, Kolkata, Mohanpur, West Bengal 741252, India*

[257] *Texas Tech University, Lubbock, TX 79409, USA*

[258] *Università degli Studi di Cagliari, Via Università 40, 09124 Cagliari, Italy*

[259] *Inje University Gimhae, South Gyeongsang 50834, Republic of Korea*

[260] *Technology Center for Astronomy and Space Science, Korea Astronomy and Space Science Institute (KASI), 776 Daedeokdae-ro, Yuseong-gu, Daejeon 34055, Republic of Korea*

[261] *NAVIER, École des Ponts, Univ Gustave Eiffel, CNRS, Marne-la-Vallée, France*

[262] *National Center for High-performance Computing, National Applied Research Laboratories, No. 7, R&D 6th Rd., Hsinchu Science Park, Hsinchu City 30076, Taiwan*

[263] *NASA Marshall Space Flight Center, Huntsville, AL 35811, USA*

[264] *Institut fuer Theoretische Astrophysik, Zentrum fuer Astronomie Heidelberg, Universitaet Heidelberg, Albert Ueberle Str. 2, 69120 Heidelberg, Germany*

[265] *School of Physics Science and Engineering, Tongji University, Shanghai 200092, China*

[266] *Institut d'Estudis Espacials de Catalunya, c. Gran Capità, 2-4, 08034 Barcelona, Spain*

[267] *Columbia University, New York, NY 10027, USA*

[268] *Institucio Catalana de Recerca i Estudis Avançats (ICREA), Passeig de Lluís Companys, 23, 08010 Barcelona, Spain*

[269] *Leibniz Universität Hannover, D-30167 Hannover, Germany*

[270] *Research Center for Space Science, Advanced Research Laboratories, Tokyo City University, 3-3-1 Ushikubo-Nishi, Tsuzuki-Ku, Yokohama, Kanagawa 224-8551, Japan*

[271] *Tsinghua University, Beijing 100084, China*

[272] *Institute for Photon Science and Technology, The University of Tokyo, 2-11-16 Yayoi, Bunkyo-ku, Tokyo 113-8656, Japan*

[273] *School of Physical & Chemical Sciences, University of Canterbury, Private Bag 4800, Christchurch 8041, New Zealand*

[274] *GRAPPA, Anton Pannekoek Institute for Astronomy and Institute for High-Energy Physics, University of Amsterdam, 1098 XH Amsterdam, Netherlands*

[275] *Institut des Hautes Etudes Scientifiques, F-91440 Bures-sur-Yvette, France*

[276] *Faculty of Law, Ryukoku University, 67 Fukakusa Tsukamoto-cho, Fushimi-ku, Kyoto City, Kyoto 612-8577, Japan*

[277] *Department of Physics and Astronomy, University of Notre Dame, 225 Nieuwland Science Hall, Notre Dame, IN 46556, USA*

[278] *Phenikaa Institute for Advanced Study (PIAS), Phenikaa University, To Huu street Yen Nghia Ward, Ha Dong District, Hanoi, Vietnam*

[279] *University of Stavanger, 4021 Stavanger, Norway*

[280] *Department of Astronomy, The University of Tokyo, 7-3-1 Hongo, Bunkyo-ku, Tokyo 113-0033, Japan*

[281] *Physics Program, Graduate School of Advanced Science and Engineering, Hiroshima University, 1-3-1 Kagamiyama, Higashihiroshima City, Hiroshima 903-0213, Japan*

[282] *Observatoire Astronomique de Strasbourg, 11 Rue de l'Université, 67000 Strasbourg, France*

[283] *Observatoire de Paris, 75014 Paris, France*

[284] *Laboratoire Univers et Théories, Observatoire de Paris, 92190 Meudon, France*

[285] *National Institute for Mathematical Sciences, Daejeon 34047, Republic of Korea*

[286] *Graduate School of Science and Technology, Niigata University, 8050 Ikarashi-2-no-cho, Nishi-ku, Niigata City, Niigata 950-2181, Japan*

[287] *Niigata Study Center, The Open University of Japan, 754 Ichibancho, Asahimachi-dori, Chuo-ku, Niigata City, Niigata 951-8122, Japan*

[288] *University of Maryland, Baltimore County, Baltimore, MD 21250, USA*

[289] *CSIR-Central Glass and Ceramic Research Institute, Kolkata, West Bengal 700032, India*

[290] *Consiglio Nazionale delle Ricerche - Istituto dei Sistemi Complessi, I-00185 Roma, Italy*

[291] *Department of Physics, Aristotle University of Thessaloniki, 54124 Thessaloniki, Greece*

[292] *Department of Astronomy, Yonsei University, 50 Yonsei-Ro, Seodaemun-Gu, Seoul 03722, Republic of Korea*

[293] *Department of Physics, University of Guadalajara, Av. Revolucion 1500, Colonia Olimpica C.P. 44430, Guadalajara, Jalisco, Mexico*

[294] *Hobart and William Smith Colleges, Geneva, NY 14456, USA*

[295] *Dipartimento di Ingegneria, Università del Sannio, I-82100 Benevento, Italy*

[296] *Museo Storico della Fisica e Centro Studi e Ricerche "Enrico Fermi", I-00184 Roma, Italy*

[297] *Dipartimento di Ingegneria Industriale, Elettronica e Meccanica, Università degli Studi Roma Tre, I-00146 Roma, Italy*

[298] *Subatech, CNRS/IN2P3 - IMT Atlantique - Nantes Université, 4 rue Alfred Kastler BP 20722 44307 Nantes CÉDEX 03, France*

[299] *Universidad de Antioquia, Medellín, Colombia*

[300] *Departamento de Física - ETSIDI, Universidad Politécnica de Madrid, 28012 Madrid, Spain*

[301] *Department of Electronic Control Engineering, National Institute of Technology, Nagaoka College, 888 Nishikatakai, Nagaoka City, Niigata 940-8532, Japan*

[302] *Dipartimento di Fisica e Scienze della Terra, Università Degli Studi di Ferrara, Via Saragat, 1, 44121 Ferrara FE, Italy*

[303] *Faculty of Science, Toho University, 2-2-1 Miyama, Funabashi City, Chiba 274-8510, Japan*

[304] *Indian Institute of Technology, Palaj, Gandhinagar, Gujarat 382355, India*

[305] *Department of Physics, Indian Institute of Technology Gandhinagar, Gujarat 382055, India*

[306] *Laboratoire MSME, Cité Descartes, 5 Boulevard Descartes, Champs-sur-Marne, 77454 Marne-la-Vallée Cedex 2, France*

[307] *University of Tsukuba, 1-1-1, Tennodai, Tsukuba, Ibaraki 305-8573, Japan*

[308] *Institute for Quantum Studies, Chapman University, 1 University Dr., Orange, CA 92866, USA*

[309] *Faculty of Information Science and Technology, Osaka Institute of Technology, 1-79-1 Kitayama, Hirakata City, Osaka 573-0196, Japan*

[310] *NASA Goddard Space Flight Center, Greenbelt, MD 20771, USA*

[311] *iTHEMS (Interdisciplinary Theoretical and Mathematical Sciences Program), RIKEN, 2-1 Hirosawa, Wako, Saitama 351-0198, Japan*

[312] *Scuola Internazionale Superiore di Studi Avanzati, Via Bonomea, 265, I-34136, Trieste TS, Italy*

[313] *Laboratoire de Physique de l'École Normale Supérieure, ENS, (CNRS, Université PSL, Sorbonne Université, Université Paris Cité), F-75005 Paris, France*

[314] *Institut für Theoretische Physik, Johann Wolfgang Goethe-Universität, Max-von-Laue-Str. 1, 60438 Frankfurt am Main, Germany*

[315] *INAF, Osservatorio di Astrofisica e Scienza dello Spazio, I-40129 Bologna, Italy*

[316] *Universidade Estadual Paulista, 01140-070 São Paulo, Brazil*

[317] *School of Physics, Georgia Institute of Technology, Atlanta, Georgia 30332, USA*

[318] *Faculty of Physics, University of Warsaw, Ludwika Pasteura 5, 02-093 Warszawa, Poland*

[319] *Department of Physics, Kyoto University, Kita-Shirakawa Oiwake-cho, Sakyou-ku, Kyoto City, Kyoto 606-8502, Japan*

[320] *Yukawa Institute for Theoretical Physics (YITP), Kyoto University, Kita-Shirakawa Oiwake-cho, Sakyou-ku, Kyoto City, Kyoto 606-8502, Japan*

[321] *Carleton College, Northfield, MN 55057, USA*

[322] *University of Catania, Department of Physics and Astronomy, Via S. Sofia, 64, 95123 Catania CT, Italy*

[323] *National Institute of Technology, Fukui College, Geshi-cho, Sabae-shi, Fukui 916-8507, Japan*

[324] *Department of Communications Engineering, National Defense Academy of Japan, 1-10-20 Hashirimizu, Yokosuka City, Kanagawa 239-8686, Japan*

[325] *Texas A&M University, College Station, TX 77843, USA*

[326] *Eindhoven University of Technology, 5600 MB Eindhoven, Netherlands*

[327] *The University of Utah, Salt Lake City, UT 84112, USA*

[328] *Kavli Institute for the Physics and Mathematics of the Universe, WPI, The University of Tokyo, 5-1-5 Kashiwa-no-Ha, Kashiwa City, Chiba 277-8583, Japan*

[329] *Department of Astronomy, Beijing Normal University, Xinjiekouwai Street 19, Haidian District, Beijing 100875, China*